\documentclass[10pt]{article}
\pdfoutput=1
\usepackage{jheppub}

\hypersetup{
  pdfencoding=unicode,
  colorlinks=true,
  urlcolor=Maroon,
  linkcolor=RoyalBlue,
  citecolor=Maroon,
  pdftitle={Worldsheet string theory on \texorpdfstring{$\mathrm{AdS}_3 \times \mathbf{S}^3 \times \mathrm{K3}$}{AdS3 x S3 x K3}},
  pdfauthor={Lorenz Eberhardt, Matthias Gaberdiel, Sameer Murthy, Mukund Rangamani},
  pdfdisplaydoctitle=true,
  pdfstartview=FitH,
  linktocpage=true
}
\usepackage[usenames,dvipsnames]{xcolor}
\usepackage{amssymb}
\usepackage{amsthm}
\usepackage{mathtools}
\usepackage{dsfont}
\usepackage{pdfpages}
\usepackage{verbatim}
\usepackage{tensor}
\usepackage{slashed}
\usepackage{physics}
\usepackage{bm}
\usepackage[mathscr]{euscript} 
\usepackage{tikz}
\usepackage{float}
\usetikzlibrary{decorations.pathreplacing, decorations.markings,calc,shapes.misc,decorations.pathmorphing,patterns.meta, math}
\usepackage{listings}
\usepackage{enumitem}
\usepackage[most]{tcolorbox}
\RequirePackage{cleveref}
\crefname{section}{\S\!\!}{\S\S\!\!}
\Crefname{section}{\S}{\S\S}
\crefname{appendix}{Appendix}{Appendices}
\crefname{figure}{Fig.\!}{Figs.\!}
\crefname{table}{Table}{Tables}
\crefname{equation}{}{Eqs.\!}

\newcommand{\bd}{\begin{tikzpicture}}
\newcommand{\ed}{\end{tikzpicture}}

\renewcommand\Im{\mathop{\text{Im}}}

\renewcommand\Re{\mathop{\text{Re}}}
\allowdisplaybreaks[1]

\newcommand\SO{\text{SO}}
\newcommand\SL{\text{SL}}
\newcommand\SU{\text{SU}}

\newcommand\CC{\mathbb{C}}
\newcommand\ZZ{\mathbb{Z}}
\newcommand\RR{\mathbb{R}}
\newcommand\HH{\mathbb{H}}
\newcommand\NN{\mathbb{N}}

\renewcommand\dd{\text{d}}
\newcommand\sgn{\text{sgn}}
\newcommand\LargeRes{\mathop{\text{Res}}}

\newcommand{\dualtau}{\bm{\tau}}
\newcommand{\dualq}{\bm{q}}
\newcommand{\dualqb}{\overline{\bm{q}}}
\newcommand{\dualy}{\bm{y}}
\newcommand{\dualyt}{\widetilde{\bm{y}}}
\newcommand{\dualz}{\bm{\xi}}
\newcommand{\dualF}{\bm{F}}
\newcommand{\dualPF}{\bm{Z}}
\newcommand{\dualfE}{\bm{f}}
\newcommand{\dualEG}{\bm{E}}
\newcommand{\dualEGsp}{\bm{\mathscr{E}}}

\newcommand{\dualG}{\bm{G}}
\newcommand{\dualL}{\bm{L}}
\newcommand{\dualJ}{\bm{K}}
\newcommand{\dualh}{\bm{h}}
\newcommand{\dualj}{\bm{j}}
\newcommand{\dualhb}{\bar{\bm{h}}}
\newcommand{\dualjb}{\bar{\bm{j}}}
\newcommand{\intStrip}{\int_{\amalg}}

\newcommand{\chsl}{\mathcal{C}^{\smash[b]{^{\mathfrak{sl}(2)_{k+2}}}}}
\newcommand{\chsu}{\mathcal{C}^{\smash[b]{^{\mathfrak{su}(2)_{k-2}}}}}
\newcommand{\chpsu}{\bm{\chi}^{\smash[b]{^{\mathfrak{psu}(1,1|2)}}}}
\newcommand{\chNfour}{\bm{\chi}^{\smash[b]{^{\mathcal{N}=4}}}}
\newcommand{\nchsusl}{\bm{\slashed{\chi}}^{\smash[b]{^{\mathfrak{sl}(2)\oplus \mathfrak{su}(2)}}}}
\newcommand{\nchsu}{\slashed{\mathcal{C}}^{\smash[b]{^{\mathfrak{su}(2)_{k-2}}}}}
\newcommand{\Zsl}[1]{\mathcal{Z}^{\smash[b]{^{\mathfrak{sl}(2)}}}_{#1}}
\newcommand{\Zsu}[1]{\mathcal{Z}^{\smash[b]{^{\mathfrak{su}(2)}}}_{#1}}
\newcommand{\ZK}[1]{\mathcal{Z}^{\smash[b]{^{\scriptscriptstyle\mathrm{K3}}}}_{#1}}
\newcommand{\EK}[1]{\varphi^{\smash[b]{{\scriptscriptstyle\mathrm{K3}}}}_{#1}}

\newcommand{\llm}{L_{\ell,\, m}}
\newcommand{\llmt}{L_{\ell,\, \widetilde{m}}}

\title{Worldsheet string theory on $\boldsymbol{\mathrm{AdS}_3 \times \mathbf{S}^3 \times \mathrm{K3}}$}

\author{Lorenz Eberhardt${}^{a}$, Matthias R.~Gaberdiel${}^{b,c}$, Sameer Murthy${}^{d}$,
  Mukund Rangamani${}^{e}$}

\affiliation[a]{Institute for Theoretical Physics, University of Amsterdam, Amsterdam, 1098XH, NL}
\affiliation[b]{Institut f\"ur Theoretische Physik, ETH Z\"urich,\\
  Wolfgang-Pauli-Strasse 27, 8093 Z\"urich, Switzerland}
\affiliation[c]{
  Kavli Institute for Theoretical Sciences (KITS) and School of Quantum,\\
  University of the Chinese Academy of Sciences, Beijing 100190, China}
\affiliation[d]{Department of Mathematics, King’s College London, The Strand, London WC2R 2LS, UK}
\affiliation[e]{Center for Quantum Mathematics and Physics (QMAP)\\
  Department of Physics \& Astronomy, University of California, Davis, CA 95616 USA}

\emailAdd{gaberdiel@itp.phys.ethz.ch}
\emailAdd{l.eberhardt@uva.nl}
\emailAdd{sameer.murthy@kcl.ac.uk}
\emailAdd{mukund@physics.ucdavis.edu}

\abstract{
  The RNS formalism describing string theory on $\mathrm{AdS}_3 \times \mathbf{S}^3 \times \mathrm{K3}$ supported by pure NS-NS flux is in principle straightforward to write down. One can directly check that the worldsheet theory reproduces the correct chiral spectrum expected from supergravity by analyzing the BRST cohomology. However, when one attempts to construct the worldsheet partition function using the standard rules of the RNS formalism by appropriately summing over spin structures to ensure worldsheet modular invariance, one finds a result in tension with spacetime supersymmetry. We explain why this is the case, and, more importantly, propose a resolution. Specifically, we fix the physical state projection for the partition function uniquely by requiring spacetime supersymmetry. Perhaps surprisingly, the projected worldsheet partition function does not factorize into a bosonic and a fermionic part and involves functions that are not modular covariant. As a consistency check we show that our prescription for the worldsheet partition function reproduces the supergravity spectrum for both Type IIB and IIA theories. Furthermore, we compute the elliptic genus of the dual CFT from the worldsheet, which has both a holomorphic short string and a non-holomorphic long string contribution. The latter displays characteristics of a mock modular function, albeit with some key differences, while the former has the tantalizing structure of a symmetric orbifold.   We also point out  the worldsheet construction of a non-supersymmetric theory, whose spectrum is free of tachyons.
}

\begin{document}

\maketitle

\makeatletter
\g@addto@macro\bfseries{\boldmath}
\makeatother

\section{Introduction}

String perturbation theory with pure NS-NS background is well known to admit a tractable description in the RNS formalism. Many such backgrounds have been explored, e.g., the phenomenologically relevant $\RR^{1,3} \times \mathcal{M}_6$, or the $\mathrm{AdS}_3 \times \mathbf{S}^3 \times \mathcal{M}_4$ geometries arising in AdS$_3$/CFT$_2$ holography~\cite{deBoer:1998kjm}.
The latter is most frequently analyzed for $\mathcal{M}_4=\mathbb{T}^4$, which is well known to support small $\mathcal{N}=(4,4)$ superconformal symmetry and is on the same conformal manifold as the symmetric orbifold $\text{Sym}^N(\mathbb{T}^4)$~\cite{Maldacena:1997re, Larsen:1999uk, Eberhardt:2018ouy, Eberhardt:2019ywk, Dei:2020zui}. The last decade has also seen renewed interest in the case $\mathcal{M}_4=\mathbf{S}^3 \times \mathbf{S}^1$ that features large $\mathcal{N}=(4,4)$ superconformal symmetry~\cite{Elitzur:1998mm, deBoer:1999gea, Gukov:2004ym, Tong:2014yna, Eberhardt:2017pty,Witten:2024yod, Murthy:2025moj}.
The final maximally supersymmetric example is the case $\mathcal{M}_4=\text{K3}$, which is similarly conjectured to be on the same moduli space as $\text{Sym}^N(\text{K3})$~\cite{Maldacena:1997re, Dijkgraaf:1998gf,Aharony:1999ti,  Seiberg:1999xz}. Its worldsheet theory is much less studied than that of $\mathcal{M}_4=\mathbb{T}^4$. Since a particular realization of $\text{K3}$ is given by the orbifold $\mathbb{T}^4/\ZZ_2$, it is commonly assumed that the passage to $\text{K3}$ is straightforward. Indeed, we are not aware of dedicated papers on the worldsheet theory of this background.

There is, however, a marked difference between $\mathbb{T}^4$ and $\mathrm{K3}$. For $\mathbb{T}^4$, one can T-dualize on any of the four torus directions and therefore Type IIA and IIB string theory on this background are equivalent. Not so for $\mathrm{K3}$, where the Type IIB background has $\mathcal{N}=(4,4)$ supersymmetry, while the Type IIA background only has $\mathcal{N}=(0,4)$ supersymmetry.
The difference between the Type IIA and IIB theories lies in the worldsheet GSO projection, so a choice must be made for the $\mathrm{AdS}_3 \times \mathbf{S}^3 \times \mathrm{K3}$ background.

One of the simplest observables of the background $\mathrm{AdS}_3 \times \mathbf{S}^3 \times \mathrm{K3}$ for which the difference between IIA and IIB string theory matters is the thermal partition function, which via holography is identified with the
partition function of the dual CFT. It can be computed in two ways: (i) extracting the physical string spectrum on the global $\text{AdS}$ background by imposing physical state conditions on the worldsheet, or (ii) computing the torus partition function in the thermal $\text{AdS}$ background.
Both computations have been carried out in the literature for the cases $\mathcal{M}_4=\mathbb{T}^4$ and $\mathcal{M}_4=\mathbf{S}^3 \times \mathbf{S}^1$, see~\cite{Giveon:1998ns, Maldacena:2000hw, Eberhardt:2018ouy} for the former and~\cite{Gawedzki:1991yu, Maldacena:2000kv, Ashok:2020dnc, Eberhardt:2020bgq,Ferko:2024uxi,Murthy:2025moj} for the latter.
They both become, however, subtle in the present setting. For (i), one has to give a precise meaning to the left- and right-moving worldsheet fermion number operators $(-1)^{\text{F}_\text{L}}$ and $(-1)^{\text{F}_\text{R}}$, which are a priori not canonically defined on the physical spectrum. For (ii), one naively writes down the product of matter and ghost contributions to the worldsheet path integral and sums over spin structures. This, however, fails in the Ramond sector with $(-1)^\text{F}$ insertions (for either left- or right-movers), since the $\beta\gamma$ ghost contribution is formally $\big|\frac{\eta(\tau)}{\vartheta_1(\tau)}\big|^2$, corresponding to the fact that the $\beta\gamma$ ghosts should cancel two real fermionic oscillator degrees of freedom. However, $\vartheta_1(\tau)=0$ as a consequence of the zero-modes of the $\beta\gamma$ ghosts and thus the contribution is ill-defined. It is tempting to cancel the problematic $\vartheta_1$ factor with the corresponding factor that appears in the numerator for the free fermions, since this matches the intuition from light-cone gauge fixing. It is, however, a priori not clear that this is correct since in $\mathrm{AdS}_3 \times \mathbf{S}^3 \times \mathrm{K3}$, it is not possible to fix light-cone gauge.
In fact, as we will show in this paper, such a naive treatment of the ghosts is \emph{incorrect}! The physical worldsheet partition function does not and cannot factorize into contributions from the different factors making up the worldsheet CFT.\footnote{The one-loop partition function is obviously ill-defined on backgrounds where the worldsheet trace in the Ramond sector does not have fermion zero modes. As recently discussed in~\cite{Sen:2026kyn,Sen:2026xfy}, strings with target space $\mathcal{X}_4 \times \mathcal{M}_6$ with $\mathcal{X}_4$ a four-dimensional black hole background and $\mathcal{M}_6$ having non-vanishing Euler characteristic suffer from this phenomenon. We thank Ashoke Sen for highlighting this and for helpful discussions.}
\medskip

This problem makes it non-trivial to determine the correct physical spectrum of the background. The relevance of the background to AdS/CFT holography then makes it particularly pressing to revisit the problem, especially since it possesses rich protected invariants such as the elliptic genus~\cite{Eguchi:1988vra}. In this paper, we make a proposal for consistent physical state conditions. Our motivation for this proposal is two-fold: we evaluate the covariant physical state conditions from BRST-closedness on the massless spectrum, and we demand consistency with spacetime supersymmetry (for Type IIB). Either input alone already yields a unique answer, given in Eq.~\eqref{eq:final-type-IIB-partition-function}. Perhaps surprisingly, the worldsheet partition function features objects that are individually not modular covariant. Modular invariance of the final expression is only guaranteed because the worldsheet partition function includes a sum over winding sectors of the worldsheet torus around the spacetime thermal cycle.

We discuss a number of checks and applications of our proposal. First, we extract the $\frac{1}{2}$-BPS spectrum from the formula and demonstrate that it does agree with the expectations from supergravity, as computed in~\cite{deBoer:1998kjm, Deger:1998nm, Eloy:2020uix}. This would already fail for the `naive' physical state projection (cf., comments in~\cite{Ferko:2024uxi}). We also extract the divergent part of the thermal partition function that originates from the long-string sector and can be directly matched to the long-string CFT~\cite{Seiberg:1999xz, Eberhardt:2019qcl}. This CFT describes the free propagation of long strings, and its marginal non-normalizable deformation gives the full dual CFT~\cite{Balthazar:2021xeh, Eberhardt:2021vsx}. The long-string CFT is insensitive to the distinction between Type IIA and IIB; nevertheless, the computation shows that our proposal preserves this sector and gives direct evidence for the duality in this case.

Finally, and most non-trivially, we compute the full string-theory elliptic genus of the background. This had so far only been computed in supergravity~\cite{deBoer:1998us}. As expected, the elliptic genus receives holomorphic contributions from short string states, which arrange themselves into a formula that looks tantalizingly close to the DMVV formula for the elliptic genus of the symmetric orbifold~\cite{Dijkgraaf:1996xw}. In addition, we find contributions from the long-string sector. Even though the long-string sector contains no BPS states, it still contributes a finite amount to the elliptic genus: the boson--fermion cancellation expected for a continuum of non-BPS states is spoiled by the divergence of the density of states, leaving a finite remainder. This leads to the appearance of mock modular quantities in the final result~\cite{Zwegers:2002kt}.
A similar structure is well known to emerge in the elliptic genus of $\mathcal{N}=2$
Liouville theory and non-compact SCFTs~\cite{Troost:2010ud, Eguchi:2010cb, Ashok:2011cy,
  Murthy:2013mya,Ashok:2013pya,Harvey:2014nha}.
The elliptic genus cannot be directly compared with a dual CFT because the computation is done at the singular NS-NS point in moduli space and a small perturbation is expected to lift the long-string continuum~\cite{Seiberg:1999xz, Eberhardt:2018vho, Cho:2018nfn}. It is currently unknown how to perform a similar computation for the conjectural CFT dual~\cite{Eberhardt:2021vsx, Yu:2025qnw} away from the tensionless limit~\cite{Eberhardt:2018ouy, Eberhardt:2019ywk}.
\medskip

The paper is organized as follows. In~\cref{sec:background}, we begin with a lightning review of the background and establish our notation and conventions. We then recall the work of~\cite{Maldacena:2000hw, Giveon:1998ns} on the canonical quantization of the background, discuss the setup for the thermal worldsheet partition function, and finally write down the full partition function as suggested by the light-cone-gauge intuition, which is shown to be inconsistent. In~\cref{sec:proposal}, we make a corrected proposal, informed by canonical quantization and the preservation of spacetime supersymmetry. Subsequently, in~\cref{sec:checks}, we discuss the checks mentioned above and also generalize the analysis to the Type IIA background. \cref{sec:NS-elliptic-genus} is devoted to the computation of the spacetime elliptic genus. It is the most technically involved part of the paper. Finally, we end with a short discussion. Some of the more technical aspects are relegated to three appendices.

\paragraph{Notation:}
To orient the reader, we collect in~\cref{tab:ws-notation,tab:st-notation} the notation used throughout this paper. Worldsheet quantities are set in standard font, while spacetime/dual CFT quantities are distinguished by bold italic symbols.

\begin{table}[H]
  \centering
  \caption{Worldsheet quantities appearing in our discussion. \medskip}
  \label{tab:ws-notation}
  \renewcommand{\arraystretch}{1.25}
  \small
  \begin{tabular}{@{}lp{10.5cm}@{}}
    \hline\hline
    $\tau=\tau_1+i\,\tau_2$ & Worldsheet modular parameter \\
    $q=\mathrm{e}^{2\pi i\tau}$ & Worldsheet nome \\
    $k$ & WZW level \\
    $a,\,\widetilde{a}\in\{1,2,3,4\}$ & Spin structure labels: $\{1,2,3,4\}\leftrightarrow\{\widetilde{\mathrm{R}},\mathrm{R},\mathrm{NS},\widetilde{\mathrm{NS}}\}$ \\
    $\alpha,\,\widetilde{\alpha}$ & GSO projection signs (left-/right-movers) \\[4pt]
    \multicolumn{2}{c}{\textit{Fields and primaries}} \\[2pt]
    $J^A,\;\mathcal{J}^A,\;\psi^A$ & $\mathfrak{sl}(2,\RR)$ super-WZW, bosonic ($\mathfrak{sl}(2,\RR)_{k+2}$), and fermionic currents ($A\in\{\pm,3\}$) \\
    $K^a,\;\mathcal{K}^a,\;\chi^a$ & $\mathfrak{su}(2)$ super-WZW, bosonic ($\mathfrak{su}(2)_{k-2}$), and fermionic currents ($a\in\{\pm,3\}$) \\
    $Y^i,\;\lambda^i$ & K3 ($\mathbb{T}^4$) bosons and fermions ($i\in\{1,2,3,4\}$) \\
    $H_i\;(i=1,\dots,5)$ & Bosonized fermion fields \\
    $\varphi$ & Bosonized $\beta\gamma$ ghost \\
    $b,c;\;\beta,\gamma$ & Reparameterization and superconformal ghosts \\
    $\Phi_j(z,x)$ & $\mathfrak{sl}(2,\RR)_{k+2}$ primary of spin $j$ \\
    $V_\ell(z,y)$ & $\mathfrak{su}(2)_{k-2}$ primary of spin $\ell$ \\
    $\sigma,\;\overline{\sigma}$ & K3 bosonic twist fields ($h=\bar{h}=\tfrac{1}{4}$) \\
    $S_\pm(x,y)$ & $\mathfrak{sl}(2,\RR)\times\mathfrak{su}(2)$ spin fields \\
    $\mathsf{s}_\pm^{\mathsf{A}}$ & K3/$\mathbb{T}^4$ spin fields \\
    $x,\overline{x};\;y,\tilde{y}$ & $\mathfrak{sl}(2,\RR)$ and $\mathfrak{su}(2)$ fugacity variables \\[4pt]
    \multicolumn{2}{c}{\textit{Characters and partition functions}} \\[2pt]
    $\chsl_j(\dualtau\mid\tau)$ & $\mathfrak{sl}(2,\RR)_{k+2}$ affine character \\
    $\chsu_\ell(\dualz\mid\tau)$ & $\mathfrak{su}(2)_{k-2}$ affine character \\
    $\nchsu_\ell(\dualz\mid\tau)$ & Non-supersymmetric $\mathfrak{su}(2)_{k-2}$ character \\
    $\Zsl{a,\widetilde{a}}(\dualtau\mid\tau)$ & $\mathfrak{sl}(2,\RR)$ super-WZW partition function \\
    $\Zsu{a,\widetilde{a}}(\dualz\mid\tau)$ & $\mathfrak{su}(2)$ super-WZW partition function \\
    $\ZK{a,\widetilde{a}}(\tau)$ & K3 sigma model partition function \\
    $\mathcal{Z}_{a,\widetilde{a}}$ & Full worldsheet partition function (fixed spin structure) \\[4pt]
    \multicolumn{2}{c}{\textit{Special functions and shorthands}} \\[2pt]
    $\vartheta_a(z\mid\tau)$ & Jacobi theta functions ($a=1,2,3,4$) \\
    $\eta(\tau)$ & Dedekind eta function \\
    $\Theta_{k,\ell}(\dualz\mid\tau)$ & Level-$k$ theta function \\
    $\Psi_{k,\ell}(\dualz\mid\tau)$ & False theta function \\
    $\llm\equiv\ell+k\,m+\tfrac{1}{2}$ & Shorthand for level-$k$ theta function index \\
    $\EK{}(\dualz\mid\tau)$ & K3 elliptic genus \\
    \multicolumn{2}{c}{\textit{Notational shorthands}} \\[2pt]
    $\intStrip$ & Integration over the strip $\Re(\tau) \in [-\frac{1}{2},\frac{1}{2}]$
    and $\Im(\tau) \in [0,\infty)$\\
    $\sum_{\ell = 0}^{\frac{k}{2} -1}$ & Summation over allowed $\mathfrak{su}(2)$ spins with $\ell \in \{0, \frac{1}{2}, \cdots , \frac{k}{2} -1\}$\\
    \hline\hline
  \end{tabular}
\end{table}

\begin{table}[!thbp]
  \centering
  \caption{Spacetime (dual CFT) quantities of interest set in bold italic symbols ($\bm{\cdot}$).}
  \label{tab:st-notation}
  \renewcommand{\arraystretch}{1.25}
  \small
  \begin{tabular}{@{}lp{10.5cm}@{}}
    \hline\hline
    $\dualtau$ & Spacetime modular parameter \\
    $\dualq=\mathrm{e}^{2\pi i\,\dualtau}$,\; $\dualqb=\mathrm{e}^{-2\pi i\,\overline{\dualtau}}$ & Spacetime nome and conjugate \\
    $\dualz,\;\widetilde{\dualz}$ & $\SU(2)_L$ and $\SU(2)_R$ chemical potentials \\
    $\dualy=\mathrm{e}^{2\pi i\,\dualz}$,\; $\dualyt=\mathrm{e}^{-2\pi i\,\widetilde{\dualz}}$ & Fugacities for $\mathrm{Spin}(4)$  \\
    $\dualF$ & Spacetime fermion number \\
    $\dualh,\;\dualhb$ & Dual CFT (anti-)holomorphic conformal weights \\
    $\dualj,\;\dualjb$ & Dual CFT (anti-)holomorphic R-charges \\
    $\dualG^{\alpha\,A}_{r}$ & Spacetime supercharges \\
    $\dualL_n$ & Spacetime Virasoro generators ($n\in\{0,\pm 1\}$) \\
    $\dualJ^a$ & Spacetime $\SU(2)$ R-symmetry currents \\
    $\dualPF$ & Multi-string spacetime partition function \\
    $\dualfE$ & Single-string spacetime free energy \\
    $\dualEG$ & Spacetime elliptic genus \\
    $\dualEGsp$ & Single-string contribution to the spacetime elliptic genus \\
    $\chpsu_j(\dualz,\,\dualtau)$ & Short character of $\mathfrak{psu}(1,1|2)$ \\
    $\nchsusl_j(\dualz,\dualtau)$ & Non-supersymmetric character combining $\mathfrak{sl}(2)$ and
    $\mathfrak{su}(2)$ representations \\
    $\chi_j(\dualz)$ & Spin-$j$ $\mathfrak{su}(2)$ character \\
    \hline\hline
  \end{tabular}
\end{table}

\section{Background: RNS strings in \texorpdfstring{$\mathrm{AdS}_{3} \times \mathbf{S}^3 \times \mathrm{K3}$}{AdS3 x S3 x K3}}\label{sec:background}

\subsection{The ingredients of the worldsheet theory}\label{subsec:worldsheet-theory}

To set up our conventions, we begin in this section by reviewing the ingredients needed to define the worldsheet theory in the RNS formalism on $\mathrm{AdS}_3 \times \mathbf{S}^3 \times \mathrm{K3}$ with pure NS-NS flux. Most of the following is standard material; we refer to~\cite{Giveon:1998ns, Maldacena:2000hw, Maldacena:2000kv}  for additional background.

\paragraph{Worldsheet fields:} The worldsheet theory is a product of the following conformal field theories:
\begin{itemize}[wide,left=0pt]
  \item[1.] The $\mathfrak{sl}(2,\mathbb{R})$ super-WZW model at level $k$. The supersymmetric currents $J^A$ can be decoupled from the fermions into the following commuting fields: bosonic currents $\mathcal{J}^A$ defining an $\mathfrak{sl}(2,\mathbb{R})_{k+2}$ algebra and fermions $\psi^A$ in the adjoint representation, whose bilinears define an $\mathfrak{sl}(2,\mathbb{R})_{-2}$ current algebra. Here, $A\in \{\pm,3\}$ denote the $\mathfrak{sl}(2,\mathbb{R})$ adjoint indices.
  \item[2.] The $\mathfrak{su}(2)$ super-WZW model at level $k$, whose currents $K^a$ at level $k$ can similarly be decoupled from the fermions $\chi^a$. The decoupled currents $\mathcal{K}^a$ define an $\mathfrak{su}(2)_{k-2}$ current algebra. Here, $a\in \{\pm,3\}$ denote the $\mathfrak{su}(2)$ adjoint indices.
  \item[3.] A $\mathrm{K3}$ sigma model with $\mathcal{N}=(4,4)$ supersymmetry. We will be able to work at a generic point in the $\mathrm{K3}$ moduli space, although it is occasionally useful to keep in mind that a specific example can be constructed by taking orbifolds of $\mathbb{T}^4$, most notably $\mathbb{T}^4/\ZZ_2$. The $\mathbb{T}^4$ CFT is characterized by free bosons $Y^i$ and fermions $\lambda^i$ with $i\in \{1,2,3,4\}$. The  $\mathbb{Z}_2$ action is $Y^i \to - Y^i$ and $\lambda^i \to - \lambda^i$.
  \item[4.] The standard $bc$ and $\beta\gamma$ ghosts of the RNS formalism.
\end{itemize}
The total central charge adds up to zero as follows,
\begin{equation}
\pqty{\dfrac{3(k+2)}{k}+\dfrac{3}{2} }+ \pqty{\dfrac{3(k-2)}{k}+\dfrac{3}{2}}  +6+\big(-26+11\big)=0\,,
\end{equation}
with the four terms corresponding to the four constituents described above.

\paragraph{Worldsheet primaries:}  The primaries of interest on the worldsheet are
\begin{itemize}[wide,left=0pt]
  \item  The $\mathfrak{sl}(2,\mathbb{R})_{k+2}$ primaries are\footnote{For simplicity, we largely suppress the dependence on both the worldsheet antiholomorphic coordinate and the spacetime right-moving fugacities. We also work with the conventions that $\dualL_0 = -J^3_0$, $\dualL_{\pm 1} = -J^\pm_{0}$.}
  $\Phi_j(z,x)$ with worldsheet conformal dimension $\Delta = -\frac{j(j-1)}{k}$. In the NS sector the fermion operator $\psi(z,x) = -\psi^+(z) + 2\, x\, \psi^3(z) - x^2\, \psi^-(z)$ transforms in the adjoint representation of  $\mathfrak{sl}(2,\mathbb{R})$. It has worldsheet conformal weight equal to $\frac{1}{2}$, and spacetime conformal weight $h=-1$, where $h$ is the eigenvalue of $-J^3_0$ at $x=0$. Here the spacetime $\mathfrak{sl}(2,\mathbb{R})$ generators act as
  $[J^A_0,\psi(z,x)] = D^A_{h}\, \psi(z,x)$ with
  \begin{align}\label{sl2R}
  D^-_h = - \partial_x \ , \quad D^3_h =  - x\partial_x  - h  \ , \quad D^+_h = - x^2\partial_x - 2 h x \ , \qquad (h=-1)  \ .
  \end{align}
  The combination $\psi(z,x)\, \Phi_{j+1}(z,x)$ then transforms in the spin-$j$ representation of the  $\mathfrak{sl}(2,\mathbb{R})$ algebra generated by $J^A_0$.
  \item Similarly, $\mathfrak{su}(2)_{k-2}$ has primaries $V_\ell(z,y)$ with worldsheet conformal dimension $\Delta = \frac{\ell\,(\ell+1)}{k}$, and the super-WZW model has fermionic primaries $\chi(z,y) = -\chi^+(z) + 2\,y\, \chi^3(z) + y^2\, \chi^-(z) $ that transform in the adjoint representation of $\mathfrak{su}(2)$ and have  spacetime R-charge $m=+1$, where $m$ is the eigenvalue of $K^3_0$ at $y=0$. Here the spacetime $\mathfrak{su}(2)$ generators act as $[K^A_0,\chi(z,y)] = P^A_{m}\, \chi(z,y)$, cf.\ (\ref{sl2R})
  \begin{align}\label{su2}
  P^-_m = \partial_y \ ,\quad P^3_m = - y \partial_y + m \ , \quad P^+_m = - y^2\partial_y + 2 m y \ , \qquad (m=+1)  \ .
  \end{align}
  The combination $\chi(z,y) \, V_\ell(z,y)$ then transforms in the spin-$(\ell +1)$ representation of the $\mathfrak{su}(2)$ algebra generated by $K^A_0$.
  \item   The K3 CFT has operators inherited from the untwisted sector of the $\mathbb{T}^4$ CFT. In addition, we have bosonic $\mathbb{Z}_2$ twist fields $\sigma$, $\overline{\sigma}$ of weights $h = \frac{1}{4}$ and $\overline{h} = \frac{1}{4}$, respectively. These twist fields can be dressed with the fermionic zero modes which are present in the twisted NS sector, to give weight $\frac{1}{2}$ primaries (the fermionic part of the vertex operator can be built from the bosons $H_{4,5}$ introduced below).
  \item To describe the spacetime supercharges and the R-sector vertex operators, we introduce the following bosonic fields
  \begin{equation}
  \begin{aligned}
  \partial H_1
   & = \frac{2}{k}\, \psi^2\, \psi^1  \,,
   & \qquad
  \partial H_2
   & = \frac{2}{k}\, \chi^2\, \chi^1  \,,
   & \qquad
  \partial H_3
   & =  \frac{2i}{k}\, \chi^3\, \psi^3  \,,
  \\
  \partial H_4
   & = \lambda^2\, \lambda^1  \,,
   & \qquad
  \partial H_5
   & =  \lambda^4\, \lambda^3 \,.
   & \qquad
   &
  \end{aligned}
  \end{equation}
  These combinations have free field OPEs, $H_i(z)\, H_j(w) \sim - \delta_{ij}\, \log(z-w)$. We refer the reader to~\cite{Dabholkar:2007ey} for cocycle factors, which won't play a role in our analysis.
  \item We use $\varphi$ for the bosonized description of the $\beta\gamma$ system to construct vertex operators in the appropriate picture.
\end{itemize}

\subsection{Physical states in the BRST cohomology}\label{subsec:BRST}

The worldsheet spectrum can be derived by imposing the standard physical state conditions as described in~\cite{Giveon:1998ns,Maldacena:2000hw}. We will work out the case for the Type IIB theory below, and defer discussion of other chiral GSO projections to later. To be specific, let us consider the case where K3 is the $\mathbb{Z}_2$ orbifold of $\mathbb{T}^4$.

\paragraph{The spacetime superisometry algebra:}
To describe the spectrum succinctly it is helpful to record the generators of the spacetime superisometry algebra on the worldsheet. First, the spacetime supercharges are
\begin{equation}
Q = \oint \, dz\, e^{-\frac{\varphi}{2}}\,   \exp(\frac{i}{2}\, \sum_{i=1}^5\, \epsilon_i\, H_i) \,,
\end{equation}
constrained by
\begin{equation}\label{eq:gksgso}
\prod_{i=1}^5\, \epsilon_i =  \prod_{i=1}^3\, \epsilon_i  = 1\,.
\end{equation}
The first of these conditions is the standard GSO projection on the spacetime supercharges. The second arises by considering the locality of the OPE between the supercharges and the worldsheet superconformal current~\cite{Giveon:1998ns}. This additional projection ensures that the target spacetime preserves 8 of the total 32 supercharges.

Of the 8 supercharges, the states with $\epsilon_1 = -1$ correspond to the spacetime lowering generators $\dualG^{\alpha\,\mathsf{A}}_{-\frac{1}{2}}$. Here $\alpha\in\{\pm\}$ is a spinor index with respect to spacetime R-symmetry $\mathrm{SU}(2)_L$, while $\mathsf{A}\in \{\pm\}$  is a spinor index with respect to the outer-automorphism symmetry $\mathrm{SU}(2)_{\mathrm{aut}}$.
\begin{equation}\label{eq:dual-Gsusy}
\begin{aligned}
\dualG_{-\frac{1}{2}}^{\alpha\,\mathsf{A}}
 & =
\oint\,dz\, e^{-\frac{\varphi}{2}} \,
e^{\frac{i}{2} (-H_1 + \alpha\, H_2 -\alpha H_3)}\, e^{\frac{i}{2}\,\mathsf{A} \, (H_4 + H_5)} \,,
\\
\dualG_{\frac{1}{2}}^{\alpha\,\mathsf{A}}
 & =
\oint\,dz\, e^{-\frac{\varphi}{2}} \,
e^{\frac{i}{2} (H_1 + \alpha\, H_2 +\alpha H_3)}\, e^{\frac{i}{2}\,\mathsf{A} \, (H_4 + H_5)} \,.
\end{aligned}
\end{equation}
The spacetime $\mathfrak{psu}(1,1|2)$ NS superalgebra is generated by $\dualG_r^{\alpha\, \mathsf{A}}$ given above, along with
\begin{equation}
\begin{aligned}
\dualL_0 = -\oint\, dz\ \, J^3(z)\,,
\qquad \dualL_{\pm 1} = - \oint\, dz\, J^{\pm}(z) \,, \qquad
\dualJ^a = -\oint\, dz\, K^a(z)\,.
\end{aligned}
\end{equation}
Notice that the supercharges are identical for the $\mathbb{T}^4$ and K3 compactification---they survive the
orbifold projection, which we take to act as
$H_4 \to H_4 + \pi$ and $H_5 \to H_5 - \pi$ for the two bosonized spin fields.

\paragraph{The chiral spectrum:}
Having identified the spacetime superisometry algebra, we may restrict ourselves to the highest weight states with respect to this (global) superisometry algebra;
the remaining states can then be obtained from them by the action of the supercharge creation operators $\dualG_r^{\alpha\,\mathsf{A}}$. For brevity, we shall also focus on the spacetime chiral primaries in the following, i.e., the states that  are annihilated by $\dualG_{-\frac{1}{2}}^{+\, \mathsf{A}}$ and that
give rise to short representations of $\mathfrak{psu}(1,1|2)$. We will concentrate on the K3 case (which is largely left implicit in the literature); for the case of $\mathbb{T}^4$, many details are spelled out in~\cite{Kutasov:1998zh} and~\cite{Dabholkar:2007ey}.

Imposing the physical state conditions, the spacetime chiral primaries originate from the worldsheet sectors associated to
\begin{equation}\label{eq:worldsheetsector}
\mathcal{O}_j(x,\overline{x},y,\overline{y}) = \Phi_{j+1}(x,\overline{x})\, V_j(y,\overline{y}) \,, \qquad
\text{ for }
j =  0, \frac{1}{2},1,\cdots, \frac{k}{2}-1  \,.
\end{equation}
These operators can be dressed with the fermionic fields $\psi$ and $\chi$ (and their antiholomorphic counterparts) from the $\mathfrak{sl}(2,\mathbb{R})$ or $\mathfrak{su}(2)$ current algebras in the NS sector, or with spin fields in the Ramond sector. For example, in the untwisted NS sector, they are explicitly described by
\begin{equation}\label{eq:worldsheet}
\begin{aligned}
e^{-\varphi - \widetilde{\varphi}} \,
\mathcal{O}_{j}(x,\overline{x},y,\overline{y})\, \chi(y)\, \overline{\psi}(\overline{x})\,,
\quad
 &
\quad
e^{-\varphi - \widetilde{\varphi}} \,
\mathcal{O}_{j}(x,\overline{x},y,\overline{y})\, \psi(x)\, \overline{\chi}(\overline{y})\,,
\\
e^{-\varphi - \widetilde{\varphi}} \,  \mathcal{O}_{j}(x,\overline{x},y,\overline{y})\, \psi(x)\, \overline{\psi}(\overline{x})\,,
\quad
 &
\quad
e^{-\varphi - \widetilde{\varphi}} \,
\mathcal{O}_{j}(x,\overline{x},y,\overline{y})\, \chi(y)\, \overline{\chi}(\overline{y})\,.
\end{aligned}
\end{equation}
Here the states in the first line fill out a spacetime spin-$2$ supermultiplet, with the $j=0$ states comprising the boundary supergraviton multiplet. The states in the second line fill out spin-$1$ supermultiplets.

In the twisted NS sector, the K3 factor contributes at least conformal dimension $h=\frac{1}{2}$---the twisted sector ground states associated to the $16$ fixed points of $\mathbb{T}^4/\mathbb{Z}_2$ each have conformal dimension $\frac{1}{2}$. (Since there are now fermionic zero modes for the fermions associated to $\mathbb{T}^4/\mathbb{Z}_2$, there are in fact four such ground states for each fixed point.) These $h=\frac{1}{2}$ states already saturate the mass-shell condition, and we now cannot dress the ground states of~\eqref{eq:worldsheetsector} with the $\mathfrak{sl}(2,\mathbb{R})$ or $\mathfrak{su}(2)$ fermions as in~\eqref{eq:worldsheet}. As a consequence, there are no chiral primary states arising from the twisted NS sector.
\smallskip

Let us now turn to the Ramond sector states. For their description, it is useful to introduce the spin field combinations
\begin{equation}\label{eq:spinslsu}
\begin{aligned}
S_+(x,y)
 & =
- i\, x\,y\, e^{\frac{i}{2} (-H_1 - H_2 + H_3)}
- x\,  e^{\frac{i}{2} (-H_1 + H_2 - H_3)}
+ i\,y\, e^{\frac{i}{2} (H_1 - H_2 - H_3)}
+ e^{\frac{i}{2} (H_1 + H_2 + H_3)}
\\
S_-(x,y)
 & =
i\, x\,y\, e^{\frac{i}{2} (-H_1 - H_2 - H_3)}
+ x\,  e^{\frac{i}{2} (-H_1 + H_2 + H_3)}
+ i\,y\, e^{\frac{i}{2} (H_1 - H_2 + H_3)}
+ e^{\frac{i}{2} (H_1 + H_2 - H_3)} \,.
\end{aligned}
\end{equation}
The fields $S_\pm$ satisfy $\prod_{i=1}^3 \epsilon_i = \pm 1$, respectively, and for fixed $\pm$, transform in the $(\mathbf{2},\mathbf{2})$ representation with respect to (\ref{sl2R}) and (\ref{su2}), where now $h=-\frac{1}{2}$ and $m=\frac{1}{2}$.\footnote{At $x=y=0$ they define the highest weight state of this representation with eigenvalue $J^3_0 = K^3_0 = \frac{1}{2}$.}
Similarly, for the K3/$\mathbb{T}^4$ fermions let
\begin{equation}\label{eq:spinint}
\mathsf{s}_+^\mathsf{A}
= e^{\frac{i}{2}\, \mathsf{A}\,  (H_4 + H_5)}
\,, \qquad
\mathsf{s}_-^\mathsf{A}
= e^{\frac{i}{2}\, \mathsf{A}\,  (H_4 - H_5)}  \,.
\end{equation}
In the untwisted R sector we then have the physical chiral primary states
\begin{equation}\label{eq:wsRuntwist}
e^{-\frac{1}{2}\,\varphi - \frac{1}{2}\, \widetilde{\varphi}} \,
\mathcal{O}_{j}(x,\overline{x},y,\overline{y})\,
S_-(x,y)\, \overline{S}_-(\overline{x},\overline{y})\,
\mathsf{s}_-^\mathsf{A} \, \overline{\mathsf{s}}_-^\mathsf{B}\,.
\end{equation}
The choice of the polarization in the spin fields is dictated by locality with the worldsheet supercurrent generator.

In the twisted R sector, the ground states $\sigma_i$, $i=1,\ldots, 16$ of the $16$ fixed points of $\mathbb{T}^4/\mathbb{Z}_2$ all have $h=\frac{1}{4}$. (In the twisted R sector there are no fermionic zero modes associated to $\mathbb{T}^4/\mathbb{Z}_2$; thus there is a unique ground state in each twisted sector.) These states can now be combined as in (\ref{eq:wsRuntwist}), leading to the chiral primary states of the form
\begin{equation}\label{eq:wsRtwist}
e^{-\frac{1}{2}\,\varphi - \frac{1}{2}\, \widetilde{\varphi}} \,
\mathcal{O}_{j}(x,\overline{x},y,\overline{y})\,
S_-(x,y)\, \overline{S}_-(\overline{x},\overline{y})\;
\sigma_i \,,
\end{equation}
where $i=1,\ldots,16$, one for each fixed point. Together with the 4 states in~\eqref{eq:wsRuntwist} they make up the expected states from the middle dimension cohomology of K3. Once again these states are annihilated by  $\dualG_{-\frac{1}{2}}^{+\,\mathsf{A}}$. At this point, one can check that this worldsheet spectrum agrees with that obtained from Type IIB supergravity compactified on K3 (in the large $k$ limit).

\subsection{Partition functions}\label{subsec:pfns}

The physical states on the worldsheet are straightforward to describe using the decoupled description of the super-WZW models, as illustrated above. We now turn to the question of deriving this spectrum from a path integral perspective, assembling the necessary pieces first.

\paragraph{Thermal AdS$_3$:} The worldsheet fields introduced so far describe string theory on global $\mathrm{AdS}_3 \times \mathbf{S}^3 \times \mathrm{K3}$. To compute the spacetime partition function, we will need to Wick rotate to Euclidean $\mathrm{AdS}_3$ and take a quotient of the target space geometry that describes thermal $\mathrm{AdS}_3$. On the worldsheet side, this quotient is implemented by a $\ZZ$-orbifold. Therefore, we should replace $\mathrm{AdS}_3$ above by $\mathrm{EAdS}_3/\ZZ$, whose conformal boundary is a torus where the dual CFT is located. Thus, this setup will compute the dual CFT partition function.\footnote{For $k>1$ a one-loop worldsheet analysis around the thermal AdS$_3$ geometry only computes a particular contribution to the dual CFT partition function. In particular, it will not lead to a result that is manifestly target space modular invariant.  Nevertheless, we refer to the result we derive as the `dual CFT partition function'. We also note that we use dual CFT and spacetime synonymously.}

Since the target space now has a non-contractible thermal circle, we can also turn on
an $\mathrm{SU}(2)$ gauge field around this cycle. This corresponds to letting the $\ZZ$-orbifold also act by a rotation on the $\mathbf{S}^3$. We can also choose the spin structure, i.e., whether spacetime fermions are periodic or antiperiodic around this cycle. For concreteness, we will make the periodic choice, which corresponds to including $(-1)^{\dualF}$ in the partition function, where $\dualF$ is the spacetime fermion number.

The $\ZZ$-orbifold will introduce twisted sectors on the worldsheet. The twisted sectors of the worldsheet Hilbert space are described by a pair of integers $(m,n)$, labeling the winding numbers of the spatial and thermal circles of the worldsheet torus around the target space cycle. From a computational point of view, it will suffice to determine the partition function of the sector $(0,n)$ where $n>0$, i.e., the case where only the thermal worldsheet circle wraps the target space cycle. This is because $(m,n)$ transforms as an $\SL(2,\ZZ)$ doublet under worldsheet modular transformations, and we can always find an $\SL(2,\ZZ)$ matrix that maps $(m,n)$ to $(0,n')$ with $n'=\gcd(m,n)$. The sector $(0,0)$ is the global AdS sector. It is already modular invariant on its own. Its treatment is a bit delicate since it involves divergent factors. For the purpose of computing the string partition function, it will however not be very important since it is insensitive to the boundary moduli. It computes the one-loop shift of the Casimir energy of the boundary CFT, i.e., $\dualq^{-\frac{\delta c}{24}}\,\dualqb^{-\frac{\delta c}{24}}$, where $\delta c$ is the $\mathcal{O}(1)$ correction of the central charge originating from the worldsheet torus diagram. Here and in the following, we use bold italic fonts to denote quantities associated to the dual CFT (cf.,~\cref{tab:st-notation}). In particular, we will write $\dualq=\mathrm{e}^{2\pi i \dualtau}$, with $\dualtau$ the spacetime modular parameter.

To proceed we can now apply the unfolding trick~\cite{Polchinski:1985zf,McClain:1986id} where we have used that $(m,n)$ transforms as an $\SL(2,\ZZ)$ doublet.\footnote{This is an application of the Rankin-Selberg unfolding method, which is explained for example in~\cite{Zagier:RanSel}.}
From a modular bootstrap standpoint, we should note that this means that none of  the sectors described by $(m,n)$ are guaranteed to have good modular properties.
However, in all other thermal string backgrounds in the literature that we are aware of---such as when the thermal circle is described by a compact free boson---it always happens that each of the sectors $(m,n)$ still has modular covariance properties, see e.g.,~\cite{Seiberg:1986by}. In fact, this is guaranteed by general 2d CFT arguments if one can fix light-cone gauge, since the thermal parameter transforms like a chemical potential of a $\mathrm{U}(1)$ symmetry describing the spacetime circle, see e.g.,~\cite{Benjamin:2016fhe}.
In curved backgrounds and the Ramond sector this argument can fail.\footnote{The standard argument dating back to~\cite{Seiberg:1986by} involves constructing modular invariant combinations of characters separately for even and odd spin structures and then assembling them into a worldsheet partition function. The modern understanding is that the choice of phases is controlled by the choice of anomaly free $\mathbb{Z}_2$ involutions on the worldsheet~\cite{Kaidi:2019pzj}. In most familiar examples the odd spin structure contribution vanishes owing to the existence of fermion zero modes. The situation we describe will correspond to one where it doesn't and this leads to the aforementioned complication.} Indeed, as we shall see, the worldsheet partition function for $\mathrm{AdS}_3 \times \mathbf{S}^3 \times \mathrm{K3}$ in a given sector has no modular covariance property.

From the worldsheet, this orbifold is in principle straightforward to implement on the level of the torus partition function. Consider first the sector $(m,n)=(0,1)$. Its torus partition function is computed by taking the trace $\Tr_{\mathcal{H}}(g \, q^{L_0}\,\bar{q}^{\overline{L}_0})$ over the worldsheet Hilbert space $\mathcal{H}$ before orbifolding (which has vanishing central charge), where the group element $g \in \text{Isom}$ takes values in the target space isometry group that describes the action of the $\ZZ$-quotient. For Euclidean target space signature, this amounts to turning on a background $\SL(2,\CC) \times  \SO(4)$ gauge field on the worldsheet. Noticing that the zero modes of the worldsheet currents generate the corresponding global symmetries, we set
$g=\dualq^{J_0^3}\, \dualqb^{\bar{J}_0^3} \, \dualy^{K_0^3}\, \dualyt^{\bar{K}_0^3}$, where $\dualy=\mathrm{e}^{2\pi i \,\dualz}$ and $\dualyt=\mathrm{e}^{-2\pi i \,\widetilde{\dualz}}$ are the chemical potentials for the
$\mathrm{Spin}(4)$ corresponding to the symmetries
of the three-sphere $\mathbf{S}^3$.
The reality conditions imply that $\dualz$ and $\widetilde{\dualz}$ are real and independent, but we can analytically continue the definition to any independent complex numbers. To compute the partition function of the sector $(m,n)=(0,n)$, we only have to replace $g \to g^n$.

Let us describe the partition functions of each factor on the worldsheet:
\begin{itemize}[wide,left=0pt]
  \item[1.] The partition function of the $\SL(2,\mathbb{R})$ WZW model and its analytic continuation to Euclidean target space signature was discussed in~\cite{Gawedzki:1991yu, Maldacena:2000kv}.  The Euclidean model is also known as the $\HH_3^+$ or $\SL(2,\CC)/\SU(2)$ model. Its partition function can be computed from the path integral, or by combining characters of the current algebra.

  Let us explain the latter route. The character of a lowest-weight representation of spin $j$ of the worldsheet current algebra $\mathfrak{sl}(2,\RR)_{k+2}$ takes the form
  \begin{equation}
  \begin{aligned}
  \chsl_j(\dualtau\mid \tau)
   & =
  \frac{q^{\Delta(j)-\frac{c}{24}}      \,  \dualq^j}{(1-\dualq)\;
    \prod_{n=1}^\infty\,  (1-q^n)\, (1-\dualq\, q^n)\, (1-\dualq^{-1} \,q^n)}
  \\
   &
  =-\frac{q^{-\frac{1}{k}(j-\frac{1}{2})^2}\, \dualq^{j-\frac{1}{2}}}{\vartheta_1(\dualtau \mid \tau)}\,,
  \end{aligned}
  \end{equation}
  where
  \begin{equation}
  \vartheta_1(\dualtau \mid \tau)
  = q^{\frac{1}{8}} \, (\dualq^{\frac{1}{2}}-\dualq^{-\frac{1}{2}}) \,
  \prod_{n=1}^\infty \, (1-q^n) \, (1-\dualq\, q^n)\, (1-\dualq^{-1} \,q^n)
  \end{equation}
  is the Jacobi theta function.\footnote{We use a non-standard normalization of $\vartheta_1$ throughout the paper in which it is more natural to state the characters.} We have also used that the worldsheet conformal dimension takes the form $\Delta(j)=-\frac{j(j-1)}{k}$. The partition function of the full $\mathfrak{sl}(2,\mathbb{R})_{k+2}$ WZW model is given by
  \begin{equation}
  \int \dd j\, \abs{\chsl_j}^2
  \;\propto\;
  \sqrt{\frac{k}{\Im \tau}}\;
  \frac{\mathrm{e}^{-\frac{k\, \pi\, (\Im \dualtau)^2}{\Im \tau}}}{\abs{\vartheta_1(\dualtau \mid \tau)}^2}\,,
  \label{eq:sl2R-WZW-partition-function}
  \end{equation}
  which is indeed modular covariant. We have used the $\propto$ symbol, since we have omitted an infinite normalization factor, coming from the normalization of the measure over $j$. In the Euclidean model, one does not have to sum over spectrally flowed sectors. Instead, one integrates over the principal series $j \in \frac{1}{2}+i \RR$. Spectrally flowed sectors appear automatically when computing the string partition function because of the presence of poles at $\dualtau\in \ZZ\, \tau\oplus \ZZ$ of the partition function~\cite{Maldacena:2000kv}.

  Since the fermions are decoupled, we can multiply~\eqref{eq:sl2R-WZW-partition-function} by the free fermion answer to obtain the partition function of the super-WZW model. One obtains
  \begin{equation}
  \Zsl{a,\widetilde{a}}(\dualtau \mid \tau)
  \propto
  \sqrt{\frac{k}{\Im \tau}}\,
  \mathrm{e}^{-\frac{k \pi\, (\Im \dualtau)^2}{\Im \tau}} \,
  \abs{\frac{\vartheta_a(\dualtau \mid \tau)\, \vartheta_a(\tau)^{1/2}}{\vartheta_1(\dualtau \mid \tau)\, \eta(\tau)^{3/2}} }^2 \,.
  \label{eq:super-sl2Rk-WZW-partition-function}
  \end{equation}
  The parameter $a,\, \widetilde{a} \in \{1,2,3,4\}$ corresponds to the spin structures (in the same order) $\{\widetilde{\text{R}},\text{R},\text{NS},\widetilde{\text{NS}}\}$, where the tilde denotes periodic boundary conditions around the thermal circle, i.e., a $(-1)^\text{F}$ insertion. We choose the spin structure for left- and right-movers independently, which will be necessary to define the GSO projection later on. Furthermore, we use the slight abuse of notation, where $\big| \bullet\big|^2$ denotes the product of the left- and right-moving quantity, even though the independent spin structures mean that they are not necessarily complex conjugates. Finally, we should also highlight the fact that~\eqref{eq:super-sl2Rk-WZW-partition-function} and similar expressions that follow fail to be invariant under $\tau \to \tau+2$ in the NS-R sector because the Ramond sector has ground state energy $\frac{3}{16} \not \in \frac{1}{2}\ZZ$. This failure will be cured once we combine all factors of the worldsheet theory.

  \item[2.] The $\mathfrak{su}(2)$ super-WZW at level $k$ can be treated similarly, see~\cite{DiFrancesco:1997nk} for a standard reference. A crucial difference is that the $\mathfrak{su}(2)_{k-2}$ representations have null vectors. Consider the spin-$\ell$ representation
  $\mathscr{R}_\ell$ where $\ell \in \{0,\frac{1}{2},\dots,\frac{k-2}{2}\}$. The simplest such null vector is inherited from the finite $\mathfrak{su}(2)$ character. Denoting by $\ket{\ell}$ the highest weight state, it has the form $(\mathcal{K}^-_0)^{2\ell+1} \ket{\ell}$ and has spin $-\ell-1$. A second independent null vector is generated by the vector $(\mathcal{K}^+_{-1})^{k-1-2\ell} \ket{\ell}$ and has spin $k-\ell-1$.
  The full module has an infinite nested structure of null vectors. More formally, this means that $\mathscr{R}_\ell$ fits into a long exact sequence of modules,
  \begin{equation}
  \dots \to \mathscr{V}_{-k+\ell} \oplus \mathscr{V}_{k+\ell} \to \mathscr{V}_{-\ell-1} \oplus \mathscr{V}_{k-\ell-1} \to \mathscr{V}_\ell \to \mathscr{R}_\ell \to 0\,,
  \label{eq:resolution-spin-l-representation}
  \end{equation}
  where $\mathscr{V}_\ell$ denotes the Verma module of spin $\ell$. This allows one to compute the $\mathfrak{su}(2)_{k-2}$ character as the following infinite alternating sum:
  \begin{align}
  \chsu_\ell(\dualz\mid \tau)
   & =
  \frac{\sum_{m \in \ZZ} \,
    \big(q^{\Delta\, (\ell+k m)-\frac{c}{24}} \, \dualy^{\ell+k m}
    - q^{\Delta\, (-\ell-1-k m)-\frac{c}{24}}\, \dualy^{-\ell-1-k m}\big)}{(1-\dualy^{-1})\,
    \prod_{n=1}^\infty\, (1-q^n) \, (1-\dualy\, q^n) \, (1-\dualy^{-1}\, q^n)}
  \notag \\
   & =\frac{\Theta_{k,\ell}(\dualz \mid \tau)-\Theta_{k,\ell}(-\dualz \mid \tau)}{\vartheta_1(\dualz \mid \tau)} \,.
  \label{eq:su2-affine-character}
  \end{align}
  Here
  \begin{equation}
  \Theta_{k,\ell}(\dualz \mid \tau) =
  \sum_{m \in \ZZ}\, q^{\frac{1}{k}(\ell+k \,m+\frac{1}{2})^2} \,
  \dualy^{\ell+k \,m+\frac{1}{2}}
  \equiv
  \sum_{m\in \ZZ}\, q^{\frac{1}{k}\, \llm^2}\, \dualy^{\llm}
  \label{eq:levelk-theta}
  \end{equation}
  is a level-$k$ theta function. For notational simplicity we introduced
  \begin{equation}
  \llm \equiv \ell+k\, m+\frac{1}{2}\,.
  \label{eq:Ldefn}
  \end{equation}
  The worldsheet conformal dimension takes the form $\Delta(\ell)=\Delta(-\ell-1)=\frac{\ell(\ell+1)}{k}$. In the second line, Weyl symmetry, that is invariance under $\dualz \mapsto -\dualz$, is manifest. This form is also dictated by the spectral flow operation, which together with the Weyl group generates the affine Weyl group $\ZZ_2 \ltimes \ZZ$.

  The partition function of the full $\SU(2)$ super-WZW model is given by
  \begin{equation}
  \Zsu{a,\widetilde{a}}(\dualz \mid \tau)
  = \pqty{\sum_{\ell =0}^{\frac{k}{2} - 1} \,
    \abs{ \chsu_\ell( \dualz \mid \tau)}^2 }\,
  \times  \abs{\,\frac{\vartheta_a(\dualz \mid \tau)\,
    \vartheta_a(\tau)^{1/2}}{\eta(\tau)^{3/2}} \, }^2\,,
  \label{eq:super-su2-partition-function}
  \end{equation}
  where we have also now accounted for the contribution from the fermions. As indicated earlier, we are abusing notation and using $\abs{\bullet}^2$ to denote the product of left and right movers; in the present case this is especially germane since $\dualz$ and $\widetilde{\dualz}$ are not complex conjugates.

  \item[3.] The partition function of the supersymmetric K3 sigma model will be denoted by $\ZK{a,\widetilde{a}}(\dualz, \, \widetilde{\dualz} \mid \tau)$,
  where $a$ and $\widetilde{a}$ are the same labels for the spin structure as above. It has $\mathcal{N}=(4,4)$ supersymmetry, which implies that the partition functions with different spin structures are related by spectral flow symmetry, meaning that they can be obtained from one another by applying the shifts $\dualz \mapsto \dualz+1$ and $\dualz \mapsto \dualz+\tau$ or $\widetilde{\dualz} \mapsto \widetilde{\dualz}+1$ and $\widetilde{\dualz} \mapsto \widetilde{\dualz}-\bar{\tau}$.
  In the realization $\mathrm{K3}=\mathbb{T}^4/\ZZ_2$~\cite{Dixon:1985jw}, it is simple to write down explicit formulae for the partition function. For example, in the $\widetilde{\text{R}}$-$\widetilde{\text{R}}$ sector,\footnote{Notice that our normalizations differ by a factor of 2 in the chemical potential relative to what is commonly found in the literature. In our conventions, fermions have half-integer charge.}
  \begin{equation}
  \mathcal{Z}^{^{\mathbb{T}^4/\ZZ_2}}_{1,1}(\dualz ,\, \widetilde{\dualz}\mid \tau)
  = \frac{ \Theta_{\mathrm{N}}(\tau)}{2} \,
  \abs{\frac{\vartheta_1(\frac{1}{2}\,\dualz\mid \tau)^2}{\eta(\tau)^6}}^2
  +8\, \sum_{\mu=2,3,4}\, \abs{\frac{\vartheta_\mu (\frac{1}{2}\,\dualz \mid \tau)^2}{\vartheta_\mu(\tau)^2}}^2\,.
  \label{eq:T4/Z2-partition-function-R-tilde-sector}
  \end{equation}
  Here, $\Theta_{\mathrm{N}}(\tau)$ is the Narain theta-function accounting for the momentum and winding modes on the four-torus in the untwisted sector.

  On the worldsheet, the K3 sigma model is actually uncharged under the spacetime $\mathrm{SU}(2) \times \mathrm{SU}(2)$ isometry (which acts only on the fields associated to $\mathbf{S}^3$). Therefore, we will only require the special case $\ZK{a,\widetilde{a}}(\tau)= \ZK{a,\widetilde{a}}(\dualz=0,\, \widetilde{\dualz}=0 \mid \tau)$ on the worldsheet for a fixed spin structure. However, as we shall see below, the chemical potentials will be activated once we sum over the worldsheet spin structures.

  From~\eqref{eq:T4/Z2-partition-function-R-tilde-sector} we can compute the elliptic genus~\cite{Eguchi:1988vra}
  \begin{equation}
  \varphi_{_\mathrm{K3}}(\dualz \mid \tau)
  =
  \mathcal{Z}_{1,1}^{^{\mathbb{T}^4/\ZZ_2}}(\dualz,\, \widetilde{\dualz}=0 \mid \tau)=8\sum_{\mu=2,3,4} \frac{\vartheta_\mu (\frac{1}{2}\,\dualz \mid \tau)^2}{\vartheta_\mu(\tau)^2}
  \label{eq:K3-elliptic-genus}
  \end{equation}
  of the theory, which is independent of the K3 moduli.

  Let us for convenience also list the corresponding Hilbert space traces for the other spin structures obtained by spectral flow
  of the elliptic genus~\eqref{eq:K3-elliptic-genus},
  given by
  \begin{subequations}
    \begin{align}
    \EK{1}(\dualz \mid \tau)
    & =
    8 \, \bigg(\frac{\vartheta_2(\frac{1}{2}\,\dualz \mid \tau)^2}{\vartheta_2(\tau)^2}+\frac{\vartheta_3(\frac{1}{2}\,\dualz \mid \tau)^2}{\vartheta_3(\tau)^2}+\frac{\vartheta_4(\frac{1}{2}\,\dualz \mid \tau)^2}{\vartheta_4(\tau)^2}\bigg)\,,
    \\
    \EK{2}(\dualz \mid \tau)
    & =
    8 \, \bigg(\frac{\vartheta_1(\frac{1}{2}\,\dualz \mid \tau)^2}{\vartheta_2(\tau)^2}-\frac{\vartheta_4(\frac{1}{2}\,\dualz \mid \tau)^2}{\vartheta_3(\tau)^2}-\frac{\vartheta_3(\frac{1}{2}\,\dualz \mid \tau)^2}{\vartheta_4(\tau)^2}\bigg)\,,
    \\
    \EK{3}(\dualz \mid \tau)
    & =
    8 \, \bigg(\frac{\vartheta_4(\frac{1}{2}\,\dualz \mid \tau)^2}{\vartheta_2(\tau)^2}-\frac{\vartheta_1(\frac{1}{2}\,\dualz \mid \tau)^2}{\vartheta_3(\tau)^2}-\frac{\vartheta_2(\frac{1}{2}\,\dualz \mid \tau)^2}{\vartheta_4(\tau)^2}\bigg)\,,
    \\
    \EK{4}(\dualz \mid \tau)
    & =
    8 \,\bigg(\frac{\vartheta_3(\frac{1}{2}\,\dualz \mid \tau)^2}{\vartheta_2(\tau)^2}+\frac{\vartheta_2(\frac{1}{2}\,\dualz \mid \tau)^2}{\vartheta_3(\tau)^2}+\frac{\vartheta_1(\frac{1}{2}\,\dualz \mid \tau)^2}{\vartheta_4(\tau)^2}\bigg) \,.
    \end{align}
    \label{eq:K3-elliptic-genera}%
  \end{subequations}
  \item[4.] The ghost partition functions with the adoption of Siegel gauge read
  \begin{equation}
  \mathcal{Z}^{^{\text{bc}}}(\tau)= \abs{\eta(\tau)^2}^2\,, \qquad
  \mathcal{Z}^{^{\beta\gamma}}_{a,\widetilde{a}}(\tau) = \abs{\frac{\eta(\tau)}{\vartheta_a(\tau)}}^2\,,
  \end{equation}
  thus canceling two uncharged bosonic or fermionic degrees of freedom, respectively. In the $\widetilde{\text{R}}$ sector, the ghost partition function is divergent and should cancel the only two uncharged fermionic oscillators describing the Cartan directions of $\SL(2,\RR)$ and $\SU(2)$. The naive cancellation turns out to be inconsistent and is at the heart of the problems that we will encounter below.
\end{itemize}

\paragraph{The worldsheet partition function for a fixed spin structure:} Let us combine the ingredients above to write down the worldsheet partition function for a fixed spin structure $a$ and $\widetilde{a}$ for left- and right-movers,\footnote{The $\mathrm{SU}(2)$ spins are quantized as usual in half-integer units, so the sum runs over $\ell \in \{0, \frac{1}{2}, \cdots , \frac{k}{2} -1 \}$.}
\begin{equation}
\begin{aligned}
\mathcal{Z}_{a,\widetilde{a}}(\dualz,\,\widetilde{\dualz},\, \dualtau \mid \tau)
 & =
\Zsl{a,\widetilde{a}}(\dualtau \mid \tau)\; \Zsu{a, \widetilde{a}}(\dualz \mid \tau)\;
\ZK{a,\widetilde{a}}(\tau)\; \mathcal{Z}^{^{\text{bc}}}(\tau)\;
\mathcal{Z}^{^{\beta\gamma}}_{a,\widetilde{a}}(\tau)
\\
 & =\sqrt{\frac{k}{\Im \tau}}\, \frac{\mathrm{e}^{-\frac{k \pi \,(\Im \dualtau)^2}{\Im \tau}}}{\abs{\vartheta_1(\dualtau \mid \tau)}^2}
\, \pqty{\sum_{\ell =0}^{\frac{k}{2} - 1} \,
    \abs{\chsu_\ell( \dualz \mid \tau)}^2 }\,
\\
 & \qquad \qquad
\times \abs{\vartheta_a(\dualtau\mid \tau)\,\vartheta_a(\dualz \mid \tau) }^2
\times \ZK{a,\widetilde{a}}(\tau)\,,
\end{aligned}
\label{eq:Znu}
\end{equation}
for $a,\,\widetilde{a} \in \{2,3,4\}$. For $a=1$ or $\widetilde{a}=1$, this is strictly speaking not legal, since we canceled the ill-defined terms $\frac{\vartheta_1(\tau)}{\vartheta_1(\tau)}$.

\paragraph{Failure of Weyl reflection symmetry:} We can now already see the basic problem if we assume that~\eqref{eq:Znu} holds also for $a=1$. It cannot be correct as written: $\vartheta_1(\dualz \mid \tau)$ is \emph{antisymmetric} under the action of the $\mathrm{SU}(2)$ Weyl reflection symmetry taking $\dualz \mapsto - \dualz$.\footnote{Since $\dualz$ and $\widetilde{\dualz}$ are not related by complex conjugation, Weyl reflection symmetry must hold independently in the two variables.} Therefore, the partition function cannot be decomposed into $\mathfrak{su}(2) \oplus \mathfrak{su}(2)$ characters and hence cannot be correct for $a=1$. Clearly, we were too fast above in setting $\frac{\vartheta_1(\tau)}{\vartheta_1(\tau)}=1$ for $a=1$. Implementing the physical state conditions is thus more subtle in the $\widetilde{\text{R}}$ sector. We will give a proposed resolution below in the case of the Type II theory. We also note that this issue does not arise for $\mathrm{AdS}_3 \times \mathbf{S}^3 \times \mathbb{T}^4$ since $\mathcal{Z}_{1,\widetilde{a}}^{^{\mathbb{T}^4}}(\tau)=0$ because of the existence of additional fermionic zero modes.

\subsection{Naive physical state projection}\label{subsec:naive-physical-state-projection}

After having reviewed the ingredients of the computation of the worldsheet partition function, we can now assemble the pieces and derive the spacetime partition function. We will at first ignore the issue of the $\widetilde{\text{R}}$ sector and give a proposed modification in~\cref{subsec:corrected-physical-state-projection}.

\paragraph{Sum over spin structures:} In order to implement the GSO projection, we should sum over the spin structures. The spin-structure-dependent part in the worldsheet partition function~\eqref{eq:Znu} when generalized to independent left- and right-moving spin structures takes the form
\begin{equation}
\frac{1}{4}\sum_{a,\widetilde{a}=1}^4 \, \alpha_{a}\, \widetilde{\alpha}_{\widetilde{a}}\;
\vartheta_a(\dualtau \mid \tau) \, \vartheta_a (\dualz \mid \tau) \,
\vartheta_{\widetilde{a}}(-\overline{\dualtau} \mid -\bar{\tau})  \,
\vartheta_{\widetilde{a}} (-\widetilde{\dualz} \mid -\bar{\tau})
\, \ZK{a,\widetilde{a}}(\tau)\,,
\label{eq:K3-sum-over-spin-structures}
\end{equation}
where $\alpha_a$ and $\widetilde{\alpha}_{\widetilde{a}}$ are appropriate signs. Note that as written the spin structure part only receives contribution from the fermion determinants, viz.,  from the second line in the last equality of~\eqref{eq:Znu}.

We take $\alpha_3=1$ and $\alpha_4=-1$, which will project out the tachyon from the NS-NS sector. The choice of $\alpha_2$ corresponds to the choice of the spacetime spin structure $(-1)^{\dualF}$, since it determines the relative sign of states in the NS-NS and R-NS sector, i.e., spacetime bosons and fermions. As mentioned above, we will make the periodic choice in target space, i.e., $\alpha_2=-1$.\footnote{This is actually necessary for the plethystic exponential in~\eqref{eq:plethystic-exponential} to capture the correct statistics of the fermions.} By the same reasoning $\widetilde{\alpha}_a=\alpha_a$ for $a=2,3,4$. In flat space, Type IIB string theory corresponds to the choice $\alpha_1=\widetilde{\alpha}_1$, while Type IIA string theory corresponds to the choice $\alpha_1=-\widetilde{\alpha}_1$. The choice of $\alpha_1$ itself doesn't matter as the sign can be flipped by reversing the target space orientation. As we shall see, this is too naive in the present context. We henceforth write $\alpha_1 \equiv -\alpha$ and $\widetilde{\alpha}_1\equiv -\widetilde{\alpha}$ and keep them arbitrary for now. In summary, the worldsheet traces are computed as
\begin{equation}
\Tr_{\mathrm{NS}}\pqty{ \frac{1 - (-1)^{F_L}}{2}} - \Tr_{\mathrm{R}}\pqty{ \frac{1 + \alpha\, (-1)^{F_L}}{2}}
\end{equation}
for the left movers (and similarly with $\widetilde{\alpha}$ for right-movers).

We claim that with these sign choices,~\eqref{eq:K3-sum-over-spin-structures} equals
\begin{equation}
\begin{aligned}
\frac{1}{4}\sum_{a,\widetilde{a}=1}^4 \, \alpha_{a}\, \widetilde{\alpha}_{\widetilde{a}}\;
\vartheta_a(\dualtau \mid \tau)\,  \vartheta_a (\dualz \mid \tau)\,
\vartheta_{\widetilde{a}}(-\overline{\dualtau} \mid -\bar{\tau}) \,
\vartheta_{\widetilde{a}} (-\widetilde{\dualz} \mid -\bar{\tau}) \, \ZK{a,\widetilde{a}}(\tau)
\\
= \vartheta_1\big(\tfrac{\dualtau-\alpha\,\dualz}{2}\mid \tau\big)^2\,
\vartheta_1\big(-\tfrac{\overline{\dualtau}-\widetilde{\alpha}\, \widetilde{\dualz}}{2}\mid -\bar{\tau}\big)^2 \,
\ZK{1,1}\big(\dualtau+\alpha\,\dualz,\, \overline{\dualtau}+\widetilde{\alpha}\, \widetilde{\dualz} \mid \tau\big)  \,.
\end{aligned}
\label{eq:K3-sum-over-spin-structures-identity}
\end{equation}
One can prove this identity by showing that it holds separately for any unitary character of the $c=6$ $\mathcal{N}=4$ superconformal algebra (associated to K3), which we do in~\cref{app:character-identities}.
Thus, after summing over spin structures and using~\eqref{eq:K3-sum-over-spin-structures-identity}, the worldsheet partition function becomes
\begin{equation}
\begin{aligned}
\mathcal{Z}(\dualz,\,\widetilde{\dualz},\, \dualtau \mid \tau)
 & =
\sqrt{\frac{k}{\Im \tau}}\, \frac{\mathrm{e}^{-\frac{k \pi (\Im \dualtau)^2}{\Im \tau}}}{\abs{\vartheta_1(\dualtau \mid \tau)}^2}
\pqty{\sum_{\ell =0}^{\frac{k}{2} - 1} \abs{ \chsu_\ell( \dualz \mid \tau)}^2}
\\
 & \qquad\times \abs{\vartheta_1(\tfrac{\dualtau-\alpha \,\dualz}{2}\mid \tau)^2 }^2 \times \ZK{1,1}\big(\dualtau+\alpha\,\dualz,\overline{\dualtau}+\widetilde{\alpha}\,\widetilde{\dualz} \mid \tau\big) \,.
\end{aligned}
\label{eq:Zworldsheet-naive}
\end{equation}
As promised, the sum over spin structures has resulted in the worldsheet K3 partition function acquiring dependence on the target space chemical potentials.

\paragraph{From the worldsheet to the spacetime partition function:} So far, we have computed the partition function of the thermal background in the sector with a single $g$ insertion. The full string partition function $\dualPF_\text{mult}^{\text{II}}$---including multi-string states---is obtained by summing over all the twisted sectors $\mathcal{Z}^{(m,n)}$, then using the unfolding trick, and finally exponentiating the result to account for multi-string states
\begin{equation}
\begin{aligned}
\dualPF^{\text{II}}_\text{mult}(\dualz,\, \widetilde{\dualz},\, \dualtau)
 & =
\exp\bigg(\Im \dualtau \int_{\mathcal{F}} \frac{\dd^2 \tau}{\Im \tau} \, \sum_{(m, n) \in \ZZ^2 \setminus \{(0,0)\}} \mathcal{Z}^{(m,n)}(\dualz,\,\widetilde{\dualz},\, \dualtau \mid \tau)\bigg)
\\
 & =
\exp\bigg(\sum_{n=1}^\infty \frac{\Im (n\,\dualtau)}{n} \intStrip\frac{\dd^2 \tau}{\Im \tau}   \, \mathcal{Z}(n\,\dualz,\, n\,\widetilde{\dualz},\,n\,\dualtau \mid \tau) \bigg) \,.
\end{aligned}
\label{eq:plethystic-exponential}
\end{equation}
To alleviate notation later, we have also introduced the shorthand
\begin{equation}
\intStrip \ := \
\int_{-\frac{1}{2} < \Re(\tau) < \frac{1}{2}}\,
\int_{\Im(\tau) = 0}^\infty \ .
\label{eq:strip-int-notation}
\end{equation}

We excluded the $(0,0)$ sector; this gives an additive constant which renormalizes the vacuum energy. The prefactor $\Im \dualtau$ corresponds to the factor $1/\text{vol}(\ZZ)$ from the worldsheet orbifold. It renders the a priori infinite normalization of~\eqref{eq:sl2R-WZW-partition-function} finite. Intuitively, the infinite normalization is proportional to the volume of $\mathrm{EAdS}_3$; dividing it by $\text{vol}(\ZZ)$ yields the volume of thermal $\mathrm{AdS}_3$, which is proportional to the size $\Im \dualtau$ of the target space thermal circle; we refer to~\cite{Maldacena:2000kv} for more precise explanations.
The final expression may be recognized as  the standard form of the plethystic exponential.
We will focus on the single-string free energy (logarithm of the partition function).
It is given by the $n=1$ term above, i.e.,
\begin{equation}
\dualfE^{\text{II}}(\dualz,\,\widetilde{\dualz},\, \dualtau)
=
\Im \dualtau \intStrip\frac{\dd^2 \tau}{\Im \tau} \,
\mathcal{Z}(\dualz,\,\widetilde{\dualz},\,\dualtau \mid \tau)  \,.
\label{eq:general-type-II-spacetime-partition-function}
\end{equation}
The normalization of the prefactor can be fixed by requiring a unit-normalized worldsheet state to correspond to a unit-normalized spacetime state, i.e., for a given term in the worldsheet $q$-expansion\footnote{This is not completely legal because the radius of convergence in the worldsheet $q$-expansion is smaller than one because of the appearance of long string states and associated poles in the integrand, but this will not influence the present argument. See~\cite{Maldacena:2000kv} for more details.}
\begin{equation}
\Im \dualtau \intStrip\frac{\dd^2 \tau}{\Im \tau}
\, \sqrt{\frac{k}{\Im \tau}} \, \mathrm{e}^{-\frac{k \pi \, (\Im \dualtau)^2}{\Im \tau}}\,
q^{\Delta} \bar{q}^{\bar{\Delta}}
= \delta_{\Delta,\bar{\Delta}}\, \dualq^{\sqrt{k \,\Delta}} \, \dualqb^{\sqrt{k \,\Delta}} \,.
\end{equation}
Thus, the integral over $\Re \tau$ imposes level-matching, while the integral over $\Im \tau$ imposes the mass-shell condition.

\paragraph{The spacetime (single-string) free energy:} Let us combine~\eqref{eq:Zworldsheet-naive} and~\eqref{eq:general-type-II-spacetime-partition-function}.
This leads to the naive spacetime single-string free energy of the Type II string,
\begin{equation}
\begin{aligned}
\dualfE^\text{II}(\dualz,\, \widetilde{\dualz},\, \dualtau)
 & \overset{?}{=}
\sqrt{k}\, \Im \dualtau \, \intStrip\,
\frac{\dd^2 \tau}{(\Im \tau)^{\frac{3}{2}}}\,
\frac{\mathrm{e}^{-\frac{k \pi (\Im \dualtau)^2}{\Im \tau}}}{\abs{\vartheta_1(\dualtau \mid \tau)}^2}
\pqty{ \sum_{\ell =0}^{\frac{k}{2} - 1} \abs{\chsu_\ell( \dualz \mid \tau)}^2}
\\
 & \qquad\times
\vartheta_1\big(\tfrac{\dualtau-\alpha\, \dualz}{2}\mid \tau\big)^2\,
\vartheta_1\big(-\tfrac{\overline{\dualtau}-\widetilde{\alpha}\,\widetilde{\dualz}}{2}\mid -\bar{\tau}\big)^2 \, \ZK{1,1}\big(\dualtau+\alpha\,\dualz,\overline{\dualtau}+\widetilde{\alpha}\,\widetilde{\dualz} \mid \tau\big) \,.
\end{aligned}
\label{eq:naive-type-II-partition-function}
\end{equation}
As already noticed below~\eqref{eq:Znu}, this expression cannot be correct because it violates Weyl symmetry. To emphasize that~\eqref{eq:naive-type-II-partition-function} cannot hold as written, we have put a question mark on the equality sign. Indeed, mapping $\dualz \mapsto -\dualz$ ($\widetilde{\dualz} \mapsto -\widetilde{\dualz}$) precisely reverses the sign of $\alpha$ ($\widetilde{\alpha}$). This is obvious from the LHS of~\eqref{eq:K3-sum-over-spin-structures-identity} because the term with $a=1$ is odd in $\dualz$, while the remaining terms are even.

Another way to see the inconsistency is to consider the zero modes of the two holomorphic theta functions in~\eqref{eq:naive-type-II-partition-function}. The background has $\mathcal{N}=(0,4)$ supersymmetry for Type IIA and $\mathcal{N}=(4,4)$ supersymmetry for Type IIB. The holomorphic zero modes in~\eqref{eq:naive-type-II-partition-function} correspond to the action of the global spacetime supercharges $\dualG^{\pm\,\mathsf{A}}_{\pm \alpha/2}$ defined in~\eqref{eq:dual-Gsusy} where $\pm$ is the R-symmetry charge and $\mathsf{A}$ is the index of the outer automorphism symmetry. However, the supercharges $\dualG^{+\,\mathsf{A}}_{-1/2}$ should vanish on BPS states, and $\dualG^{-\,\mathsf{A}}_{-1/2}$ should vanish on anti-BPS states,\footnote{Here we call a state $\phi$ BPS if it satisfies $L_0 \phi = K^3_0 \phi$, and anti-BPS if it satisfies $L_0 \phi = - K^3_0 \phi$.} while in~\eqref{eq:naive-type-II-partition-function} one type of supercharge is always non-vanishing. The partition function~\eqref{eq:naive-type-II-partition-function} is therefore inconsistent with any spacetime supersymmetry.

\section{A consistent physical state projection}\label{sec:proposal}

Thus far we have examined the worldsheet theory and concluded that while the canonically quantized spectrum reproduces target space expectations, the one-loop partition function,
which we would like to identify as a spectral trace, does not.
We will now rectify the problems with the naive partition function~\eqref{eq:naive-type-II-partition-function}.
\subsection{Corrected physical state projection}\label{subsec:corrected-physical-state-projection}

As already noted above, the problem comes from the $a=1$ or $\widetilde{a}=1$ sector, which is odd under Weyl symmetry.
Let us therefore single it out and treat the sum over $a,\widetilde{a}=2,3,4$ and the rest of the sum separately.
Note that the sum in~\eqref{eq:K3-sum-over-spin-structures-identity} restricted to $a,\widetilde{a}=2,3,4$ is given by
the average of the sign choices~$\alpha, \widetilde \alpha$ on the RHS.
Consider now the sum~\eqref{eq:K3-sum-over-spin-structures} when restricted to $a=1$.
As for the elliptic genera~\eqref{eq:K3-elliptic-genera}, but with the roles of left- and right-movers exchanged, $\ZK{a=1,\widetilde{a}}(\tau)$ is an anti-holomorphic function in $\tau$, since at vanishing chemical potential only the left-moving Ramond ground states contribute.
Therefore, the $\tau$-dependence of the worldsheet partition function dramatically simplifies; in fact
the factors of $\vartheta_1(\dualtau \mid \tau)$ and $\vartheta_1(\dualz \mid \tau)$ that feature, respectively, in the
denominators of the affine $\mathfrak{sl}(2)$ and $\mathfrak{su}(2)$ characters
cancel against the same factors in the fermionic partition function in ~\eqref{eq:K3-sum-over-spin-structures}.
This dramatic simplification is somewhat expected: the $a=1$ sector is the one that appears in the
\emph{worldsheet} elliptic genus---it is the Ramond sector with the insertion of $(-1)^F$. If we do not switch on any chemical potential, the left-moving sector should just be a constant, and the answer would be expected to be anti-holomorphic;
it is not quite anti-holomorphic because the worldsheet theory neither has
a discrete spectrum nor is unitary, and thus anti-holomorphicity can be violated~\cite{Troost:2010ud, Eguchi:2010cb}.
In fact, this sector will only contain spacetime $\frac{1}{4}$-BPS states.
Since $\vartheta_1(\dualz \mid \tau)$ is odd under $\dualz \mapsto - \dualz$, the result would again not be invariant under Weyl symmetry. The only way out is to choose the parameter $\alpha$ appearing in~\eqref{eq:K3-sum-over-spin-structures-identity} to be dependent on the state in the \emph{bosonic} part of the worldsheet theory. Such a dependence on the bosonic state is severely limited since the decoupled bosonic currents $\mathcal{J}^A$ and $\mathcal{K}^a$ commute with $(-1)^{\text{F}_\text{L}}$ on the worldsheet.
The only possibility is that we assign different fermion numbers to the null vectors appearing in the affine $\mathfrak{su}(2)_{k-2}$ representation, i.e., the different terms in~\eqref{eq:resolution-spin-l-representation}. Before the $\beta\gamma$ ghosts effectively remove the two neutral fermionic oscillators $\psi_0^3$ and $\chi_0^3$, every state in the $\widetilde{\mathrm{R}}$-sector is doubly degenerate with opposite fermion numbers. It is a priori not clear which one of the two states we should keep.

We therefore allow for a change in the signs in the numerator of~\eqref{eq:su2-affine-character}
and replace the naive product as follows,
\begin{equation}
\chi_\ell^{\mathfrak{su}(2)_{k-2}}(\dualz \mid \tau)\,
\mathcal{F}(\alpha) \,\vartheta_1(\dualz \mid \tau)\rightsquigarrow
\sum_{m \in \ZZ}\, q^{\frac{1}{k}\, \llm^2}\, \pqty{ \mathcal{F}(\alpha_{\ell,m}^1)\, \dualy^{\llm} -  \mathcal{F}(\alpha_{\ell,m}^2)\, \dualy^{-\llm}} \,,
\label{eq:character-theta-replacement}
\end{equation}
where $\mathcal{F}(\alpha)$ stands schematically for the rest of the expression~\eqref{eq:Zworldsheet-naive} at fixed $\alpha$, and $\llm$ was defined in~\eqref{eq:Ldefn}.

\paragraph{The supersymmetric chiral character:}
We will now fix the possible sign choices by imposing consistency with spacetime supersymmetry. To begin with, invariance under Weyl symmetry demands that $\alpha_{\ell,m}^2=-\alpha_{\ell,m}^1$. For $\llm  > 0$
(i.e., $m \ge 0$), the state will lead to a spacetime BPS state which means that the action of the two supercharges $\dualG_{-\frac{1}{2}}^{+A}$ should be absent.
In view of the discussion below~\eqref{eq:naive-type-II-partition-function}, this means that we should choose $\alpha_{\ell,m}^1=1$ in that case. If $L_{\ell m } < 0$
(i.e., $m< 0$), we should select $\alpha_{\ell,m}^1=-1$ instead.
Thus, guided by spacetime supersymmetry, we should replace
\begin{equation}
\begin{aligned}
\chsu_{\ell}(\dualz \mid \tau) \, \mathcal{F}(\alpha) \, \vartheta_1(\dualz \mid \tau)
 & \rightsquigarrow
\sum_{m \ge 0} \, q^{\frac{1}{k}\, \llm^2}\, \pqty{ \mathcal{F}(1) \, \dualy^{\llm}
    - \mathcal{F}(-1)\, \dualy^{-\llm}}
\\
 &
\qquad+\sum_{m \le -1}\, q^{\frac{1}{k}\, \llm^2}\, \pqty{ \mathcal{F}(-1) \, \dualy^{\llm}
    - \mathcal{F}(1)\, \dualy^{-\llm}}
\\
 & =
\sum_{\alpha=\pm 1}\,  \alpha \, \Psi_{k,\ell}(\alpha \,\dualz \mid \tau) \,
\mathcal{F}(\alpha) \,.
\end{aligned}
\label{eq:susy-reprule}
\end{equation}
Thus, the answer only features the combination (writing out $\llm$ explicitly)
\begin{equation}
\Psi_{k,\ell}(\dualz \mid \tau)
= \sum_{m\in \ZZ}\,
\mathop{\text{sgn}}(\ell+km+\tfrac{1}{2}) \;
\dualy^{|\ell+k m+\frac{1}{2}|}\; q^{\frac{1}{k}(\ell+k m+\frac{1}{2})^2}\,.
\label{eq:false-theta-function}
\end{equation}
As we explain in~\cref{app:false}, $\Psi_{k,\ell}$ can be expressed as a linear combination of a level $k$ theta function
and a false theta function discussed in~\cite{Bringmann:2019vyd}.
Similar false theta functions have also appeared in logarithmic VOA characters~\cite{Creutzig:2013zza}.

\paragraph{The non-supersymmetric chiral character:} We may also choose signs in~\eqref{eq:character-theta-replacement} to break spacetime supersymmetry, while maintaining the Weyl symmetry. Following the logic outlined above, we need $\alpha_{\ell,m}^2  = - \alpha_{\ell,m}^1$. But now we can make the opposite choice $\alpha_{\ell,m}^1 =-1$ for $m \geq 0$ and $\alpha^1_{\ell,m} = 1$ for $m < 0$.
This will give a character that breaks $\mathfrak{psu}(1,1|2)$ but preserves its bosonic subalgebras. The character replacement is then given by
\begin{equation}
\begin{aligned}
\nchsu_{\ell}(\dualz \mid \tau) \, \mathcal{F}(\alpha) \, \vartheta_1(\dualz \mid \tau)
 & \rightsquigarrow
\sum_{m \ge 0} \, q^{\frac{1}{k}\, \llm^2}\, \pqty{ \mathcal{F}(-1) \, \dualy^{\llm}
    - \mathcal{F}(1)\, \dualy^{-\llm}}
\\
 &
\qquad+\sum_{m \le -1}\, q^{\frac{1}{k}\, \llm^2}\, \pqty{ \mathcal{F}(1) \, \dualy^{\llm}
    - \mathcal{F}(-1)\, \dualy^{-\llm}}
\\
 & =
\sum_{\alpha=\pm 1}\,  \alpha \, \Psi_{k,\ell}(\alpha \,\dualz \mid \tau) \,
\mathcal{F}(-\alpha) \,.
\end{aligned}
\label{eq:no-susy-reprule}
\end{equation}
%

\subsection{The spacetime partition functions}\label{subsec:Zspacetime}

With the modified chiral characters we are now in a position to fix the spacetime partition function naively constructed in~\cref{subsec:naive-physical-state-projection}. Since we had a binary choice for the chiral characters, we have the ability to now construct spacetime partition functions that preserve the full $\mathcal{N} = (4,4)$ supersymmetry in target space, or break some or all of it. We give the results now for the three possibilities.

\paragraph{The supersymmetric partition function (IIB):}  The Type IIB compactification on
$\mathrm{AdS}_3 \times \mathbf{S}^3 \times \mathrm{K3}$ preserves the full $\mathcal{N} = (4,4)$ supersymmetry. Its partition function is given by making the supersymmetric sign choices for the chiral characters~\eqref{eq:susy-reprule} for both left and right movers on the worldsheet.  Putting everything together (viz., writing out $\mathcal{F}$ explicitly) we therefore end up with the supersymmetric answer
\begin{equation}
\begin{aligned}
\dualfE^\text{IIB}(\dualz,\, \widetilde{\dualz},\, \dualtau)
 & =
\sqrt{k}\Im \dualtau \, \intStrip\frac{\dd^2 \tau}{(\Im \tau)^{\frac{3}{2}}}\, \mathrm{e}^{-\frac{k \pi\, (\Im \dualtau)^2}{\Im \tau}}
\\
 &
\quad\times
\sum_{\alpha, \widetilde{\alpha}=\pm 1} \alpha \,\widetilde{\alpha}\,
\sum_{\ell =0}^{\frac{k}{2} - 1}\,
\Psi_{k,\ell}(\alpha\,\dualz \mid \tau)\;
\Psi_{k,\ell}(-\widetilde{\alpha}\, \widetilde{\dualz} \mid -\bar{\tau})
\\
 & \quad \times
\frac{
  \vartheta_1\big(\tfrac{\dualtau-\alpha\,\dualz}{2}\mid \tau\big)^2\,
  \vartheta_1\big(-\tfrac{\overline{\dualtau}-\widetilde{\alpha}\, \widetilde{\dualz}}{2}\mid -\bar{\tau}\big)^2 \;
  \ZK{1,1}(\dualtau+\alpha\,\dualz,\, \overline{\dualtau}+\widetilde{\alpha}\, \widetilde{\dualz} \mid \tau)}{
  \abs{\vartheta_1(\dualtau\mid\tau)\,\vartheta_1(\dualz\mid\tau)}^2} \,.
\end{aligned}
\label{eq:final-type-IIB-partition-function}
\end{equation}
This is by construction invariant under the Weyl symmetry since we can write it in a way such that $\dualz$ only appears in the combination $\alpha \,\dualz$. The spurious poles at $\dualz=m \,\tau+n$ with $m,n \in \ZZ$ cancel. We will also see momentarily that it is compatible with spacetime supersymmetry. Let us also notice that
\begin{equation}
\sum_{\alpha=\pm 1} \, \alpha \, \Psi_{k,\ell}(\alpha\,  \dualz \mid \tau)
=\sum_{\alpha=\pm 1} \,\alpha\, \Theta_{k,\ell}(\alpha \,\dualz \mid \tau)\,,
\label{eq:false-theta-asym}
\end{equation}
i.e., we can replace the false theta function by a true theta function in the worldsheet sectors $a=2,3,4$, which are independent of $\alpha$. This recovers the usual affine $\mathfrak{su}(2)$ character~\eqref{eq:su2-affine-character}. In particular, for $\mathbb{T}^4$ where we have $\mathcal{Z}_{1,\widetilde{a}}^{^{\mathbb{T}^4}}(\tau)=0$, our prescription coincides with the naive answer, which we know reproduces the correct spacetime spectrum (cf.~\cite{Ferko:2024uxi}). A similar statement applies for the $\mathbf{S}^3 \times \mathbf{S}^1$ compactification.

\paragraph{The partially supersymmetric partition function (IIA):} The Type IIA compactification on
$\mathrm{AdS}_3 \times \mathbf{S}^3 \times \mathrm{K3}$ preserves $\mathcal{N} = (0,4)$ supersymmetry in target space (we have taken supersymmetry to be preserved for right movers). Putting the chiral characters together we end up with
\begin{equation}
\begin{aligned}
\dualfE^\text{IIA}(\dualz,\, \widetilde{\dualz},\, \dualtau)
 & =
\sqrt{k}\Im \dualtau \, \intStrip\frac{\dd^2 \tau}{(\Im \tau)^{\frac{3}{2}}}\, \mathrm{e}^{-\frac{k \pi\, (\Im \dualtau)^2}{\Im \tau}}
\\
 &
\quad\times
\sum_{\alpha, \widetilde{\alpha}=\pm 1} \alpha \,\widetilde{\alpha}\,
\sum_{\ell =0}^{\frac{k}{2} - 1}\,
\Psi_{k,\ell}(\alpha\,\dualz \mid \tau)\;
\Psi_{k,\ell}(-\widetilde{\alpha}\, \widetilde{\dualz} \mid -\bar{\tau})
\\
 & \quad \times \frac{
    \vartheta_1\big(\tfrac{\dualtau + \alpha\,\dualz}{2}\mid \tau\big)^2\,
    \vartheta_1\big(-\tfrac{\overline{\dualtau}-\widetilde{\alpha}\, \widetilde{\dualz}}{2}\mid
    -\bar{\tau}\big)^2 \;
    \ZK{1,1}(\dualtau - \alpha\,\dualz,\, \overline{\dualtau}+\widetilde{\alpha}\, \widetilde{\dualz} \mid \tau)}{
    \abs{\vartheta_1(\dualtau\mid\tau)\,\vartheta_1(\dualz\mid\tau)}^2} \,.
\end{aligned}
\label{eq:final-type-IIA-partition-function}
\end{equation}
As far as we are aware, this expression has not appeared in the literature. We will verify it by computing the spectrum and checking that it agrees with the supergravity spectrum in~\cref{subsec:IIAspec}.

\paragraph{The non-supersymmetric partition function:} We may also choose to construct a third partition function where spacetime supersymmetry is broken completely. This would result in
\begin{equation}
\begin{aligned}
\dualfE^\text{non-susy}(\dualz,\, \widetilde{\dualz},\, \dualtau)
 & =
\sqrt{k}\Im \dualtau \, \intStrip\frac{\dd^2 \tau}{(\Im \tau)^{\frac{3}{2}}}\, \mathrm{e}^{-\frac{k \pi\, (\Im \dualtau)^2}{\Im \tau}}
\\
 &
\quad\times
\sum_{\alpha, \widetilde{\alpha}=\pm 1} \alpha \,\widetilde{\alpha}\,
\sum_{\ell =0}^{\frac{k}{2} - 1}\,
\Psi_{k,\ell}(\alpha\,\dualz \mid \tau)\;
\Psi_{k,\ell}(-\widetilde{\alpha}\, \widetilde{\dualz} \mid -\bar{\tau})
\\
 & \quad \times \frac{
    \vartheta_1\big(\tfrac{\dualtau + \alpha\,\dualz}{2}\mid \tau\big)^2\,
    \vartheta_1\big(-\tfrac{\overline{\dualtau}+\widetilde{\alpha}\, \widetilde{\dualz}}{2}\mid
    -\bar{\tau}\big)^2 \;
    \ZK{1,1}(\dualtau - \alpha\,\dualz,\, \overline{\dualtau} - \widetilde{\alpha}\, \widetilde{\dualz} \mid \tau)}{
    \abs{\vartheta_1(\dualtau\mid\tau)\,\vartheta_1(\dualz\mid\tau)}^2} \,.
\end{aligned}
\label{eq:final-non-susy-partition-function}
\end{equation}
We won't dwell on this example in detail below, but note here that this construction manifestly preserves the $\mathrm{SU}(2)$ symmetries. It just does so in a manner that breaks supersymmetry, so
the spacetime spectrum arising from~\eqref{eq:final-non-susy-partition-function} would be organized in $\mathrm{SL}(2,\RR)_L \times \mathrm{SL}(2,\RR)_R \times \mathrm{SU}(2)_L \times \mathrm{SU}(2)_R$ representations.  This is in contrast with the naive worldsheet characters we wrote down in~\cref{subsec:naive-physical-state-projection} which do not respect the Weyl symmetry of $\mathrm{SU}(2)$.  Finally, we note that it may be realized as the orbifold $\mathbb{T}^4/\mathbb{Z}_2 \cdot (-1)^{F_R}$, as explained in~\cite{Sen:1996na}.

\section{Basic checks of the worldsheet partition function} \label{sec:checks}

As a check of our prescription, we will begin by extracting the $\frac{1}{2}$-BPS spectrum. We shall do this first for the Type IIB compactification, see Eq.~\eqref{eq:final-type-IIB-partition-function}, and compare it with the supergravity result. The latter was obtained long ago in~\cite{Deger:1998nm} by direct Kaluza--Klein reduction of 6d supergravity, and was rederived using group-theoretic methods shortly thereafter~\cite{deBoer:1998kjm}.
Subsequently, we also obtain the Type IIA supergravity spectrum, which has recently been derived in~\cite{Eloy:2020uix}.

\subsection{\texorpdfstring{$\frac{1}{2}$}{1/2}-BPS spectrum in IIB compactification}\label{subsec:12-BPS-spectrum}

\paragraph{Supergravity spectrum:} The supergravity spectrum consists entirely of $\frac{1}{2}$-BPS multiplets, which allows for a straightforward derivation using group-theoretic methods for the KK-reduction on $\mathbf{S}^3 \times \mathrm{K3}$~\cite{deBoer:1998kjm}. Let us denote by $[\dualj; \dualjb]$  the BPS multiplet of the \emph{global} $\mathcal{N}=(4,4)$ algebra whose BPS primary has spin $\dualj$ and $\dualjb$ with respect to the left- and right-moving $\mathrm{SU}(2)$ R-symmetry. The supergravity spectrum then reads
\begin{equation}
\begin{aligned}
\bigoplus_{\dualj=\frac{1}{2},1,\dots}^\infty \bigg(
 &
\bqty{\dualj + \tfrac{1}{2};\, \dualj-\tfrac{1}{2} }
\oplus
\bqty{\dualj-\tfrac{1}{2};\, \dualj + \tfrac{1}{2} }
\oplus
\bqty{\dualj-\tfrac{1}{2}; \,\dualj-\tfrac{1}{2} }
\\
 & \qquad
\oplus
\bqty{\dualj + \tfrac{1}{2}; \,\dualj + \tfrac{1}{2} }
\oplus
20 \cdot \bqty{\dualj;\, \dualj }
\bigg) \,.
\end{aligned}
\label{eq:supergravity-spectrum}
\end{equation}

To confirm our prescription, we will now extract this result from~\eqref{eq:final-type-IIB-partition-function}. In fact, we will be slightly more general and compute the full $\frac{1}{2}$-BPS spectrum without taking $k \to \infty$ (as supergravity would necessitate).

\paragraph{$\frac{1}{2}$-BPS spectrum from the worldsheet:} The action of the global superalgebra is realized in~\eqref{eq:final-type-IIB-partition-function} by the zero modes of the Jacobi theta functions. Similarly, only Ramond sector zero modes of the K3 sigma model partition function contribute to the $\frac{1}{2}$-BPS spacetime spectrum. This means that we can replace
\[
  \vartheta_1(\dualz \mid \tau) \rightsquigarrow q^{\frac{1}{8}}\, (\dualy^{\frac{1}{2}}-\dualy^{-\frac{1}{2}}) \,,
\]
and similarly for the other theta function factors. We may also replace
\[
  \ZK{1,1}(\dualz,\widetilde{\dualz} \mid \tau)
  \rightsquigarrow
  20 + \sum_{\sigma,\widetilde{\sigma}=\pm 1} \,  \dualy^{\frac{\sigma}{2}} \,
  \dualyt^{\frac{\widetilde{\sigma}}{2}}\,.
\]

After removing the non-zero modes, the last line of~\eqref{eq:final-type-IIB-partition-function} is independent of $\tau$. Thus, we only have to perform the integral over $\tau$ of the terms in the second line of~\eqref{eq:final-type-IIB-partition-function}. We can compute this integral explicitly by writing out the false theta functions~\eqref{eq:false-theta-function}. We have
\begin{equation}
\begin{aligned}
 &
\sqrt{k} \, \Im \dualtau
\sum_{\ell =0}^{\frac{k}{2} - 1}\,
\intStrip\frac{\dd^2 \tau}{(\Im \tau)^{\frac{3}{2}}}\,
\mathrm{e}^{-\frac{k \pi (\Im \dualtau)^2}{\Im \tau}} \,
\Psi_{k,\ell}(\dualz \mid \tau)\,
\Psi_{k,\ell}(-\widetilde{\dualz} \mid -\bar{\tau})
\\
 & \qquad
= \sqrt{k} \Im \dualtau \,
\sum_{\ell =0}^{\frac{k}{2} - 1}\sum_{m,\widetilde{m} \in \ZZ}
\mathop{\text{sgn}}(\llm)\, \mathop{\text{sgn}}(\llmt) \;
\dualy^{|\llm|}\; \dualyt^{|\llmt|}
\\
 &
\qquad\qquad\times
\intStrip\frac{\dd^2 \tau}{(\Im \tau)^{\frac{3}{2}}}\,
\mathrm{e}^{-\frac{k \pi (\Im \dualtau)^2}{\Im \tau}}\,
q^{\frac{1}{k}\, \llm^2} \,
\bar{q}^{\frac{1}{k}\, \llmt^2}
\\
 & \qquad=\sum_{\ell =0}^{\frac{k}{2} - 1} \; \sum_{m \in \ZZ}
    (\dualq\, \dualqb\, \dualy\, \dualyt)^{|\llm|}  \;
\equiv
\sum_{L \in \frac{1}{2} \NN \setminus \frac{k}{2} \NN} \,
    (\dualq\,\dualqb\, \dualy\, \dualyt)^{L} \,.
\end{aligned}
\end{equation}
Here, we have used that the integral over $\Re \tau$ imposes $\llm^2=\llmt^2$ from which we infer $m = \widetilde{m}$ (for this we need to use that $\ell \in \{0,\frac{1}{2},\dots,\frac{k-2}{2}\}$).  In the final relation we also simplified notation and set  $L= \abs{\llm}$. Notice that when $\ell$ runs over $\{0,\frac{1}{2},\dots,\frac{k-2}{2}\}$ and $m$ runs over the integers, $L$ runs over $L \in \frac{1}{2}\ZZ_{\ge 0} \setminus \frac{k}{2} \ZZ_{\ge 0}$.

Thus, the $\frac{1}{2}$-BPS spectrum originating from the worldsheet zero modes reads
\begin{equation}
\begin{aligned}
\dualfE^{\text{IIB}}_{\frac{1}{2}\text{-BPS}}(\dualz,\, \widetilde{\dualz},\, \dualtau)
 & =
\sum_{L \in \frac{1}{2} \NN \setminus \frac{k}{2} \NN} \sum_{\alpha, \widetilde{\alpha}=\pm 1} \,
\alpha\, \widetilde{\alpha} \,
    (\dualq\,  \dualqb\, \dualy^{\alpha}\, \dualyt^{\widetilde{\alpha}})^{L}
\bigg(20+\sum_{\sigma,\widetilde{\sigma}=\pm 1} \,
\dualq^{\frac{\sigma}{2}} \, \dualqb^{\frac{\widetilde{\sigma}}{2}} \,
\dualy^{\frac{\sigma\,\alpha}{2}} \, \dualyt^{\frac{\widetilde{\sigma} \,\widetilde{\alpha}}{2}}
\bigg)
\\
 & \qquad
\times \frac{
    (\dualq^{-\frac{1}{4}} \, \dualy^{\frac{\alpha}{4}} -\dualq^{\frac{1}{4}} \, \dualy^{-\frac{\alpha}{4}})^2\, (\dualqb^{-\frac{1}{4}}\, \dualyt^{\frac{\widetilde{\alpha}}{4}} - \dualqb^{\frac{1}{4}} \, \dualyt^{-\frac{\widetilde{\alpha}}{4}})^2}{(\dualq^{\frac{1}{2}} -\dualq^{-\frac{1}{2}}) \,
    (\dualqb^{\frac{1}{2}}-\dualqb^{-\frac{1}{2}})\,
    (\dualy^{\frac{1}{2}}-\dualy^{-\frac{1}{2}})\,
    (\dualyt^{\frac{1}{2}}-\dualyt^{-\frac{1}{2}})} \,.
\end{aligned}
\label{eq:1/2-BPS-partition-function-1}
\end{equation}

\paragraph{Decomposition into characters:} It remains to decompose this into (short) characters of the global $\mathcal{N}=4$ superalgebra $\mathfrak{psu}(1,1|2)$, which take the form
\begin{equation}
\chpsu_j(\dualz,\, \dualtau)
=\frac{\dualq^j}{1-\dualq}\,
\bqty{\chi_j(\dualz)
  - 2\, \dualq^{\frac{1}{2}} \, \chi_{j-\frac{1}{2}}(\dualz)
  +\dualq \, \chi_{j-1}(\dualz)} \,,
\label{eq:global-N4-character}
\end{equation}
with
\begin{equation}
\chi_j(\dualz)
=\sum_{m=-j}^j \, \dualy^m
\,=\,
\frac{\dualy^{j+\frac{1}{2}}-\dualy^{-j-\frac{1}{2}}}{\dualy^{\frac{1}{2}}-\dualy^{-\frac{1}{2}}}\,,
\label{eq:su2-character}
\end{equation}
the standard $\mathfrak{su}(2)$ character.
Formally $\chi_{-\frac{1}{2}}(\dualz)=0$ and $\chi_{-1}(\dualz) = -\chi_0(\dualz)=-1$
and thus~\eqref{eq:global-N4-character} holds even for the exceptional cases
$j=0$ and $j=\frac{1}{2}$.

With these definitions, we can rewrite~\eqref{eq:1/2-BPS-partition-function-1} as
\begin{equation}\label{eq:half-BPS-IIB}
\begin{aligned}
\dualfE_{\frac{1}{2}\text{-BPS}}^{^{\mathfrak{psu}(1,1|2)}}(\dualz,\, \widetilde{\dualz},\, \dualtau)
 & =
\sum_{L \in \frac{1}{2} \NN \setminus \frac{k}{2} \NN} \,
\bigg( 20\, \chpsu_L(\dualz,\, \dualtau)\;
\chpsu_L(\widetilde{\dualz},\, -\overline{\dualtau})
\\
 & \qquad \qquad
+ \sum_{\sigma,\widetilde{\sigma}=\pm 1} \,
\chpsu_{L+\frac{\sigma}{2}}(\dualz,\, \dualtau)\;
\chpsu_{L+\frac{\widetilde{\sigma}}{2}}(\widetilde{\dualz},\, -\overline{\dualtau})\bigg) \,.
\end{aligned}
\end{equation}
In the $k \to \infty$ limit, this matches exactly the supergravity spectrum~\eqref{eq:supergravity-spectrum}. For finite $k$, one observes gaps in the spectrum, namely every
$k^{\mathrm{th}}$ Hodge diamond is missing. This phenomenon of the ``missing chiral primaries'' is well-known in the literature~\cite{Seiberg:1999xz}.  It is a consequence of the singular nature of the pure NS-NS background. Roughly speaking, the long-string sector (or the small instanton singularity in the gauge theory picture) can be thought of as having ``eaten up'' these BPS states~\cite{Seiberg:1999xz}.

\subsection{Long string spectrum}\label{subsec:long-string-spectrum}

We now extract the long-string spectrum from the worldsheet one-loop amplitude.
According to the conjectures of~\cite{Seiberg:1999xz, Eberhardt:2019qcl, Eberhardt:2021vsx, Balthazar:2021xeh} (see also~\cite{Argurio:2000tb}), it should match the spectrum of the symmetric orbifold of an $\mathcal{N}=4$ linear dilaton theory of central charge $c=6(k-1)$ tensored with the K3 sigma model.
Verifying this provides evidence both for the proposed duality and for the expression~\eqref{eq:final-type-IIB-partition-function}. A closely related analysis was performed in~\cite{Eberhardt:2020bgq} for the tensionless case $k=1$, where the linear dilaton factor is absent.

\paragraph{Long-string states as an IR-divergence:}  Long-string states originate from the poles of~\eqref{eq:final-type-IIB-partition-function}, located at
\begin{equation}
\dualtau = w \, \tau - r\,, \qquad w,\, r\in\ZZ \,.
\end{equation}
Since $\Im\tau>0$ and $\Im\dualtau>0$, only $w>0$ contributes. Working in the fundamental strip, we may furthermore restrict to $r\in\{0,\ldots,w-1\}$.

The poles lead to a logarithmic divergence of the modular integral, reflecting the continuum of long-string states.
We define the long-string contribution as the coefficient of this logarithmic divergence in a regulated integral obtained by excising small disks around the polar loci.
Concretely, near $\dualtau=w\tau-r$ we set
\begin{equation}
u := \dualtau - w\, \tau + r\,,
\end{equation}
and use the local expansion
\begin{equation}
\frac{1}{|\vartheta_1(u\mid\tau)|^2}
=\frac{1}{4\pi^2|\eta(\tau)|^6}\,\frac{1}{|u|^2}+ \cdots  \,.
\end{equation}
Regulating the integral by excising $|u|<\varepsilon$, one uses
\begin{equation}
\int_{|u|>\varepsilon}\frac{\mathrm{d}^2u}{|u|^2}
=2\pi\log\!\Big(\frac{1}{\varepsilon}\Big)+\mathcal{O}(\varepsilon^0)\,,
\label{eq:log_div_u}
\end{equation}
so the coefficient of $\log(1/\varepsilon)$ is obtained by evaluating the remaining (regular) integrand at $u=0$, i.e., at $\tau=(\dualtau+r)/w$.

Equivalently---at the level of extracting the $\log(1/\varepsilon)$ coefficient only---we may use the distributional shorthand
\begin{equation}
\frac{1}{|u|^2}\ \rightsquigarrow\ 2\pi\log\!\Big(\frac{1}{\varepsilon}\Big)\,\delta^2(u)\,,
\label{eq:delta_shorthand}
\end{equation}
which is what we implement below.\footnote{This procedure isolates the universal ``volume term'' of the long-string continuum.  A more refined treatment keeps the full regulator dependence and yields the density of states, including a finite piece controlled by the long-string phase shift; see e.g.,~\cite{Maldacena:2000kv}.}

At the tensionless point $k=1$ (where the RNS description has to be replaced by the hybrid formalism), the localization of the worldsheet integral becomes literal, so that~\eqref{eq:delta_shorthand} can be viewed as exact there~\cite{Eberhardt:2019ywk, Dei:2020zui, Eberhardt:2020bgq}.

We notice that the $\widetilde{\mathrm{R}}$ sector does not contribute to the long-string continuum: its would-be pole at $\dualtau=w\,\tau-r$ is canceled by a corresponding zero of $\vartheta_1(\dualtau\mid\tau)$ in the numerator, cf.~\eqref{eq:K3-sum-over-spin-structures}.
We may therefore work with~\eqref{eq:naive-type-II-partition-function} and set $\alpha=\widetilde{\alpha}=1$, since the final long-string answer is independent of this choice.

\paragraph{The long-string partition function:} With these replacements, the long-string sector contribution to the spacetime free energy becomes
\begin{equation}
\begin{aligned}
\dualfE^\text{IIB}_\text{long}(\dualz,\, \widetilde{\dualz},\, \dualtau)
 &
\propto\sqrt{k}\Im \dualtau \sum_{w=1}^\infty \sum_{r=0}^{w-1} \,
\intStrip\frac{\dd^2 \tau}{(\Im \tau)^{\frac{3}{2}}}\,
\mathrm{e}^{-\frac{k \pi (\Im \dualtau)^2}{\Im \tau}} \,
\delta^2(\dualtau-w\,  \tau+r)
\\
 & \qquad
\times
\pqty{\sum_{\ell =0}^{\frac{k}{2} - 1}
  \abs{\chsu_\ell( \dualz \mid \tau)}^2}
\\
 & \qquad
\times
\abs{\frac{\vartheta_1\big(\tfrac{w \,\tau-r-\dualz}{2}\mid \tau\big)^2}{
    \vartheta_1'(w\,\tau-r\mid\tau)}}^2  \,
\ZK{1,1}(w\, \tau-r+\dualz,w\, \bar{\tau}- r + \widetilde{\dualz} \mid \tau)\,,
\end{aligned}
\end{equation}
where we have used that the presence of the delta-function allows us to replace
$\dualtau \to w \,\tau-r$ everywhere. We absorbed the divergent prefactor in the proportionality sign. The last line only depends on $w \bmod 2$ and $r \bmod 2$,
which determine the spin structure.
It is convenient to encode this in a map $a(w,r)\in\{1,2,3,4\}$ defined by $a(0,0)=1$, $a(0,1)=2$, $a(1,1)=3$ and $a(1,0)=4$, where entries are viewed mod 2. Then trading the integral and the delta function, while correctly accounting for the Jacobian gives us
\begin{align}
\dualfE_\text{long}^{\text{IIB}} \propto
\sum_{w=1}^\infty \,\frac{1}{w}\,\sum_{r=0}^{w-1} \,
\mathrm{e}^{- k \pi w \, \Im \dualtau}
\mathcal{Z}_{a(w,r)}^{\smash[b]{^{\scriptscriptstyle{\mathcal{N}=4 \,\mathrm{LD} \times \mathrm{K3}}}}}
\big(\dualz,\widetilde{\dualz} \mid \tfrac{\dualtau+r}{w}\big) \,.
\label{eq:IIB-long-string-sector-partition-function}
\end{align}
Here
\begin{equation}
\mathcal{Z}_{a}^{\smash[b]{^{\scriptscriptstyle{\mathcal{N}=4 \,\mathrm{LD} \times \mathrm{K3}}}}}
\big(\dualz,\widetilde{\dualz} \mid \tau\big)
=
\underbrace{\frac{1}{\sqrt{\Im \tau}}\,\pqty{\sum_{\ell =0}^{\frac{k}{2} - 1} \,
    \abs{\chsu_{\ell}(\dualz \mid \tau)}^2  \times
    \abs{\frac{\vartheta_a\big(\tfrac{\dualz}{2}\mid \tau\big)^2}{\eta(\tau)^3}}^2}}_{\mathcal{N}=4 \,\text{linear dilaton} }
\times  \, \ZK{a}(\dualz,\widetilde{\dualz}\mid \tau)\,,
\label{eq:IIB-N4dilaton}
\end{equation}
is the partition function of the linear dilaton theory with central charge $c=6(k-1)$ times the K3 sigma model. The $\mathcal{N}=4$ linear dilaton theory can be constructed out of an $\SU(2)_{k-2}$ WZW model, a free boson with a background charge $Q=\frac{k-1}{\sqrt{k}}$ and four free fermions, see
e.g.,~\cite{Eberhardt:2019qcl} for the explicit expressions realizing the $\mathcal{N}=4$ algebra from this building block.

\paragraph{Comparison with the symmetric orbifold:} As first noticed in~\cite{Eberhardt:2019qcl} for $\mathrm{AdS}_3\times\mathbf{S}^3\times\mathbb{T}^4$,~\eqref{eq:IIB-long-string-sector-partition-function} has precisely the form of a single-particle symmetric orbifold partition function.
The factor $\mathrm{e}^{-k \pi w\,\Im\dualtau}$ accounts for the ground-state energy in the $w$-twisted sector, while $\frac{1}{w}\sum_{r=0}^{w-1}$ implements the orbifold projection.
Geometrically, the sum over $w$ and $r$ achieves a (partial) sum over connected coverings of the torus with $\tau=\frac{\dualtau+r}{w}$, the modular parameter of the covering surface.

\subsection{The Type IIA string spectrum}\label{subsec:IIAspec}

We now turn to the Type IIA string. The supergravity spectrum was derived relatively recently in~\cite{Eloy:2020uix}. As noted above we will work with conventions that the compactification preserves $(0,4)$ supersymmetry. This means that the right-movers will organize themselves into $\mathfrak{psu}(1,1|2)$ multiplets. The left-movers will just fit into $\mathfrak{sl}(2,\mathbb{R}) \oplus \mathfrak{su}(2)$ multiplets (though we can choose to further grade them by the $\mathrm{SU}(2)_{\mathrm{aut}}$ symmetry).

  It is useful to present the spectrum using graded characters as in the Type IIB example, despite the broken left-moving supersymmetry. Doing so allows for a compact representation, as part of the left-moving spectrum can be assembled into $\mathfrak{psu}(1,1|2)$ multiplets. We exploit this and present the supergravity spectrum as
  \begin{equation}
  \begin{aligned}
  \bigoplus_{\dualj=\frac{1}{2},1,\dots}^\infty \Bigg[ \bigg(
    &
    \bqty{\dualj + \tfrac{1}{2};\, \dualj-\tfrac{1}{2} }
    \oplus
    \bqty{\dualj-\tfrac{1}{2};\, \dualj + \tfrac{1}{2} }
    \oplus
    \bqty{\dualj-\tfrac{1}{2}; \,\dualj-\tfrac{1}{2}}
    \oplus
    \bqty{\dualj + \tfrac{1}{2}; \,\dualj + \tfrac{1}{2}}
    \\
    & \qquad
    \ominus\,
    2 \cdot \bqty{\dualj;\, \dualj + \tfrac{1}{2} }
    \ominus 2 \cdot \bqty{\dualj; \dualj -\tfrac{1}{2}} \bigg)
    \\
    &
    \bigoplus
    \bigg(\dualj_{\dualj-1} -2\, (\dualj+\tfrac{1}{2})_{\dualj-\frac{1}{2}} + (\dualj+1)_{\dualj} \bigg)_{\mathrm{left}}
    \otimes \bigg(2\cdot (\dualj-\tfrac{1}{2}) + 20\cdot \dualj + 2\cdot (\dualj + \tfrac{1}{2}) \bigg)_{\mathrm{right}} \Bigg] \ .
  \end{aligned}
  \label{eq:supergravity-IIA}
  \end{equation}
  In the first two lines we have presented the spectrum using the short multiplets of the
$\mathcal{N} = (4,4)$ superconformal algebra. The third line comprises the piece that
  breaks the symmetry to $\mathcal{N} = (0,4)$; here we used $h_\ell$ to denote the quantum
  numbers under the  $\mathfrak{sl}(2,\mathbb{R}) \oplus \mathfrak{su}(2)$ symmetry for the
  left, and $\dualj$ for the right-moving short multiplet.
  Compared to the IIB spectrum,
  the main change is in the nature of the left-moving multiplet, where the
$\mathfrak{su}(2)$ quantum number rises in concert with the $\mathfrak{sl}(2)$ weight,
  instead of decreasing as it does in a $\mathfrak{psu}(1,1|2)$ short multiplet.
  Let us therefore introduce
  \begin{equation}
  \nchsusl_j(\dualz,\, \dualtau)
  =\frac{\dualq^j}{1-\dualq}\,
  \bqty{\chi_{j-1}(\dualz)
    - 2\, \dualq^{\frac{1}{2}} \, \chi_{j-\frac{1}{2}}(\dualz)
    +\dualq \, \chi_{j}(\dualz)} \,,
  \label{eq:nonsusy-character}
  \end{equation}
  which will turn out to be helpful in writing the spectrum. Notice the identity
  \begin{equation}\label{eq:IIAB-charid}
  \chpsu_j(\dualz,\, \dualtau) - \nchsusl_j(\dualz,\, \dualtau) = \dualq^j \, \bigl(
  \chi_j(\dualz) - \chi_{j-1}(\dualz) \bigr)\,,
  \end{equation}
  which will be useful in the sequel.

  The main change in~\eqref{eq:final-type-IIB-partition-function} is in the sign of $\alpha$
  in the fermion and K3 partition functions.  The calculation of the short string spectrum
  works along the same lines as in~\cref{subsec:12-BPS-spectrum}; it suffices to
  replace $\alpha \to -\alpha$ in these factors.
  The states in the right-moving short multiplets arise from the zero mode piece,
  which assembles into
  \begin{equation}
  \begin{aligned}
  \dualfE^{\text{IIA}}_{\frac{1}{2}\text{-BPS, right}}(\dualz,\, \widetilde{\dualz},\,\dualtau)
   & =
  \sum_{L \in \frac{1}{2} \NN \setminus \frac{k}{2} \NN}
  \sum_{\alpha, \widetilde{\alpha}=\pm 1} \,
  \alpha\, \widetilde{\alpha} \,
    (\dualq\,  \dualqb\, \dualy^{\alpha}\, \dualyt^{\widetilde{\alpha}})^{L}
  \bigg(20+\sum_{\sigma,\widetilde{\sigma}=\pm 1} \,
  \dualq^{\frac{\sigma}{2}} \, \dualqb^{\frac{\widetilde{\sigma}}{2}} \,
  \dualy^{-\frac{\sigma\,\alpha}{2}} \, \dualyt^{\frac{\widetilde{\sigma} \,\widetilde{\alpha}}{2}}
  \bigg)
  \\
   & \qquad
  \times \frac{
    (\dualq^{-\frac{1}{4}} \, \dualy^{-\frac{\alpha}{4}} -\dualq^{\frac{1}{4}} \, \dualy^{\frac{\alpha}{4}})^2\, (\dualqb^{-\frac{1}{4}}\, \dualyt^{\frac{\widetilde{\alpha}}{4}} - \dualqb^{\frac{1}{4}} \, \dualyt^{-\frac{\widetilde{\alpha}}{4}})^2}{(\dualq^{\frac{1}{2}} -\dualq^{-\frac{1}{2}}) \,
    (\dualqb^{\frac{1}{2}}-\dualqb^{-\frac{1}{2}})\,
    (\dualy^{\frac{1}{2}}-\dualy^{-\frac{1}{2}})\,
    (\dualyt^{\frac{1}{2}}-\dualyt^{-\frac{1}{2}})} \,.
  \end{aligned}
  \label{eq:IIA-partition-function-1}
  \end{equation}
  Using the $\mathfrak{psu}(1,1|2)$ short character~\eqref{eq:global-N4-character},
  and the newly introduced combination~\eqref{eq:nonsusy-character}, we can simplify
  this expression to
  \begin{equation}\label{eq:IIA-shortstrings}
  \begin{aligned}
    &
  \dualfE^{\text{IIA}}_{\frac{1}{2}\text{-BPS, right}}(\dualz,\, \widetilde{\dualz},\, \dualtau)
  \\
  \quad & =
  \sum_{L \in \frac{1}{2} \NN \setminus \frac{k}{2} \NN} \,
  \Bigg[
    20\, \nchsusl_L(\dualz,\, \dualtau)\,
    \chpsu_L(\widetilde{\dualz},-\overline{\dualtau})
    \\
    &\qquad
    + \bigg(
    \sum_{\sigma = \pm 1} \chpsu_{L + \frac{\sigma}{2}}(\dualz, \dualtau)
    -2\chpsu_L(\dualz,\dualtau) + 2\, \nchsusl_L(\dualz,\, \dualtau)
    \bigg) \bigg(
    \sum_{\widetilde{\sigma} = \pm 1} \, \chpsu_{L+ \frac{\widetilde{\sigma}}{2}}(\widetilde{\dualz},-\overline{\dualtau})
    \bigg)
    \Bigg]\,.
  \end{aligned}
  \end{equation}
  One can check that this agrees with the supergravity spectrum~\eqref{eq:supergravity-IIA} in the $k\to \infty$ limit. The long string spectrum can similarly be worked out along the lines described in~\cref{subsec:long-string-spectrum}. Further, specializing the right-moving short character by setting $\widetilde{\dualz} = \overline{\dualtau}$ we obtain the following:
  \begin{equation}
  \left.\dualfE^{\text{IIA}}_{\frac{1}{2}\text{-BPS, right}}
  \right|_{\widetilde{\dualz}=\overline{\dualtau}}
  =\sum_{L \in \frac{1}{2} \NN \setminus \frac{k}{2} \NN}
  \left(
  2\chpsu_{L-\frac12}+2\chpsu_{L+\frac12}
  -4\chpsu_L+24\nchsusl_L
  \right).
  \end{equation}
  The difference between the IIB and IIA short string part of the spectra is captured by this expression. Explicitly,
  \begin{equation}\label{eq:AB-lmovers}
  \left[\dualfE^{\text{IIB}}_{\frac{1}{2}\text{-BPS, right}} -
    \dualfE^{\text{IIA}}_{\frac{1}{2}\text{-BPS, right}}
    \right]_{\widetilde{\dualz}=\overline{\dualtau}}
  =
  24\, \sum_{L \in \frac{1}{2}\, \NN \setminus \frac{k}{2}\NN} \, \dualq^L\,\bigg[\chi_L(\dualz)
    - \chi_{L-1}(\dualz)\bigg]\,.
  \end{equation}
  The difference is controlled by the K3 Euler characteristic, $\chi(\mathrm{K3}) =24$, obtained from the Ramond sector Witten index in the CFT.  This is a manifestation (in the partition function) of the recent observation in~\cite{Sen:2026xfy} that the IIB and IIA theories differ by a contribution to the one-loop effective action proportional to the spacetime Euler character.

  \subsection{A non-supersymmetric string spectrum}\label{subsec:nonsusy}

  As indicated in~\cref{subsec:Zspacetime} we can also construct a non-supersymmetric partition function~\eqref{eq:final-non-susy-partition-function}. Its spectrum may be analyzed similarly to the IIB/IIA theories. The discrete states lead to a single string spectrum
  \begin{equation}
  \begin{aligned}
  \dualfE^{\text{non-susy}}(\dualz,\widetilde{\dualz}, \dualtau)
   & =
  \sum_{L \in \frac{1}{2}\, \NN \setminus \frac{k}{2}\NN}\,
  \bqty{20\,\nchsusl_j(\dualz,\, \dualtau)\, \nchsusl_j(\widetilde{\dualz},\, \overline{\dualtau}) + \mathfrak{u}_L(\dualz,\dualtau)\, \mathfrak{u}_L(\widetilde{\dualz},\overline{\dualtau})}
  \\
  \mathfrak{u}_L(\dualz,\dualtau)
   & =
  \chpsu_{L-\frac12}(\dualz,\, \dualtau) + \chpsu_{L+\frac12}(\dualz,\, \dualtau) - 2\, \chpsu_{L}(\dualz,\, \dualtau) + 2\, \nchsusl_j(\dualz,\, \dualtau)\,.
  \end{aligned}
  \end{equation}
  At low value of $L$, the multiplets truncate; this is captured by our convention for $\mathfrak{su}(2)$ characters.

  The spacetime conformal dimension is bounded from below, with the lowest non-vacuum state having quantum numbers $h = \bar{h} =\frac{1}{2}$ (and $\mathrm{su}(2)$ spins $\frac{1}{2}$ as well). This operator saturates the $\mathrm{AdS}_3$ Breitenlohner-Freedman bound. The long string states also have their dimensions bounded from below $h \geq \frac{k\,w}{4} + \frac{1-2\,k}{4\,k\,w}$ in the $w^{\mathrm{th}}$ spectral flowed sector. These statements suffice to establish that the spectrum is tachyon free. It would be interesting to analyze this theory further and check whether it is perturbatively stable.

  \section{Spacetime NS-sector elliptic genus}\label{sec:NS-elliptic-genus}

  In this section we extract the spacetime NS-sector elliptic genus from the
  worldsheet partition function~\eqref{eq:final-type-IIB-partition-function}.
  The main subtlety compared to the $\frac{1}{2}$-BPS extraction in~\cref{subsec:12-BPS-spectrum}
  is that the integrand is \emph{meromorphic} on the $\tau$-strip due to the factor
$\vartheta_1(\dualtau\mid\tau)^{-1}$, and one must integrate in $\Im \tau$ piece by piece.

  \subsection{Setup}

  \paragraph{Definition:}  We define the spacetime NS elliptic genus as the specialization
  \begin{equation}
  \dualEG(\dualz,\, \dualtau) :=
  \dualPF(\dualz,\,\widetilde{\dualz}=\overline{\dualtau},\,\dualtau)=\dualPF(\dualz,\,\widetilde{\dualz}=-\overline{\dualtau},\,\dualtau)\,.
  \end{equation}
  We will keep $\dualz$ as the left-moving $\mathrm{SU}(2)$ chemical potential and set the right-moving
  chemical potential to $\widetilde{\dualz}=\pm \overline{\dualtau}$. Both sign choices give the same result thanks to the $\mathrm{SU}(2)$ Weyl reflection symmetry.
  In the following, we will find it convenient to work directly with the single-string
  free energy, and define the analogous contribution to the elliptic genus as
  \begin{equation}
  \dualEGsp(\dualz,\, \dualtau) :=
  \dualfE(\dualz,\,\widetilde{\dualz}=\overline{\dualtau},\,\dualtau)\,.
  \label{eq:singlestringEG}
  \end{equation}
  It will transpire that one can treat the Type IIB and IIA theories in parallel. We indicate the differences in the course of the discussion, and keep the discussion fairly general for the most part.

  To extract the elliptic genus from our single-particle free energy, we need to first examine the pieces that survive when we set the right-moving chemical potential equal to the antiholomorphic spacetime modular parameter.  We will then focus on how to complete the integral over the worldsheet modular parameter to extract the contribution from the short and long string sectors.\footnote{As we explain in the course of our evaluation below, it is prudent to first specialize to the locus $\widetilde{\dualz} = \overline{\dualtau}$ before evaluating the integral. Attempting to specialize the integrated expression for the single-string free energy instead results in a nonuniform expansion, and the limit is delicate to extract. We illustrate this with a simple toy model of the integral in~\cref{app:toymodels}. \label{fn:specializefirst} }

  \paragraph{Specialization to $\widetilde{\dualz}=\overline{\dualtau}$:}
  In~\eqref{eq:final-type-IIB-partition-function} the right-moving theta factor becomes
  \begin{equation}
  \vartheta_1\big(-\tfrac{\overline{\dualtau}-\widetilde{\alpha}\, \widetilde{\dualz}}{2}\mid-\bar\tau\big)^2
  =\vartheta_1\big(-\tfrac{(1-\widetilde{\alpha})\, \overline{\dualtau}}{2}\mid -\bar\tau\big)^2\,,
  \end{equation}
  hence the $\widetilde{\alpha}=+1$ contribution vanishes and only $\widetilde{\alpha}=-1$ survives.
  Moreover, for $\widetilde{\alpha}=-1$ the anti-holomorphic factor cancels against the
  anti-holomorphic part of $|\vartheta_1(\dualtau\mid\tau)\vartheta_1(\dualz \mid \tau)|^2$, so the only remaining pole
  in the strip is the \emph{holomorphic} pole of $\vartheta_1(\dualtau\mid\tau)^{-1}$.
  The K3 factor reduces to
  \begin{equation}
  \ZK{1,1}(\dualtau+\alpha\,\dualz,\,0\mid\tau)
  =\EK{1}(\dualtau+\alpha \,\dualz \mid \tau)\,,
  \end{equation}
  where $\EK{1}$ is the $\widetilde{\text{R}}$-sector K3 elliptic genus given by~\eqref{eq:K3-elliptic-genus} (see~\eqref{eq:K3-elliptic-genera} for the other spin structures).

  It is convenient to package the integrand for the worldsheet moduli integration as follows. First, introduce the function
  \begin{equation}
  \mathcal{G}_{k,\ell}^\sigma(\dualz,\,\dualtau \mid \tau)
  :=
  \sum_{\alpha=\pm} \alpha \,
  \Psi_{k,\ell}(\alpha\, \dualz \mid \tau)
  \frac{\vartheta_1\big(\tfrac{\dualtau-\sigma\,\alpha\,\dualz}{2}\mid\tau\big)^2\,
    \EK{1}(\dualtau+\sigma\,\alpha \,\dualz \mid \tau)}{\vartheta_1(\dualtau \mid \tau)\, \vartheta_1(\dualz\mid\tau)} \,,
  \label{eq:def-Galpha}
  \end{equation}
  where $\sigma=+$ for Type IIB and $\sigma=-$ for Type IIA.
  We will denote $\tau=\tau_1+i \tau_2$.
  With this preparation we can write the integral for the single-string contribution to the spacetime elliptic genus as
  \begin{equation}
  \dualEGsp_{\sigma}(\dualz,\, \dualtau)
  =
  -\sqrt{k} \Im \dualtau\,
  \int_{-\frac{1}{2}}^{\frac{1}{2}}\dd\tau_1\int_0^\infty \frac{\dd\tau_2}{\tau_2^{3/2}}\,
  \mathrm{e}^{ -\frac{k\pi (\Im \dualtau)^2}{\tau_2}}\,
  \sum_{\ell =0}^{\frac{k}{2} - 1}\
  \Psi_{k,\ell}(\overline{\dualtau}\mid-\bar\tau) \, \mathcal{G}_{k,\ell}^\sigma(\dualz, \, \dualtau\mid \tau) \,.
  \label{eq:EG-after-specialization}
  \end{equation}
  %

  \subsection{Anchoring expectations: symmetric orbifolds and supergravity}\label{sec:anchor}

  Before undertaking the evaluation of~\eqref{eq:EG-after-specialization} let us ask what we should expect the result to be. The spectrum (in the IIB case) exhibits a structure that is consistent with the symmetric orbifold interpretation~\eqref{eq:IIB-long-string-sector-partition-function}. Let us see first what this would imply for the single-particle elliptic genus, and then further refine that expectation using the large $k$ supergravity result, which is known.

  \paragraph{A symmetric orbifold perspective:}  For a symmetric orbifold SCFT, the  DMVV formula~\cite{Dijkgraaf:1996xw} predicts the following answer for the generating function of Ramond sector elliptic genera
  \begin{equation}
  \sum_{N=0}^\infty p^N \dualEG(\text{Sym}^N(\dualz,\, \dualtau))=\exp\bigg(\sum_{n=1}^\infty\, \frac{1}{n}\sum_{d=1}^\infty\, \frac{p^{nd}}{d}\,\sum_{b=0}^{d-1} \,
  \mathscr{E}_{\mathrm{seed}}(n\dualz,\, \tfrac{n\dualtau-b}{d})\bigg)\ .
  \end{equation}
  Here $\mathscr{E}_{\mathrm{seed}}(\dualz,\dualtau)$ is the seed elliptic genus, and $p$ is the grand canonical fugacity conjugate to $N$,
  the degree of the symmetric orbifold.  Consequently, the single-particle elliptic genus itself should have the form of an infinite sum
  \begin{equation}\label{eq:dmvvsp}
  \dualEGsp(\dualz,\,\dualtau)
  =
  \sum_{d=1}^\infty \,
  \frac{p^{d}}{d}\sum_{b=0}^{d-1} \mathscr{E}_{\mathrm{seed}}(\dualz,\, \tfrac{\dualtau-b}{d})\ .
  \end{equation}
  On the worldsheet there is no natural way to turn on the fugacity dual to the degree of the symmetric group. What one obtains is the $p=1$ answer; we will indeed see a structure of sums that are suggestively of the form~\eqref{eq:dmvvsp}. However, we are not a priori guaranteed convergence of the sum as $p \to 1$; it could diverge because of zero modes. To get a sense of what we should anticipate we will therefore examine the answer in the supergravity limit $k\gg 1$.

  \paragraph{The elliptic genus in supergravity:}
  The elliptic genus in supergravity was originally computed in~\cite{deBoer:1998us}. A specific notion of fugacity was introduced somewhat artificially on the supergravity side mainly as a bookkeeping device. We will review this below, and see that the single-particle elliptic genus is actually finite, while the full Fock space result after multiparticling has a divergence.

  To obtain the single-particle elliptic genus in supergravity we start from the spectrum  given in~\eqref{eq:half-BPS-IIB}. Setting $\widetilde{\dualz} = \overline{\dualtau}$ we have that the right-moving character in~\eqref{eq:global-N4-character} reduces to unity. Following~\cite{deBoer:1998us}, we introduce a degree and a related fugacity $p_{\mathrm{deg}}$ that keeps track of the symmetric orbifold structure. We have differentiated this label from the symmetric orbifold fugacity since ab initio we are not guaranteed that the two quantities are one and the same.
  Computing the degree for each of the representations and summing over the spectrum one arrives at~\cite[Eq.~(5.9)]{deBoer:1998us}, which we reproduce here:
  \begin{equation}
  \begin{aligned}
  \dualEGsp_{\mathrm{sugra}}(p_{\mathrm{deg}},\dualz, \dualtau)
  = &
  2\,p_{\mathrm{deg}} + \mathfrak{f}(p_{\mathrm{deg}}, \dualz; \dualtau)
  \\
  \mathfrak{f}(p_{\mathrm{deg}}, \dualz, \dualtau)
    & =
  \frac{1}{(1-\dualq)\,(\dualy^{\frac{1}{2}} - \dualy^{-\frac{1}{2}})}
  \, \bigg[
    (20\, p_{\mathrm{deg}} + 2\, p_{\mathrm{deg}}^2)\, \pqty{
    \dualq^{\frac{1}{2}}\, (\dualy - \dualy^{-1}) -2\, \dualq\, (\dualy^{\frac{1}{2}} - \dualy^{-\frac{1}{2}})}
    \\
    &\qquad
    + \frac{2\,p_{\mathrm{deg}} + 20\,p_{\mathrm{deg}}^2 + 2\,p_{\mathrm{deg}}^3}{1-p_{\mathrm{deg}}\, \dualq^{\frac{1}{2}}\ \dualy^{\frac{1}{2}}}
    \pqty{\dualq\, \dualy^{\frac{3}{2}} - 2\, \dualq^{\frac{3}{2}}\, \dualy + \dualq^2 \, \dualy^{\frac{1}{2}}}
    \\
    &\qquad
    - \frac{2\,p_{\mathrm{deg}} + 20\,p_{\mathrm{deg}}^2 + 2\,p_{\mathrm{deg}}^3}{1-p_{\mathrm{deg}}\, \dualq^{\frac{1}{2}}\ \dualy^{-\frac{1}{2}}}
    \pqty{\dualq\, \dualy^{-\frac{3}{2}} - 2\, \dualq^{\frac{3}{2}}\, \dualy^{-1}
    + \dualq^2 \, \dualy^{-\frac{1}{2}}}
    \bigg] \ .
  \end{aligned}
  \end{equation}
  We have retained the left-moving $\mathfrak{su}(2)_R$ chemical potential; one could rewrite this factor into a standard $\mathrm{su}(2)$ character $\chi(\dualz)$, but we won't need it at present.

  The key point we want to highlight is that this single-particle elliptic genus has a finite limit as $p_{\mathrm{deg}}\to 1$, as one can readily check.  It also presages the divergence in the multiparticle elliptic genus. The latter is obtained by standard Fock space multiparticling (plethystic exponentiation), now using $p_{\mathrm{deg}}$ as the fugacity
  \begin{equation}
  \dualEG_{\mathrm{sugra}}(p_{\mathrm{deg}},\dualz,\dualtau)
  =
  \exp(\sum_{r=1}^\infty\, \frac{1}{r}\, \dualEGsp_{\mathrm{sugra}}(p_{\mathrm{deg}}^r, r\,\dualz, r\,\dualtau ))\,.
  \end{equation}
  The presence of two NS sector ground states, viz., the  $2p_{\mathrm{deg}}$ piece in the single-particle answer, which we isolated upfront, leads to a double pole in the limit $p_{\mathrm{deg}} \to 1$:
  \begin{equation}
  \dualEG_{\mathrm{sugra}}(p_{\mathrm{deg}},\dualz,\dualtau)
  = \frac{1}{(p_{\mathrm{deg}}-1)^2} \,
  \exp(\sum_{r\geq 1}\, \frac{1}{r} \, \mathfrak{f}(p^r_{\mathrm{deg}}, r\,\dualz, r\, \dualtau))
  + \order{(p_{\mathrm{deg}} -1)^{-1}} \,.
  \end{equation}
  Our task is to understand how to compute finite $k$ (i.e., string tension) corrections to this result. What we should anticipate based on the above is that there continue to be two NS ground states at any finite $k$, leading to a finite single-particle answer, and a similar double pole for the full (multiparticled) elliptic genus.

  \subsection{Integration strategies and pitfalls}\label{sec:eg-pitfalls}

  Recall that we wish to complete the integral in~\eqref{eq:EG-after-specialization}, where we have already specialized to the locus
$\widetilde{\dualz} =\overline{\dualtau}$.  Ideally, one would like to extract a $\dualq$ series, with coefficients organized in $\mathfrak{su}(2)$ characters. However, the presence of long string states, which signals non-compactness of the spacetime CFT, suggests that we should be open to the possibility that the result may not be holomorphic. A suitable ansatz would be something like
  \begin{equation}\label{eq:eg-ansatz}
  \dualEGsp(\dualz, \dualtau)
  = \sum_{N \in \frac{1}{2}\mathbb{Z}_{\geq 0}}\, \mathfrak{a}_N(\dualz)\, \dualq^N
  + \sum_{N \in \frac{1}{2}\mathbb{Z}}\,
  \mathfrak{b}_N(\dualz,\Im\dualtau) \, \dualq^N  \,.
  \end{equation}
  We have isolated the holomorphic data into $\mathfrak{a}_N$, and let $\mathfrak{b}_N$ capture the real-analytic contribution from the long strings. These are the quantities we wish to extract.

  Consider the integrand
  \begin{align}
  \sum_{\ell=0}^{\frac{k}{2}-1} \,
  \Psi_{k,\ell}(\overline{\dualtau} \mid -\bar{\tau}) \,
  \mathcal{G}_{k,\ell}^\sigma(\dualz,\dualtau \mid \tau)
  =
  \sum_{\ell=0}^{\frac{k}{2}-1} \, \mathcal{G}_{k,\ell}^\sigma(\dualz,\dualtau \mid \tau)\,
  \sum_{L \in \ell+\frac{1}{2}+k \ZZ} \sgn(L)\, \dualqb^{-|L|} \, \bar{q}^{\frac{1}{k} L^2}\,.
  \end{align}
  The anti-holomorphic $\bar{q}$ dependence is simple, and one clearly sees the presence of $\overline{\dualtau}$ dependence.  As long as we can control
  the holomorphic piece of the integrand, the function $\mathcal{G}_{k,\ell}^\sigma$ defined in~\eqref{eq:def-Galpha}, we should be able to complete the integral. If we could expand this function in a $q$-series, then the  $\tau_1$ integral would force us to pick Fourier coefficients $\tfrac{1}{k}\, L^2$ by level matching, and leave us with an integral over $\tau_2$. The function $\mathcal{G}_{k,\ell}^\sigma$, however, has poles from the zeros of $\vartheta_1(\dualtau \mid \tau)$. This implies that the Fourier expansion is dependent on the ratio $\frac{\Im\dualtau}{\tau_2}$.

  One potential strategy is to factor out
$\frac{\eta(\tau)^3}{\vartheta_1(\dualtau\mid \tau)}$ in the integrand. This allows us to make use of the expansion due to D.~Zagier (its derivation is discussed in~\cite[Appendix A]{Ferko:2024uxi})
  \begin{equation}\label{eq:theta1Fourier}
  \frac{\eta^3(\tau)}{\vartheta_1(\dualtau|\tau) }
  = \sum_{n \in \mathbb{Z}}\, \sum_{d\geq  w +1}\, (-1)^{d}\, q^{\frac{d}{2}(d+1-2n)} \,\dualq^{n - \frac{1}{2}} \,,
  \end{equation}
  which is valid when
  \begin{equation}\label{eq:wcells}
  w < \frac{\Im(\dualtau)}{\Im(\tau)} < w+1\,.
  \end{equation}
  Using this expansion allows us to treat the poles of $\vartheta_1(\dualtau \mid \tau)$ on the integration region, which lie on the lines $\Im\dualtau = w \, \tau_2$, with $w \in \mathbb{Z}_{\geq 0}$.

  However, as we show in~\cref{app:toymodels} this strategy for evaluating the integral is fraught with pitfalls. For one, there is a piece of $\mathcal{G}_{k,\ell}^\sigma$ which is free of poles; this term is best isolated from the rest and integrated independently. For another, while the $\tau_2$ integral can be completed in terms of error functions, and decomposed to be put in the form~\eqref{eq:eg-ansatz}, the holomorphic coefficients $\mathfrak{a}_N$ get contributions from arbitrarily large values of $d$. This obscures the general structure. Introducing a fugacity by hand to Abel regulate the sum, viz., replacing $\frac{1}{d}$ in the result by $\frac{p^d}{d}$ helps. However, the resulting sum generally only converges inside the unit $p$ disk, i.e., for some $\abs{p} < 1$. The Abel limit $p\to 1^-$ may be computed and shown in the end to reproduce the answer for the elliptic genus.

  In practice, we found it useful to first separate out the contribution from the $\frac{1}{2}$-BPS states. This can be done unambiguously and only the short strings contribute. Then we can attempt to extract the coefficient of the $\dualq^N$ term using the fugacity regulated sum and then take the limit where the fugacity is removed. A discrepancy from the $\frac{1}{2}$-BPS result is helpful, for it suggests a missing boundary contribution. With careful bookkeeping one can obtain the full answer. Not isolating the term free of poles in $\mathcal{G}_{k,\ell}^\sigma$ ab initio makes the evaluation more involved.

  Any indeterminacy in the $p \to 1$ limit has to originate from short distance contributions from the region $\tau_2 \to 0$ on the worldsheet being missed. The expansion~\eqref{eq:theta1Fourier} relies on the inequality~\eqref{eq:wcells}. The introduction of a regulator $p$ suppresses the region $\tau_2 < -\log p\, \Im\dualtau$. The small $\tau_2$ region of the moduli space translates in spacetime language to control over the large $\Im\dualtau$ behavior of the elliptic genus coefficients. The above strategies suggest that potential lack of convergence of the regulated sum is tied to Cardy type growth of the spacetime elliptic genus coefficients.

  The observations above are unduly pessimistic; the integral~\eqref{eq:EG-after-specialization} can be evaluated into a convergent series after worldsheet modular integration. We motivated a fugacity regulator $p$ with delicate $p\to 1$ Abel limit, which appears to imply that the worldsheet modular integral has divergences arising from the UV region $\tau_2 \to 0$. However, for a fixed value of $\Im\dualtau$ this region is highly damped in the integrand. The IR region of the worldsheet is also under control. A saddle point analysis indicates that the contributions come from a localized region,  $\tau_2 \propto k\,\Im \dualtau$, on the worldsheet. The issue with using~\eqref{eq:theta1Fourier} is that the expansion is poorly ordered, and presents the manifestly convergent answer in a suboptimal form.
  There is a better way to expand the denominator without mixing the worldsheet and spacetime modular parameters. We will present this convergent expansion below. The reader interested in understanding the nuances is invited to consult~\cref{app:toymodels} for simple examples illustrating the issues alluded to above.

  \subsection{A convergent evaluation of the integral}\label{sec:eg-good}

  Before embarking on the evaluation let us note an essential simplification of the integrand. We claim
  \begin{equation}\label{eq:Gsigma-cancellation-split}
  \mathcal{G}_{k,\ell}^{\sigma}(\dualz, \dualtau \mid \tau)
  =
  -12\,\sigma\,\Psi^{\mathrm{sym}}_{k,\ell}(\dualz \mid \tau)
  + \frac{1}{2}\,\frac{\eta^3(\tau)}{\vartheta_1(\dualtau\mid \tau)}\,
  \Theta^{\mathrm{asym}}_{k,\ell}(\dualz \mid \tau)\,
  \bigg[\mathcal{F}_{+}(\dualz, \dualtau \mid \tau)
    +\mathcal{F}_{-}(\dualz, \dualtau \mid \tau) \bigg] \,.
  \end{equation}
  To see this set
  \begin{align}
  \mathcal{F}_{\alpha}(\dualz,\dualtau\mid\tau)
  =
  \frac{
    \vartheta_1\big(\frac{\dualtau-\alpha\,\dualz}{2}\mid\tau\big)^2\,
    \EK{1}(\dualtau+\alpha\,\dualz\mid\tau)}
  {\eta(\tau)^3\, \vartheta_1(\dualz\mid\tau)}\,,
  \qquad \alpha=\pm1\,,
  \end{align}
  and notice that
  \begin{equation}\label{eq:K3-cancellation-H}
  \mathcal{F}_{+}(\dualz,\dualtau\mid\tau)
  -\mathcal{F}_{-}(\dualz,\dualtau\mid\tau)
  =-24\,\frac{\vartheta_1(\dualtau\mid\tau)}{\eta(\tau)^3}\,.
  \end{equation}
  Therefore, we obtain an exact identity~\eqref{eq:Gsigma-cancellation-split} upon symmetrizing the two factors in the $\alpha$ sum. In writing~\eqref{eq:Gsigma-cancellation-split} we defined the symmetric and antisymmetric combinations of the false theta functions as
  \begin{equation}
  \begin{aligned}
  \Psi_{k,\ell}^{\mathrm{sym}}(\dualz\mid \tau)
   & = \Psi_{k,\ell}(\dualz\mid \tau) + \Psi_{k,\ell}(-\dualz\mid \tau)
  \,, \qquad
  \Psi_{k,\ell}^{\mathrm{asym}}(\dualz\mid \tau)
   & = \Psi_{k,\ell}(\dualz\mid \tau) - \Psi_{k,\ell}(-\dualz\mid \tau)\,.
  \end{aligned}
  \end{equation}

  The first term is bereft of any pole and is the sole difference between the IIB and the IIA theory. The second term contains the genuine pole, is common to the two theories, and involves the
  antisymmetric combination of the false theta function, which however reduces to the standard level-$k$ theta function~\eqref{eq:false-theta-asym} (see also~\cref{app:false}) as indicated.  In the final result, the term containing the pole has its $k$-dependence solely concentrated in the level-$k$ theta function. Therefore, by a suitable rearrangement we can reduce the core part of the problem to evaluation of the integral for $k=2$, for
  \begin{equation}
  \mathcal{G}_{k,\ell}^\sigma(\dualz, \dualtau \mid \tau) + 12\, \sigma\, \Psi^{\mathrm{sym}}_{k,\ell}(\dualz \mid \tau)
  =
  \chsu_\ell( \dualz \mid \tau)\, \bigg(\mathcal{G}_{2,0}^\sigma(\dualz, \dualtau \mid \tau) + 12\, \sigma\, \Psi^{\mathrm{sym}}_{2,0}(\dualz \mid \tau) \bigg)\,.
  \end{equation}
  The general $k$ integrand is thus obtained by suitably combining the integrand for $k=2$ with an $\mathfrak{su}_{k-2}(2)$ character. The latter is regular in the worldsheet modular parameter. As a consequence of the above, we first focus on the expansion of $\mathcal{G}_{2,0}^\sigma(\dualz,\dualtau\mid \tau)$. Once this is under control, we simply convolve with the expansion of the $\chsu_\ell(\dualz\mid \tau)$ to obtain the holomorphic part of the integrand.

  \subsubsection{The holomorphic part of the integrand}

  We want to write the single-particle spacetime elliptic genus as
  \begin{equation}
  \dualEGsp_\sigma(\dualz, \dualtau)
  = \sum_{N \in \frac{1}{2}\mathbb{Z}}\, \mathfrak{A}^\sigma_N(\dualz, \Im\dualtau)\, \dualq^N\,,
  \end{equation}
  and seek an expression for $\mathfrak{A}_N^\sigma$ that is manifestly convergent. Owing to the non-holomorphic nature of the answer, it is useful to pick out Fourier modes in $\Re\dualtau$, leaving the $\Im\dualtau$ dependence explicit. The worldsheet $\tau$ integral will then be performed by exploiting properties of these Fourier modes; in doing so, we shall take care not to expand series outside their natural domain of convergence.

  \paragraph{Ellipticity of the integrand:} Consider the  function $\mathcal{G}_{k,\ell}^\sigma(\dualz,\dualtau \mid \tau)$ defined in~\eqref{eq:def-Galpha} and set $k=2$, whence $\ell =0$ is the only allowed value.   It is an elliptic function of $\dualtau$ of index-$0$ on the lattice $\Lambda_2 = 2\, \mathbb{Z} + 2\,\tau \, \mathbb{Z}$; one can check from its definition that
  \begin{equation}\label{eq:k2-Galpha-elliptic}
  \mathcal{G}_{2,0}^\sigma(\dualz,\dualtau +2 \mid \tau)
  = \mathcal{G}_{2,0}^\sigma(\dualz,\dualtau + 2\,\tau \mid \tau)
  = \mathcal{G}_{2,0}^\sigma(\dualz,\dualtau \mid \tau)\,.
  \end{equation}
  The integrand $\mathcal{G}_{2,0}^\sigma$ has poles at the zeros of $\vartheta_1(\dualtau \mid \tau)$, which number four per unit lattice cell.

  The periodicity of the integrand w.r.t.~$\Re\dualtau$ suggests passing to a Fourier domain. Define rescaled Fourier coefficients by transforming with an additional factor, using $\dualq^{-n}$ at fixed $\Im\dualtau \notin \tau_2 \, \mathbb{Z}$ as
  \begin{equation}\label{eq:Galpha-Fourier}
  \mathfrak{G}^\sigma_n(\dualz, \Im\dualtau\mid\tau)
  = \frac{1}{2}\, \int_0^2\, du\, \mathcal{G}_{2,0}^\sigma(\dualz, u + i\, \Im\dualtau \mid \tau)\, \mathrm{e}^{-2\pi i\, n (u + i\, \Im\dualtau)}\,, \qquad
  n \in \frac{1}{2} \mathbb{Z} \,.
  \end{equation}
  The Fourier modes are half-integer quantized thanks to the doubled period. The dependence on $\Im\dualtau$, which we hold fixed,  is actually locally constant
  as long as $\Im\dualtau \notin \tau_2 \mathbb{Z}$ (hence the rescaling). This can be deduced using a contour integral; pick a contour with sides set by the lattice period anchored on a fixed $\Im\dualtau$ line, i.e., the loop $(i\, \Im\dualtau, 2 + i\, \Im\dualtau, 2+ 2\,\tau+ i\, \Im\dualtau, 2\,\tau + i\, \Im\dualtau)$.
  The horizontal segments at bottom and top contribute  $2\,  \mathfrak{G}^\sigma_n(\Im\dualtau)$ and $-2\, q^{-2n}\, \mathfrak{G}^\sigma_n(\Im\dualtau)$, respectively, upon using~\eqref{eq:k2-Galpha-elliptic}, while the vertical segments cancel. The dependence on $\Im\dualtau$ is directly given in terms of the residues of poles in the enclosed region, and depends on the height of $\Im\dualtau$.

  For $n \neq 0$, one finds upon computing the residues
  \begin{equation}
  \begin{aligned}
  \mathfrak{G}^\sigma_n(\dualz, \Im\dualtau\mid \tau)
  =
  \begin{dcases}
  \frac{\mathcal{K}_{\mathrm{e}}^\pm(\dualz\mid \tau) + q^{n}\, \mathcal{K}_{\mathrm{o}}^\pm(\dualz\mid \tau)}{1-q^{2n}}
   & \qquad 0< \Im\dualtau < \tau_2 \,,
  \\
  \frac{\mathcal{K}^\pm_{\mathrm{e}}(\dualz\mid \tau) + q^{-n}\, \mathcal{K}_{\mathrm{o}}^\pm(\dualz\mid \tau)}{1-q^{2n}}
   & \qquad \tau_2< \Im\dualtau < 2\tau_2 \,,
  \end{dcases}\,,
  \end{aligned}
  \end{equation}
  in the fundamental unit cell. The superscript signs are correlated with the parity of $n$: the $\mathcal{K}_\epsilon^+$ is relevant for $2n$ odd, while $\mathcal{K}_\epsilon^-$ is operative for $2n$ even. This formula can be translated to other cells  using
  \begin{equation}\label{eq:Galpha-Fourier-period}
  \mathfrak{G}^\sigma_n(\dualz, \Im\dualtau + 2\,\tau_2  \mid \tau)
  =
  q^{-2n}\, \mathfrak{G}^\sigma_n(\dualz, \Im\dualtau\mid \tau)\,.
  \end{equation}
  We have written this expression in terms of standard theta functions and K3 elliptic genus, by introducing
  \begin{equation}
  \mathcal{K}_a(\dualz \mid \tau)  = (-1)^{\lceil a/2 \rceil}\,
  \frac{\vartheta_a(\tfrac{1}{2}\, \dualz \mid \tau)^2\, \EK{a}(\dualz \mid \tau)}{\eta(\tau)^3}\,.
  \end{equation}

  The Fourier zero mode at $n =0$ is not determined by the above argument as  it arises from the constant piece of the elliptic function which is regular. We can fix it by using a Weierstrass decomposition,\footnote{ An elliptic function is a sum of Weierstrass zeta functions centered at the poles and weighted by the residues. This fixes the function up to the constant term, which can then be deduced by evaluating the decomposition ansatz at a suitable location away from the poles.}  which consistently recovers the constant expected from~\eqref{eq:Gsigma-cancellation-split} as a nice sanity check. The end result is
  \begin{equation}
  \mathfrak{G}_0^\sigma(\dualz,\Im\dualtau\mid \tau)
  =
  -12\, \sigma\, \Psi^{\mathrm{sym}}_{2,0}(\dualz \mid \tau)
  \pm  \frac{1}{4} \, \bqty{\mathcal{K}_1(\dualz \mid \tau) -
    \mathcal{K}_2(\dualz \mid \tau)}\,.
  \end{equation}
  The upper sign pertains to $0 < \Im\dualtau < \tau_2$, and the lower sign to the region
$\tau_2 < \Im \dualtau < 2\,\tau_2$.

  The Fourier modes only have factors of $(1-q^{2n})^{-1}$, which are independent of $\dualq$. When we evaluate the integral over the worldsheet, we can expand this factor around $q=0$ appropriately depending on $\mathop{\text{sgn}}(n)$. The key feature is that the rewriting has decoupled the dependence in $\Im\dualtau$, and the integral retains its convergence property at each step. This is the central difference from using~\eqref{eq:theta1Fourier} directly to expand the worldsheet integrand.

  \paragraph{Spin decomposition:} The residues effectively pick out different spin sectors, which arise from shifts of $\dualtau$. The final assembly of the integrand is easier if we combine them appropriately. Define, therefore, the combinations
  \begin{equation}\label{eq:k2-parity-kernels}
  \begin{aligned}
  \mathcal{K}_{\mathsf{e}}^{\pm}(\dualz \mid \tau)
  =\frac{1}{2}\left(\mathcal{K}_{1} (\dualz \mid \tau) \pm\mathcal{K}_{2}(\dualz \mid \tau) \right),
  \qquad
  \mathcal{K}_{\mathsf{o}}^{\pm}
  =\frac{1}{2}\left(\mathcal{K}_{4}(\dualz \mid \tau) \pm\mathcal{K}_{3} (\dualz \mid \tau)\right),
  \end{aligned}
  \end{equation}
  so that with $\epsilon(d)=\mathsf{e}$ or $\mathsf{o}$ for $d$ even
  or odd, respectively, one has
  \begin{equation}\label{eq:k2-spin-kernel-projector}
  \mathcal{K}_{a(d,b)}(\dualz \mid \tau)
  =\mathcal{K}_{\epsilon(d)}^+(\dualz \mid \tau)
  +(-1)^b\, \mathcal{K}_{\epsilon(d)}^-(\dualz \mid \tau) \,.
  \end{equation}
  The spin parity function $a(d,b)$ is defined above~\eqref{eq:IIB-long-string-sector-partition-function}. These specific combinations
  have a clean Fourier expansion
  \begin{equation}
  \mathcal{K}_{\epsilon}^{\pm}(\dualz\mid \tau)
  =\sum_{r\in\ZZ_{\geq 0}}\, k_{\epsilon,r}^{\pm}(\dualy) \,
  q^{\frac{1}{8}\,r + \delta^{\pm}_{\epsilon}}\,, \qquad
  \delta^\pm_{\mathsf{e}} =
  \delta^-_{\mathsf{o}} = \frac{1}{8}\,,\quad
  \delta^+_{\mathsf{o}} = -\frac{3}{8}\,.
  \label{eq:k2-channel-coefficients}
  \end{equation}
  with only the odd plus combination $\mathcal{K}_{\mathsf{o}}^+$ having a polar term. The first few terms in their expansion are
  \begin{equation}
  \begin{aligned}
  \mathcal{K}^+_{\mathsf{e}}(\dualz\mid\tau)
   & =
  (38- 2\,\chi_1(\dualz))\, q^{\frac{1}{8}}
  + (198\,\chi_1(\dualz) + 590 )\, q^{\frac{9}{8}}
  + \cdots \,,
  \\
  \mathcal{K}^-_{\mathsf{o}}(\dualz\mid\tau)
   & =
  - \mathcal{K}^-_{\mathsf{e}}(\dualz\mid\tau)
  \\
   & =
  16\, \chi_{\frac{1}{2}}(\dualz)\, q^{\frac{1}{8}}
  + (16\,\chi_{\frac{3}{2}}(\dualz) + 560\, \chi_{\frac{1}{2}}(\dualz))\, q^{\frac{9}{8}} + \cdots \,,
  \\
  \mathcal{K}^+_{\mathsf{o}}(\dualz\mid\tau)
   & =
  2 \, q^{-\frac{3}{8}}
  - (36\,\chi_1(\dualz) + 162 )\, q^{\frac{5}{8}}
  \\
   & \qquad
  + (2\,\chi_2(\dualz) - 790\, \chi_1(\dualz) - 1948 )\, q^{\frac{13}{8}}
  + \cdots \,.
  \end{aligned}
  \end{equation}
  Notice that
  \begin{equation}\label{eq:k2-channel-support}
  k_{\epsilon,r}^{\pm} =0
  \qquad\text{unless } r \in 8\ZZ_{\geq0}\,,
  \qquad
  k_{\mathsf{o},r}^{-}=-k_{\mathsf{e},r}^{-}.
  \end{equation}

  \subsubsection{The worldsheet modular integral}\label{sec:wsintegral}

  Assembling the integrand, the coefficient of $\dualq^N$ in $\dualEGsp_\sigma$ is given by the following:
  \begin{equation}\label{eq:EGcfs-integralrep}
  \begin{aligned}
  \mathfrak{A}^\sigma_N(\dualz, \Im \dualtau)
   & =
  \sum_{w=0}^\infty\, \int_{w}^{w+1}\,
  \dd\zeta\, \sqrt{\frac{k\, \Im\dualtau}{\zeta}}\, e^{-k\pi \, \Im\dualtau\, \zeta}\,
  \int_{-\frac{1}{2}}^{\frac{1}{2}}\, \dd\tau_1\,
  \mathfrak{I}^\sigma_N(\dualz, \Im\dualtau \mid \tau)\,.
  \end{aligned}
  \end{equation}
  We introduced
  \begin{equation}
  \mathfrak{I}_N^\sigma(\dualz, \Im\dualtau\mid\tau)
  = -\frac{e^{2\pi N \Im \dualtau}}{2}\,
  \int_0^2\, \dd u\,e^{-2\pi iN\, u} \, \sum_{\ell=0}^{\frac{k}{2}-1}\,
  \Psi_{k,\ell}(\overline{\dualtau}\mid-\bar{\tau}) \,
  \mathcal G_{k,\ell}^\sigma(\dualz, \dualtau\mid\tau) \,.
  \end{equation}
  In writing the integral we changed variables to $\zeta = \frac{\Im\dualtau}{\tau_2}$, and split the integral over it across the horizontal lines where the integrand has poles $\zeta \in \mathbb{Z}_{\geq 0}$. We have refrained from a priori reordering the sums and integrals in the above expression for the moment.

  The anti-holomorphic false theta function has a series expansion in $\bar{q}$, which we can read off from~\eqref{eq:false-theta-function}. To extract coefficients, let us rewrite $\llmt$ by introducing
  \begin{equation}\label{eq:lr-def}
  \lambda_r =
  \begin{dcases}
  k\,\widetilde{m} + \ell + \frac{1}{2} \,,
   & \qquad r = 2\,\widetilde{m} \,, \ \widetilde{m}\geq 0\,,
  \\
  k\,\widetilde{m} - \ell - \frac{1}{2} \,,
   & \qquad r = 2\,\widetilde{m} -1 \,, \ \widetilde{m}\geq 1\,,
  \end{dcases}
  \qquad
  \operatorname{\text{sgn}}(\llmt) = (-1)^r\,.
  \end{equation}
  This says
  \begin{equation}
  \Psi_{k,\ell}(\overline{\dualtau}\mid -\bar{\tau})
  = \sum_{r=0}^\infty \, (-1)^r\, \overline{\dualq}^{-\lambda_r}\, \bar{q}^{\frac{1}{k}\, \lambda_r^2}
  \end{equation}
  To obtain the contribution to the coefficient of $\dualq^N$ on the holomorphic side we should isolate the coefficient of $\dualq^{N-\lambda_r}$, which compensates the right-moving factor
$\overline{\dualq}^{-\lambda_r} = \dualq^{\lambda_r}\, e^{4\pi\, \Im\dualtau\,\lambda_r}$. Then since the $\tau_1$ integral enforces level matching, we end up projecting onto the mode $q^{\frac{1}{k}\,\lambda_r^2}$.

  Working in the cell $w\, \tau_2 < \Im\dualtau < (w+1)\,\tau_2$ we can therefore write
  \begin{equation}
  \int_{-\frac{1}{2}}^{\frac{1}{2}}\, \dd \tau_1\,
  \mathfrak{I}^\sigma_N(\dualz, \Im\dualtau \mid \tau)
  =
  \sum_{\ell=0}^{\frac{k}{2}-1} \,\sum_{r=0}^\infty\,
  (-1)^r\, \mathfrak{g}_{N,\ell, r, (w)}^\sigma(\dualz)\,
  \exp(4\pi\,\Im\dualtau\, \bqty{\lambda_r  - \frac{\lambda_r^2}{k\,\zeta}})\,.
  \end{equation}
  The coefficients $\mathfrak{g}_{N,\ell, r, (w)}^\sigma(\dualz)$ can be determined to be
  \begin{equation}
  \begin{aligned}
   &
  \mathfrak{g}_{N,\ell, r, (w)}^\sigma(\dualz)
  \\
   & =
  \operatorname{Coeff}\bqty{q^{\frac{\lambda_r^2}{k}}}\,
  \begin{dcases}
  -\,q^{-2\,(N-\lambda_r)\, \lfloor \frac{w}{2} \rfloor}\,
  \chsu_\ell(\dualz\mid\tau)\,
  \, \frac{\mathcal{K}_{\mathrm{e}}^\pm(\dualz\mid\tau) + q^{(-1)^w\, (N-\lambda_r)} \, \mathcal{K}_{\mathrm{o}}^\pm(\dualz\mid\tau)}{1-q^{2\,(N-\lambda_r)}} \,,
   & \quad N- \lambda_r \neq 0
  \\
  12 \,\sigma\, \Psi_{k,\ell}^{\mathrm{sym}}
  - \frac{(-1)^w}{2}\, \chsu_\ell(\dualz\mid\tau)\, \mathcal{K}_{\mathrm{e}}^-(\dualz\mid\tau)  \,,
   & \quad N- \lambda_r = 0
  \end{dcases}
  \end{aligned}
  \end{equation}
  The contribution from the zero mode is localized in the terms where the power of $\dualq$ is suitably matched with the worldsheet $q$ power. Hence, only $\lambda_r = N$ terms differentiate the IIB and IIA elliptic genus.

  The integral over $\zeta$ can be completed in terms of error functions (see~\eqref{eq:error-function-integral} for a quick derivation)
  \begin{equation}
  \begin{aligned}
  \mathfrak{A}^\sigma_N(\dualz, \Im\dualtau)
   & =
  \sum_{\ell=0}^{\frac{k}{2}-1}\, \sum_{r=0}^\infty\, \sum_{w=0}^\infty\,
  (-1)^r\, \mathfrak{g}_{N,\ell, r, (w)}^\sigma(\dualz)
  \int_w^{w+1}\dd \zeta\,\sqrt{\frac{k \Im\dualtau}{\zeta}}\,
  \exp(-\frac{\pi \Im\dualtau}{k\, \zeta}(k\,\zeta-2\,\lambda_r)^2) \,,
  \\
   & =
  \sum_{\ell=0}^{\frac{k}{2}-1}\, \sum_{r=0}^\infty\, \sum_{w=0}^\infty\,
  (-1)^r
  \mathfrak{g}_{N,\ell, r, (w)}^\sigma(\dualz)\,e^{4\pi \Im\dualtau\,\lambda_r}
  \bqty{ \mathsf{E}_{k,\lambda_r^2/k}(w+1, \Im\dualtau)
    -\mathsf{E}_{k,\lambda_r^2/k}(w, \Im\dualtau)}\,.
  \end{aligned}
  \label{eq:EGk-full-qNcf}
  \end{equation}
  Furthermore, the combinations of error functions appearing above are defined as
  \begin{equation}
  \begin{aligned}
  \mathsf{E}_{k,n}(d;\Im\dualtau)
   & = \frac{1}{2} \sum_{\eta = \pm 1} \, \eta\,
  e^{-4\pi\,\eta \,\Im\dualtau\, \sqrt{k\,n}} \,
  \operatorname{\text{erfc}} \pqty{\sqrt{\frac{\pi\,n\,\Im\dualtau}{d}} \pqty{2- \eta\, \sqrt{\frac{k}{n}}\,d} } \,.
  \end{aligned}
  \label{eq:errorfn-def}
  \end{equation}
  In addition, $\mathsf{E}_{k,n}(0;\Im\dualtau)=0$ by continuity.
  This is an exact representation in the domain of absolute
  convergence of the worldsheet $\tau_2$ integral. The convergence can be deduced by
  noting that the integrand is a Gaussian centered at
$\zeta =\frac{2}{k}\, \lambda_r$, within the cell with $w=r$.
  As $\Im\dualtau \to \infty$, this integral localizes onto $w=r$ and vanishes otherwise.
  Therefore, the holomorphic piece can be extracted by simply taking the $\Im\dualtau \to \infty$ limit, leading to
  \begin{equation}
  \begin{aligned}
  \mathfrak{a}^\sigma_N(\dualz)
   & =
  \sum_{\ell=0}^{\frac{k}{2}-1}\, \sum_{r=0}^\infty\,
    (-1)^{r}\,\mathfrak{g}_{N,\ell, r, (r)}^\sigma(\dualz) \,.
  \\
   & =
  \sum_{\ell=0}^{\frac{k}{2}-1}
  \bigg[
    -\sum_{r:\,\lambda_r < N} \,
    \sum_{d\leq r} \, +
    \sum_{r:\,\lambda_r > N} \,
    \sum_{d \geq  r+1} \bigg] (-1)^r\,
  \operatorname{\text{Coeff}}\bqty{q^{\frac{\lambda_r^2}{k} + d\, (N-\lambda_r)}}
  \pqty{\chsu_\ell(\dualz\mid \tau)\, \mathcal{K}^{\pm}_{\epsilon(d)}(\dualz\mid \tau)}
  \\
   & \qquad +\sum_{\ell=0}^{\frac{k}{2}-1}
  \sum_{r:\,\lambda_r=N}
  \bqty{
    12\,\sigma\, \bigg(\chi_N(\dualz)-\chi_{N-1}(\dualz)\bigg)\,
    -\frac{1}{2}\, \operatorname{\text{Coeff}}\bqty{q^{N^2/k}}\,
    \pqty{\chsu_\ell(\dualz\mid \tau)\, \mathcal{K}^-_{\mathrm{e}}(\dualz\mid\tau)} }.
  \end{aligned}
  \label{eq:EGk-full-holqN}
  \end{equation}
  At fixed $k$ and $N$ only finitely many values of $r$ and $d$ contribute. Specifically, the allowed values are constrained to be
  \begin{equation}
  0\leq r \leq 2\, \left\lfloor \frac{2\,N}{k} \right\rfloor \,.
  \end{equation}
  Given this the range of $d$ is also bounded to lie in the following ranges
  \begin{equation}
  \begin{aligned}
  \lambda_r < N:  & \qquad
  - \left\lfloor \frac{\lambda_r^2 - \pqty{\ell +\frac{1}{2}}^2 + \frac{k}{2}}{k\, (N-\lambda_r)} \right\rfloor
  \leq d \leq r \,,
  \\
  \lambda_r > N : & \qquad
  r+1 \leq d \leq
  \left\lfloor \frac{\lambda_r^2 - \pqty{\ell +\frac{1}{2}}^2 + \frac{k}{2}}{k\, (\lambda_r - N )} \right\rfloor
  \end{aligned}
  \end{equation}
  The sums over $d$ run over integers constrained by
  \begin{equation}
  \frac{1}{k} \bqty{\lambda_r^2 - \pqty{\ell + \tfrac{1}{2}}^2} + d\, (N - \lambda_r) \geq -\frac{1}{2}\,.
  \end{equation}
  Outside this range the coefficients vanish.

  To obtain this result use the zero modes of each of the factors: the $\mathfrak{su}_{k-2}$ factor gives $q^{\frac{1}{k}\, (\ell+\frac{1}{2})^2 -\frac{1}{8}}$, while the $\mathcal{K}_\epsilon^\pm$ start out with $q^{-\frac{3}{8}}$ at worst. To obtain the desired coefficient in the first line of~\eqref{eq:EGk-full-holqN} we need
  \begin{equation}
  \frac{\lambda_r^2}{k} + d\, (N-\lambda_r) - \frac{1}{k}\, (\ell+\tfrac{1}{2})^2 +\frac{1}{8} + \frac{3}{8} \geq 0 \,,
  \end{equation}
  which constrains $d$ as above. Furthermore, if $\lambda_r >N$ the inequality has to hold when  $d= r+1$ (there is no constraint when $\lambda_r < N$ since $\lambda_r \geq (\ell+\tfrac{1}{2})$ always). For the two parity choices of $r$ using~\eqref{eq:lr-def}
  \begin{equation}
  \begin{aligned}
  \text{even}\ r:
   & \qquad
  (r+1)\, N - \tfrac{1}{4} \,k\, r\,(r+2) - (\ell+\tfrac{1}{2}) +\tfrac{1}{2} \geq 0 \,,
  \\
  \text{odd}\ r:
   & \qquad
  \frac{1}{2}\, (r+1)\, \pqty{2\,N - k \, \frac{r+1}{2}}\  +\frac{1}{2} \geq 0 \,,
  \end{aligned}
  \end{equation}
  From the odd $r$ constraint we learn $r \leq 2 \lfloor \tfrac{2\,N}{k}\rfloor -1$. Assuming $r > 2\, \lfloor \tfrac{2\,N}{k}\rfloor$ for even $r$, the l.h.s. of our inequality can be shown to be bounded from above by $0$, which is a contradiction. This bounds $r$ as stated above.

  The non-holomorphic remainder piece is obtained by subtracting out the holomorphic part from the full expression. One obtains
  \begin{equation}\label{eq:EGsp-nonholo-bN}
  \begin{aligned}
  \mathfrak{b}^\sigma_N(\dualz,\Im\dualtau)
   & =
  \sum_{\ell=0}^{\frac{k}{2}-1}\, \sum_{r =0}^\infty\, \sum_{w=0}^\infty\, (-1)^r
  \, \mathfrak{g}^\sigma_{N,\ell,r,(w)}(\dualz)
  \\
   & \qquad\qquad
  \times \bqty{ e^{4\pi \,\Im\dualtau\,\lambda_r}
    \left[\mathsf{E}_{k,\lambda_r^2/k}(w+1;\Im\dualtau) -\mathsf{E}_{k,\lambda_r^2/k}(w;\Im\dualtau)\right] -\delta_{w,r} }.
  \end{aligned}
  \end{equation}
  We will interpret the two contributions in turn below.

  \subsection{Interpreting the spacetime elliptic genus}\label{sec:EG-symorb}

  First notice that the IIA and IIB elliptic genera are very simply related. Their holomorphic terms differ, but the non-holomorphic terms agree. Explicitly,
  \begin{equation}
  \begin{aligned}
  \mathfrak{a}_N^+(\dualz) - \mathfrak{a}_N^-(\dualz)
   & = 24\,
  \pqty{\chi_N(\dualz)-\chi_{N-1}(\dualz)}\,,
  \quad N > 0 \ \&\  k\nmid 2\,N\,,
  \\
  \mathfrak{b}_N^+(\dualz, \Im \dualtau)
   & = \mathfrak{b}_N^-(\dualz, \Im \dualtau)\,.
  \end{aligned}
  \end{equation}
  The difference in the holomorphic part arises from the short string part of the spectrum as noted earlier in~\eqref{eq:AB-lmovers}. Henceforth, we focus exclusively on the IIB theory, and thus drop the $\sigma$ superscript for convenience.

  \begin{table}[H]
  \centering
  \renewcommand{\arraystretch}{1.6}
  \resizebox{\textwidth}{!}{$
\begin{array}{c|| l | l| l| l|}
  N & k=2 & k=3 & k=4 & k=5
  \\ \hline\hline
  0
  & 2
  & 2
  & 2
  & 2
  \\ \hline
  \frac12
  & 20\,\chi_{\frac{1}{2}}
  & 22\,\chi_{\frac{1}{2}}
  & 22\,\chi_{\frac{1}{2}}
  & 22\,\chi_{\frac{1}{2}}
  \\ \hline
  1
  & 4\,\chi_1-80
  & 22\,\chi_1-44
  & 24\,\chi_1-44
  & 24\,\chi_1-44
  \\ \hline
  \frac32
  & 20\,\chi_{\frac{3}{2}}+300\,\chi_{\frac{1}{2}}
  & 4\,\chi_{\frac{3}{2}}-44\,\chi_{\frac{1}{2}}
  & 22\,\chi_{\frac{3}{2}}-26\,\chi_{\frac{1}{2}}
  & 24\,\chi_{\frac{3}{2}}-26\,\chi_{\frac{1}{2}}
  \\ \hline
  2
  & \begin{aligned}
   & 4\,\chi_2-1256\,\chi_1 \\[-2pt]
   & \qquad-3036
  \end{aligned}
  & 22\,\chi_2+104\,\chi_1-198
  & 4\,\chi_2-40\,\chi_1-40
  & 22\,\chi_2-24\,\chi_1-20
  \\ \hline
  \frac52
  & \begin{aligned}
   & 20\,\chi_{\frac{5}{2}}+4268\,\chi_{\frac{3}{2}} \\[-2pt]
   & \qquad+30080\,\chi_{\frac{1}{2}}
  \end{aligned}
  & \begin{aligned}
   & 22\,\chi_{\frac{5}{2}}-94\,\chi_{\frac{3}{2}} \\[-2pt]
   & \qquad+1260\,\chi_{\frac{1}{2}}
  \end{aligned}
  & \begin{aligned}
   & 22\,\chi_{\frac{5}{2}}+138\,\chi_{\frac{3}{2}} \\[-2pt]
   & \qquad+196\,\chi_{\frac{1}{2}}
  \end{aligned}
  & 4\,\chi_{\frac{5}{2}}-40\,\chi_{\frac{3}{2}}-4\,\chi_{\frac{1}{2}}
  \\[1ex] \hline
  3
  & \begin{aligned}
   & 4\,\chi_3-12744\,\chi_2 \\[-2pt]
   & \qquad-195580\,\chi_1-268008
  \end{aligned}
  & \begin{aligned}
   & 4\,\chi_3-468\,\chi_2 \\[-2pt]
   & \qquad-3856\,\chi_1-10512
  \end{aligned}
  & \begin{aligned}
   & 24\,\chi_3-308\,\chi_2 \\[-2pt]
   & \qquad-1292\,\chi_1-2032
  \end{aligned}
  & \begin{aligned}
   & 22\,\chi_3+140\,\chi_2 \\[-2pt]
   & \qquad-2\,\chi_1-232
  \end{aligned}
  \\\hline
\end{array}
$}
  \caption{The holomorphic coefficients $\mathfrak{a}_N^+$ in the string theory elliptic genus for Type IIB theory. The argument of the $\mathfrak{su}(2)$ characters is $\dualz$ and has been suppressed for clarity. The constant terms are coefficients of the singlet character $\chi_0(\dualz) =1$ in the elliptic genus.}
  \label{tab:eg-holomorphic-coefficients-k2-k5}
  \end{table}

  \paragraph{The holomorphic discrete states:}
  We can extract algorithmically the coefficients $\mathfrak{a}_N(\dualz)$ for fixed values of $k$. For some low-lying cases the results of this extraction are tabulated in~\cref{tab:eg-holomorphic-coefficients-k2-k5}. Notice the stabilizing pattern in the coefficients. At fixed $N$, the coefficient is independent of $k$ once $k \geq 2\,N+2$, and the stable values, which we denote as $\mathfrak{a}_N^{(\infty)}$ are
  \begin{equation}
  \mathfrak{a}_0^{(\infty)}
  = 2,\qquad
  \mathfrak{a}_N^{(\infty)} =
  \begin{dcases}
  24\,\bqty{\chi_N(\dualz) -\chi_{N-1}(\dualz)} -20 \,, & \qquad
  N\in\mathbb{Z}_{>0}\,,
  \\
  24\,\bqty{\chi_N(\dualz)-\chi_{N-1}(\dualz)}
  - 2\, \chi_{\frac{1}{2}}(\dualz)\,,
    & \qquad
  N\in\mathbb{Z}_{\geq0}+\frac{1}{2}\,.
  \end{dcases}
  \end{equation}
  This is precisely the supergravity result; with these coefficients the holomorphic series can be resummed into a generating series
  \begin{equation}
  \begin{aligned}
  \sum_{N}\, \mathfrak{a}_N^{(\infty)}(\dualz)\, \dualq^N
   & =
  24\,
  \sum_{\ell \in \frac{1}{2}\, \mathbb{Z}_{\geq 0}} \,
  \chpsu_\ell(\dualz, \dualtau) - \frac{1}{1-\dualq} \,\bqty{22 + 2\, \chi_{\frac{1}{2}}(\dualz)\, \dualq^{\frac{1}{2}} - 26\, \dualq }\,,
  \\
   & =
  2 + 24\, \bqty{\frac{1-\dualq}{
    \pqty{1- \dualq^{\frac{1}{2}}\, \dualy^{\frac{1}{2}}}  \,
    \pqty{1-\dualq^{\frac{1}{2}}\, \dualy^{-\frac{1}{2}}}} -1}
  - \frac{20\, \dualq + 2\, \dualq^{\frac{1}{2}}\, \chi_{\frac{1}{2}}(\dualz)}{1-\dualq} \,.
  \end{aligned}
  \end{equation}
  The second line can be obtained from the first using~\eqref{eq:global-N4-character}, and remembering our $\mathfrak{su}(2) $ character conventions stated below~\eqref{eq:su2-character}.

  At fixed finite $k$, the above generating function gives the coefficients $\mathfrak{a}_N$ accurately up to $N = \frac{k}{2} -1$. One reason for the simplicity of these states is that they do not get contributions from the string oscillators. They are largely obtained by taking the zero modes on $\mathrm{K3}$ and dressing them with the $\mathfrak{su}(2)$ primaries.

  For $N> \frac{k}{2}$ we will start picking up oscillator contributions. The coefficients, consequently, should be extracted from the $q$-series expansion of the function $\mathcal{G}(\dualz ,\dualtau \mid \tau)$. One can similarly work out the contribution in the intervals $m\,\frac{k}{2} < N<  (m+1)\,\frac{k}{2}$, for $m=1,2,\cdots$, etc., for fixed $k$, obtaining results that stabilize at fixed $N$ as we increase $k$.

  An efficient encoding of the coefficients $\mathfrak{a}_N$ is to present a single  generating function, whose Fourier series extracts them. A  useful guess is that such a generating function ought to be built out of the worldsheet characters themselves suitably projected onto a specified power of $q$.
  Introduce the combination
  \begin{equation}
  \mathcal{H}^\pm_d(\dualz\mid \tau) =
  - \frac{1}{\vartheta_1(\dualz\mid \tau)}\,
  \begin{dcases}
  \frac{1+ q^{\frac{d}{2}}\, \dualy^{\frac{1}{2}}}{1-q^{\frac{d}{2}}\, \dualy^{\frac{1}{2}}}
  \, \pqty{
    \mathcal{K}^+_{\mathrm{o}}(\dualz\mid \tau) + \mathcal{K}_o^-(\dualz\mid \tau)} \,,
   & \qquad \text{odd}\ d\,,
  \\
  \frac{1+ q^d\, \dualy}{1-q^d \dualy}\, \mathcal{K}_{\mathrm{e}}^\pm(\dualz\mid \tau) +
  \frac{2\, q^{\frac{d}{2}}\, \dualy^{\frac{1}{2}}}{1-q^d\, \dualy}\, \mathcal{K}_{\mathrm{e}}^\mp(\dualz\mid \tau) \,,
   & \qquad \text{even}\ d\,.
  \end{dcases}
  \end{equation}
  Note that for odd $d$ the two superscripts give the same function.
  Then, the formal series in $\dualq^{\frac{1}{2}}$ can be obtained as
  \begin{equation}
  \begin{aligned}
   & \sum_{N \in \frac{1}{2} \mathbb{Z}_{\geq 0}}\,
  \pqty{\mathfrak{a}_N - \mathfrak{a}_N^{(\infty)}}\, \dualq^N
  \\
   & =
  \operatorname{\text{Coeff}}\bqty{q^0}\left\{
  \,\sum_{\substack{b,d \geq 1 \\b \equiv d \!\pmod{2}}}\,
  \frac{q^{\frac{k}{4}\, (b^2 + d^2)}}{(1-q^{-b}\, \dualq)\, }
  \bqty{
    \sum_{\eta = \pm}\, \dualy^{\frac{\eta}{2}\, k\,d} \pqty{
    \dualq^{\frac{1}{2}}\,  q^{-\frac{b}{2}}\,
    \mathcal{H}_d^+\big(\eta\, \dualz\mid \tau \big)
    + \dualq\, q^{-b}\,
    \mathcal{H}_d^-\big(\eta\,\dualz\mid \tau \big)
    } }\right.
  \\
   & \left.
  \qquad \qquad\qquad
  + \frac{1}{\dualy^{\frac{1}{2}}- \dualy^{-\frac{1}{2}}}\, \frac{1}{\vartheta_1(\dualz\mid \tau)}\, \sum_{m\geq 1}\, \frac{q^{k\,m^2}}{1-\dualq\, q^{-2m}}\,
  \bigg[
    \dualq^{\frac{1}{2}}\, q^{-m}\, \pqty{2\, \mathcal{K}_{\mathrm{e}}^-(\dualz\mid \tau)
    + \chi_{\frac{1}{2}}(\dualz)\, \mathcal{K}_{\mathrm{e}}^+(\dualz\mid \tau) }
    \right.
    \\
    & \left. \qquad \qquad \qquad\qquad
    +\, \dualq\, q^{-2\,m} \pqty{2\, \mathcal{K}_{\mathrm{e}}^+
    + \chi_{\frac{1}{2}}(\dualz)\, \mathcal{K}_{\mathrm{e}}^-(\dualz\mid \tau) }
    \bigg] \right\} .
  \end{aligned}
  \end{equation}
  Note that one formally expands first in $\dualq$, and then takes the coefficient of the constant term in $q$. This coefficient extraction can be done efficiently numerically and can be checked against the direct integral representation.

  \paragraph{The non-holomorphic completion:} To understand the non-holomorphic contribution encoded in
$\mathfrak{b}_N(\dualz,\Im\dualtau)$ it will help to first rewrite the result in~\eqref{eq:EGsp-nonholo-bN}. One can check that upon using error function identities that it can be reexpressed as
  \begin{equation}\label{eq:EG-nonholo-Tdef}
  \begin{aligned}
  \mathfrak{b}^\sigma_N(\dualz,\Im\dualtau)
   & =
  \frac{1}{2}\, \sum_{\ell=0}^{\frac{k}{2}-1}\, \sum_{r =0}^\infty\, \sum_{d=1}^\infty\,
  (-1)^r \pqty{
    \mathfrak{g}^\sigma_{N,\ell,r,(d)}(\dualz)
    -  \mathfrak{g}^\sigma_{N,\ell,r,(d-1)}(\dualz)}\, \mathcal{T}^\sigma_{k,d,r}(\Im\dualtau)
  \\
  \mathcal{T}^\sigma_{k,d,r}(y)
   & \equiv
  \operatorname{\text{sgn}}(k\,d-2\,\lambda_r)\, \operatorname{\text{erfc}}
  \pqty{\sqrt{\frac{\pi\,y}{k\,d}} \,\abs{k\,d - 2\, \lambda_r}}
  + \;e^{8\pi\,y\, \lambda_r}
  \operatorname{\text{erfc}}
  \pqty{\sqrt{\frac{\pi\,y}{k\,d}} \,(k\,d + 2\, \lambda_r)}\,.
  \end{aligned}
  \end{equation}

  The first term in $\mathcal{T}^\sigma_{k,d,r}$ has the form of a standard error function smoothing of sharp cut-offs characteristic of mock modular forms~\cite{Zwegers:2002kt}. To see this use the identity $\operatorname{\text{sgn}}(x) \, \operatorname{\text{erfc}}(\abs{x}) = \operatorname{\text{sgn}}(x) - \operatorname{\text{erfc}}(x)$. Phrased this way the error function provides a Gaussian smooth of a sharp step.
  This term captures the continuum modes arising from the non-compactness of the target, whose onset is set by the location of the step (occurring here at  $\lambda_r > \frac{k\,d}{2}$).
  Therefore, one ought to have anticipated this piece based on investigations of elliptic genera of non-compact CFTs~\cite{Troost:2010ud, Eguchi:2010cb, Ashok:2011cy,Murthy:2013mya,Ashok:2013pya,Harvey:2014nha}.

  The second term in $\mathcal{T}^\sigma_{k,d,r}$ seems to be different, and does not appear to have been encountered in the context of superconformal field theories. Despite appearances, this term is exponentially suppressed (at the same rate as the first term), with the Gaussian damping of the error function overwhelming the prefactor exponential. Its presence is mandated by the  integral we need to compute on the worldsheet.

  One way to see the difference between the two terms is to compute the derivative of the elliptic genus with respect to $\Im\dualtau$. This captures the anti-holomorphic anomaly, and one finds
  \begin{equation}
  \partial_y \mathcal{T}^\sigma_{k,d,r}(y)
  = -\sqrt{\frac{k\,d}{y}}\, e^{-\frac{\pi \,y}{k\,d}\, (k\,d-2\,\lambda_r)^2}
  + 4\pi\, \lambda_r\, e^{8\pi\,\lambda_r\,y}\,  \operatorname{\text{erfc}}
  \pqty{\sqrt{\frac{\pi\,y}{k\,d}} \,(k\,d + 2\, \lambda_r)}\,.
  \end{equation}
  This differs from the conventional shadow encountered in mock modular forms owing to the survival of the error function contribution.

  The non-holomorphic contribution to the elliptic genus originates from the difference of the spectral density of bosonic and fermionic excitations~\cite{Ashok:2011cy}. The basic mechanism can be inferred by examining the computation of Witten index in supersymmetric quantum mechanics with a non-compact target~\cite{Akhoury:1984pt,Pioline:2015wza}. One can equivalently
  extract the factor from a reflection phase shift calculation.
  To expose this, write $\mathcal{T}^\sigma_{k,d,r}$ back in integral representation as
  \begin{equation}
  \mathcal{T}^\sigma_{k,d,r}(y)
  = \int_0^\infty\, \frac{ds}{\pi}\, e^{-4\pi \, y\, (s^2 + a_-^2)}\,
  \bqty{\frac{a_-}{s^2 + a_-^2} + \frac{a_+}{s^2+a_+^2} }\,, \qquad
  a_\pm = \frac{k\,d \pm 2\,\lambda_r}{2\,\sqrt{k\,d}}\,.
  \end{equation}
  The factor $\frac{a_-}{s^2 + a_-^2} = \frac{1}{2\,i}\, \partial_s\, \log \frac{s-i\,a_-}{s+ i\,a_-}$ leads to the conventional error function smoothed mock modular completion, and captures exactly the spectral asymmetry between bosons and fermions. The second term has a similar form, but at a different scale $a_+$---this offset leads to the exponential prefactor, $a_+^2 - a_-^2 = 2\,\lambda_r$. Understanding the answer then boils down to asking why does one have two independent scales in the continuum part of the spectrum.  It would be interesting to understand the origin of the two scales, and find other models that manifest a similar phenomenon.\footnote{It is amusing that the second factor appears in a one-dimensional quantum mechanical scattering problem~\cite{Bondurant:2004sqg}. Let $H = -\partial_s^2 + a_-^2$ for $s\geq 0$, and consider two sets of boundary conditions for the wavefunctions at the origin: (a) Robin with $\psi'(0) = a_+ \psi(0)$ and (b) Dirichlet $\psi(0)=0$. Computing the difference of the Hilbert space traces with these two boundary conditions produces the second term $\frac{1}{2}\, e^{4\pi(a_+^2-a_-^2)\,y}\,\operatorname{\text{erfc}}(2\, a_+\, \sqrt{\pi\,y})$. One can equivalently verify this by a scattering calculation. }

  \paragraph{A symmetric orbifold interpretation(?):} In the supergravity limit $k\to \infty$ the Type IIB elliptic genus agrees with that of the symmetric product orbifold of K3~\cite{deBoer:1998us}. The degree of the symmetric orbifold is $N_1\, k$, where $N_1$ is the number of fundamental strings, which are bound to $k$ NS5-branes.  The perturbative worldsheet theory  does not see $N_1 = g_s^{-2}\, k^{-\frac{1}{2}}$ with $g_s$ the three-dimensional string coupling. Since both $k$ and $N_1$ are assumed large in the supergravity limit, one does not resolve the factor $N_1 \,k$. At finite $k$, however, one could ask if there is a seed CFT whose spectrum can be constrained by the single-particle elliptic genus we have just computed. A natural guess is based on the proposal of~\cite{Eberhardt:2019qcl}: the $N_1$ fold symmetric product of a seed theory deformed by a double twist operator. The seed theory is the $\mathcal{N} = 4$ linear dilaton theory with central charge $c= 6\,(k-1)$ tensored with the K3 sigma model, making up a seed with central charge $6\,k$ discussed in~\cref{subsec:long-string-spectrum}. In the seed theory the spectrum has a gap of order $\frac{k}{4}$, beyond which one only encounters the continuum states. As we noted, this matches with the long-string part of the spectrum.

  The discrete states that are present in the elliptic genus, and that contribute to the holomorphic part, ought to arise as bound states in the deformed theory. However, there is as yet no quantitative check that this part of the spectrum is reproduced. Our results should therefore be taken as a target for such an endeavor. As a simpler check one could have hoped that the
  long-string continuum gives rise to the non-holomorphic part. Unfortunately, the presence of the second term in $\mathcal{T}^\sigma_{k,d,r}$ noted above makes this interpretation problematic.

  \section{Discussion}

  In this paper we have proposed a consistent GSO projection for the RNS string on $\mathrm{AdS}_3 \times \mathbf{S}^3 \times \mathrm{K3}$. We uncovered that there is an interesting tension between the GSO projection and the physical state projection. Even though the worldsheet theory essentially factorizes into bosonic theories and free fermions (at least for the $\mathrm{AdS}_3 \times \mathbf{S}^3$ factor), the GSO projection couples them non-trivially. Our prescription was forced by consistency conditions on the theory, such as spacetime supersymmetry. As an application, we reproduced the expectations on the spectrum, both from supergravity and the correspondence with the dual CFT. Finally, we computed the full spacetime elliptic genus of string theory on the background. It features both discrete contributions originating from the short string sector and error function terms from the long string sector that are characteristic of elliptic genera of non-compact 2d CFTs.

  Let us discuss some of the most salient features further.
  One puzzling aspect of our proposal is that the string integrand loses its modular covariance after the GSO projection. This is not really a problem since the full partition function involves a sum over all winding sectors as in~\eqref{eq:plethystic-exponential}, which restores modular invariance. Since the sectors $(m,n)$ with $m \neq 0$ are defined as the modular images of the sectors $(0,n)$, this holds by construction; it would be interesting to verify directly that the resulting sectors admit a consistent interpretation as traces over the twisted-sector Hilbert spaces.
  The reason for this non-covariance can be traced back to the fact that the worldsheet integrand does not factorize into the matter and the ghost sector. Such a factorization is delicate at best since the combined $bc$ and $\beta\gamma$ ghost partition function in the $\widetilde{\mathrm{R}}$ sector is formally $\left|\frac{\eta(\tau)^3}{\vartheta_1(0\mid\tau)}\right|^2$, which diverges. The origin of this divergence is the zero modes of the $\beta\gamma$ ghosts in the odd spin structure, which lead to an infinitely degenerate ground state. Equivalently, in the supermoduli space picture, the odd spin structure sector has a fermionic conformal Killing spinor as well as an odd modulus, corresponding to the zero modes of $\gamma$ and $\beta$, respectively~\cite{Witten:2012bh}. The treatment of the former is delicate since it has to be soaked up and there is no a priori prescription for how to do this. The problem disappears as soon as one computes a one-point function on the torus, since the odd coordinate of the vertex operator can then be used to fix the fermionic conformal Killing spinor. It would be interesting to test our prescription independently in this way, e.g., by computing a suitable torus one-point function.

  Apart from the recent discussion of related issues in~\cite{Sen:2026kyn,Sen:2026xfy}, we are not aware of any background where such a phenomenon has been observed and the resulting physical state projection has been worked out explicitly. Indeed, any possible example must not be treatable in light-cone gauge, where the non-factorization does not occur by construction.

  It is clear that one can make the problem that we encountered in this paper even worse. In the extreme case, one can e.g., consider string theory on the Euclidean target space $\mathrm{CY}_5$, where the complex five-dimensional Calabi--Yau manifold has non-zero Euler characteristic $\chi$. The Euler characteristic coincides with the Witten index of the worldsheet matter theory and thus the worldsheet partition function is non-vanishing in the $\widetilde{\mathrm{R}}$--$\widetilde{\mathrm{R}}$ sector. Taking the divergent contribution of the $\beta\gamma$ ghosts into account implies that the string one-loop partition function is divergent. As explained recently in~\cite{Sen:2026xfy}, this divergent contribution is part of the logarithmic contribution $\mp \frac{1}{24} \chi \, \log g_\text{s}$ to the one-loop partition function (with the upper sign for Type IIB and the lower sign for Type IIA), with the infinite additive constant presumably corresponding to the scheme dependence of the logarithm. We thus see that the phenomenon observed in this paper is a marginal case: a finite torus partition function can still be derived since the Euler characteristic of $\mathrm{AdS}_3 \times \mathbf{S}^3 \times \mathrm{K3}$ vanishes, but the dangerous $\widetilde{\mathrm{R}}$ spin structure contributes non-trivially.

  \section*{Acknowledgements}
  It is a pleasure to thank Jan de Boer, Andrea Dei, Matilda Delgado and Ashoke Sen for helpful discussions.
  L.E.~was supported by  the European Research Council (ERC) under the European Union’s Horizon 2020 research  and innovation programme (grant agreement No 101115511).
  The work of M.R.G.~is supported by the Simons
  Foundation grant 994306 (Simons Collaboration on Confinement and QCD Strings), as
  well as the NCCR SwissMAP that is funded by the Swiss National Science Foundation.
  S.M.~acknowledges the support of the STFC UK grant ST/X000753/1.
  M.R.~was supported by
  U.S.~Department of Energy grant DE-SC0009999 and by funds from the University of California.  M.R.G.~and M.R.~also thank the Aspen Center for Physics, which is supported by National Science Foundation
  grant PHY-2210452, for hospitality during the course of this project.

  \medskip
  \noindent\textbf{Statement of AI use:} We acknowledge the use of GPT Codex models 5.5/5.6/6.0 (Open AI) and Claude Opus 4.8/5.0 and Fable 5.0 (Anthropic) for assistance with computations and checks. Specifically, the AI models were helpful in improving the computation of the elliptic genus (Section 5), and for tabulating the notation used in the paper. The results and arguments reported in the manuscript were independently derived and verified by the authors, who take sole responsibility for the material presented.

  \appendix
  \section{An identity for the K3 partition function}\label[appendix]{app:character-identities}

  \subsection{\texorpdfstring{$\mathcal{N}=4$}{N=4} characters}

  To derive the identity~\eqref{eq:K3-sum-over-spin-structures-identity}, we need the form of the characters of the $\mathcal{N}=4$ superconformal algebra of central charge $c=6k$. We are interested in the case $k=1$. Its unitary representations have been classified in~\cite{Eguchi:1987sm, Eguchi:1987wf}. For all characters, the different spin structures are as usual related by spectral flow, shifting $\dualz \mapsto \dualz+1$ and $\dualz \mapsto \dualz+\tau$. The algebra has long and short representations whose characters we will now discuss.

  \begin{itemize}[wide,left=0pt]
  \item A long representation is specified by its conformal dimension $h$. Unitarity of the representation requires the highest-weight state to have vanishing $\mathfrak{su}(2)_\text{R}$ spin. The descendant obtained by acting with $\dualG_{-\frac{1}{2}}^{\alpha A}$ on the superprimary is still affine primary and unitarity of the R-symmetry currents $\mathfrak{su}(2)_1$ requires it to have spin at most $\frac{1}{2}$. Therefore, the affine character takes the form
  \begin{equation}
  \chNfour_{a,h}(\dualz\mid \tau)=q^{h-\frac{1}{8}} \,
  \frac{\vartheta_a(\frac{\dualz}{2} \mid \tau)^2}{\eta(\tau)^3}\,.
  \end{equation}
  The term $\vartheta_a(\tfrac{\dualz}{2} \,|\, \tau)^2\,\eta(\tau)^{-2}$ describes the action of the four supercharges, while $\eta(\tau)^{-1}$ accounts for the Virasoro descendants. For $k=1$, this automatically accounts for the current descendants thanks to the existence of null vectors in the module.\footnote{For general $k$, the character is multiplied by an $\mathfrak{su}(2)_{k-1}$ character, which, however, trivializes for $k=1$.}
  \item For a short representation, the spin of the superprimary state can be $0$ or $\frac{1}{2}$, since the descendants $\dualG_{-\frac{1}{2}}^{\alpha A}$ discussed for the long representation above are absent and do not impose the same unitarity constraint. We denote the corresponding character by $\chNfour_{\text{S},a,\ell}(\dualz \mid \tau)$.
  The characters satisfy the identity
  \begin{equation}
  \chNfour_{a,h=0}(\dualz\mid \tau) =
  2\, \chNfour_{\text{S},a,0}(\dualz\mid \tau)
  + \chNfour_{\text{S},a,\frac{1}{2}}(\dualz\mid \tau)\,,
  \label{eq:N=4-characters-BPS-identities}
  \end{equation}
  expressing the decomposition of the non-BPS multiplet at threshold into BPS multiplets.
  Because of this identity, it is enough to consider the short representation with $\ell=0$; we shall consequently drop the label $\ell=0$. It is known that the character can further be simplified to~\cite{Eguchi:1988af}
  \begin{equation}
  \chNfour_{\text{S},a}(\dualz\mid \tau)=
  \frac{\vartheta_a(\frac{\dualz}{2} \mid \tau)^2}{\eta(\tau)^3} \,
  \mu_a(\tfrac{\dualz}{2} \mid \tau)\,,
  \end{equation}
  where
  \begin{subequations}
    \begin{align}
    \mu_1(\dualz \mid \tau)
    & =
    \frac{1}{\vartheta_1(\dualz \mid \tau)} \,
    \sum_{n \in \ZZ}
    \frac{(-1)^n q^{\frac{1}{2}n(n+1)}\, \dualy^{n+\frac{1}{2}}}{1-\dualy\, q^n}\,,
    \label{eq:lerch1}
    \\
    \mu_2(\dualz \mid \tau)
    & =
    \frac{1}{\vartheta_2(\dualz \mid \tau)} \,
    \sum_{n \in \ZZ}
    \frac{q^{\frac{1}{2}n(n+1)}\, \dualy^{n+\frac{1}{2}}}{1+\dualy\, q^n}\,,
    \label{eq:lerch2}
    \\
    \mu_3(\dualz \mid \tau)
    & =
    \frac{1}{\vartheta_3(\dualz \mid \tau)} \,
    \sum_{n \in \ZZ}
    \frac{q^{\frac{1}{2}(n^2-\frac{1}{4})}\, \dualy^{n}}{1+\dualy\, q^{n-\frac{1}{2}}}\,,
    \label{eq:lerch3}
    \\
    \mu_4(\dualz \mid \tau)
    & =
    \frac{1}{\vartheta_4(\dualz \mid \tau)} \sum_{n \in \ZZ}
    \frac{(-1)^n q^{\frac{1}{2}(n^2-\frac{1}{4})}\dualy^{n}}{1-\dualy q^{n-\frac{1}{2}}} \,,
    \label{eq:lerch4}
    \end{align}
    \label{eq:lerchsums}
  \end{subequations}
  are so-called Appell--Lerch sums. These explicitly enter into the rewriting of the K3 elliptic genus.
  \end{itemize}

  \subsection{An abstruse K3 identity}

  We want to prove~\eqref{eq:K3-sum-over-spin-structures-identity}. By decomposing $Z_{a,\widetilde{a}}^\mathrm{K3}$ into both left- and right-moving $\mathcal{N}=4$ characters,
  it suffices to do this character by character. For a long character, this is the identity (with $\alpha_1=-\alpha$,  $\alpha_2=-1$, $\alpha_3=1$, $\alpha_4=-1$)
  \begin{equation}
  \frac{1}{2} \, \sum_a \alpha_a \, \vartheta_a(\dualtau \mid \tau) \, \vartheta_a(\dualz \mid \tau)
  \, \chNfour_{a,h}(0 \mid \tau)
  =\vartheta_1(\tfrac{1}{2}(\dualtau-\alpha\, \dualz) \mid \tau)^2 \,
  \chNfour_{1,h}(\dualtau+\alpha\, \dualz\mid \tau)\,,
  \end{equation}
  which holds as a direct consequence of the Jacobi abstruse identity
  \begin{equation}
  \frac{1}{2}\, \sum_{a=1}^4 \,
  \alpha_a \, \vartheta_a(\dualtau \mid \tau) \, \vartheta_a(\dualz \mid \tau)\,
  \vartheta_a(\tau)^2
  = \vartheta_1(\tfrac{1}{2}(\dualtau-\dualz) \mid \tau)^2\,
  \vartheta_1(\tfrac{1}{2}(\dualtau+\dualz) \mid \tau)^2 \,.
  \label{eq:Jacobi-abstruse-identity}
  \end{equation}
  The case of the short representation amounts to the identity
  \begin{multline}
  \frac{1}{2}\, \bigg(\alpha \, \vartheta_1(\dualz \mid \tau) \,
  \vartheta_1(\dualtau \mid \tau)\, \eta(\tau)^{-6}
  + \sum_{a=2}^4\,  \alpha_a\,  \vartheta_a(\dualtau \mid \tau)\,
  \vartheta_a(\dualz \mid \tau) \vartheta_a(\tau)^2\, \mu_a(\tau)\, \eta(\tau)^{-9}\bigg) \\
  =\vartheta_1(\tfrac{1}{2}(\dualtau-\dualz) \mid \tau)^2 \,
  \vartheta_1(\tfrac{1}{2}(\dualtau+\dualz) \mid \tau)^2\,
  \mu_1(\tfrac{1}{2}(\dualtau+\alpha\, \dualz)  \mid \tau)\, \eta(\tau)^{-9} \,.
  \label{eq:short-N=4-character-abstruse-identity}
  \end{multline}
  Here, we used that $\lim_{\dualz \to 0} \chi_{\text{S},a=1}^{\mathcal{N}=4}(\dualz \mid \tau)=-1$. We multiplied by a power of $\eta(\tau)$ which will be convenient in the following. We can also set $\alpha=1$ in the following, since the case $\alpha=-1$ is obtained by replacing $\dualz \mapsto -\dualz$.

  One can prove~\eqref{eq:short-N=4-character-abstruse-identity} from Mock modularity of the short $\mathcal{N}=4$ character. This amounts to the following fact. There exists a non-holomorphic function $R(\tau)$ such that $\hat{\mu}_1(\dualz \mid \tau)=\mu_1(\dualz\mid \tau)-\frac{1}{2}R(\tau)$ transforms like a weak Jacobi form of weight $\frac{1}{2}$ and index $0$~\cite{Zwegers:2002kt}. Explicitly,
  \begin{equation}
  R(\tau) =
  \sum_{n \in \ZZ}\, (-1)^n\, \Big[ \mathop{\text{sgn}}
    (n+\tfrac{1}{2}) - E\big((n+\tfrac{1}{2})\, \sqrt{2 \Im \tau}\big)\Big] \,
  \mathrm{e}^{-\pi i \tau (n+\frac{1}{2})^2}\,,
  \end{equation}
  with $E(z)=1-\mathop{\text{erfc}}(\sqrt{\pi} z)$.
  Let $f(u,\, v \mid \tau)$ be the difference of the LHS and the RHS of~\eqref{eq:short-N=4-character-abstruse-identity}, where $u=\frac{1}{2}(\dualtau+\dualz)$ and $v=\frac{1}{2}(\dualtau-\dualz)$. Since the additive piece $R(\tau)$ would cancel out of this difference thanks to the Jacobi abstruse identity~\eqref{eq:Jacobi-abstruse-identity}, we could also replace $\mu_a \mapsto \hat{\mu}_a$ in the difference. This shows that the difference is both holomorphic and has good transformation properties. Explicitly, by using the quasiperiodicity properties of Jacobi theta functions, as well as the double periodicity of $\hat{\mu}_1$, we have for $m,\,n,\,r,\,s \in \ZZ$,
  \begin{align}
  f(u+n\tau+m,v+r \tau+s \mid \tau) & =\mathrm{e}^{-2\pi i (n^2\tau+2n u)-2\pi i (r^2\tau+2r v)}f(u,v \mid \tau) \,. \label{eq:f-ellipticity}
  \end{align}
  The power of $\eta(\tau)$ in~\eqref{eq:short-N=4-character-abstruse-identity} was chosen such that only integer exponents of $q$ appear in the expression and thus under $\tau \to \tau+1$, the difference is clearly invariant. Finally, under an S-modular transformation, one has thanks to $\hat{\mu}_1(\frac{z}{\tau} \mid -\frac{1}{\tau})=-\sqrt{-i \tau} \hat{\mu}_1(z \mid \tau)$, $\eta(-\frac{1}{\tau})=\sqrt{-i \tau} \eta(\tau)$ and $\vartheta_1(\frac{z}{\tau},-\frac{1}{\tau})=-i \mathrm{e}^{\frac{\pi i z^2}{\tau}} \sqrt{-i \tau} \vartheta_1(z, \tau)$ as well as similar identities for the other theta functions and $\hat{\mu}_a$ obtained by shifting the arguments by half-units,
  \begin{equation}
  f(\tfrac{u}{\tau},\tfrac{v}{\tau} \mid -\tfrac{1}{\tau})= \tau^{-2} \mathrm{e}^{\frac{2\pi i}{\tau} (u^2+v^2)} f(u,v \mid \tau)\,,
  \end{equation}
  which shows that
  \begin{equation}
  f(\tfrac{u}{c\tau+d},\tfrac{v}{c\tau+d} \mid \tfrac{a \tau+b}{c\tau+d})= (c\tau+d)^{-2} \mathrm{e}^{\frac{2\pi ic}{c\tau+d} (u^2+v^2)} f(u,v \mid \tau) \,. \label{eq:f-modular-transformation}
  \end{equation}
  Let $J^{\text{wk}}_{-2,(1,1)}$ denote the space of weak Jacobi forms of weight $-2$ and index~$1$ in each of the two elliptic variables separately, i.e., of holomorphic functions $f(u,v \mid \tau)$ obeying~\eqref{eq:f-ellipticity} and~\eqref{eq:f-modular-transformation} whose Fourier expansion involves only non-negative powers of $q$. In the terminology of the mathematics literature, these are the weak Jacobi forms of lattice index $L \cong \ZZ^2$ and associated quadratic form $Q(u,v)=u^2+v^2$, which appears in the exponents of~\eqref{eq:f-ellipticity} and~\eqref{eq:f-modular-transformation}~\cite{Gritsenko:1988}.

  The space $J^{\text{wk}}_{-2,(1,1)}$ can be determined explicitly. Recall first the Eichler--Zagier structure theorem for weak Jacobi forms in a single elliptic variable~\cite{Eichler:1985ja}: the bigraded ring of weak Jacobi forms of even weight is the polynomial ring $J^{\text{wk}}_{*,*}=\CC[E_4,E_6,\phi_{-2,1},\phi_{0,1}]$, with the two generators
  \begin{equation}
  \phi_{-2,1}(z \mid \tau)=\frac{\vartheta_1(z \mid \tau)^2}{\eta(\tau)^6}\,, \qquad
  \phi_{0,1}(z \mid \tau)=4 \sum_{a=2}^4 \frac{\vartheta_a(z \mid \tau)^2}{\vartheta_a(\tau)^2}
  \label{eq:EZ-generators}
  \end{equation}
  of index~$1$ and weight $-2$ and $0$, respectively.\footnote{The sign of $\phi_{-2,1}$ differs from the formula usually found in the literature because of our convention for $\vartheta_1$.} In particular, $J^{\text{wk}}_{*,1}=M_* \, \phi_{-2,1} \oplus M_*\, \phi_{0,1}$ is free of rank two over the ring of modular forms $M_*=\CC[E_4,E_6]$. Recall second that for an index which is a direct sum, weak Jacobi forms simply factorize~\cite{WangWilliams:2021},
  \begin{equation}
  J^{\text{wk}}_{*,1} \otimes_{M_*} J^{\text{wk}}_{*,1} \xrightarrow{\ \sim\ } J^{\text{wk}}_{*,(1,1)} \,, \qquad
  (f \otimes g)(u,\, v\mid \tau) := f(u \mid \tau)\, g(v \mid \tau) \,.
  \label{eq:tensor-decomposition}
  \end{equation}
  Since there are no modular forms of weight $2$, the only elements of total weight $-2$ are the two products of a generator of weight $-2$ with a generator of weight $0$, and hence
  \begin{equation}
  J^{\text{wk}}_{-2,(1,1)}=\CC\, \phi_{-2,1}(u \mid \tau)\, \phi_{0,1}(v \mid \tau)\ \oplus\
  \CC\, \phi_{0,1}(u \mid \tau)\, \phi_{-2,1}(v \mid \tau)
  \label{eq:two-dimensional-space}
  \end{equation}
  is two-dimensional.
  It remains to check that $f$ is weak, and not merely weakly holomorphic. This is manifest from~\eqref{eq:short-N=4-character-abstruse-identity}: with the powers of $\eta(\tau)$ as chosen there, each individual term starts at order $q^0$.

  We may therefore write $f=c_+\, \phi_{-2,1}(u \mid \tau)\, \phi_{0,1}(v \mid \tau)+c_-\, \phi_{0,1}(u \mid \tau)\, \phi_{-2,1}(v \mid \tau)$ for two constants $c_\pm \in \CC$. By comparing two suitable Fourier coefficients, one sees that $c_+=c_-=0$ and thus $f \equiv 0$.  This proves
  ~\eqref{eq:short-N=4-character-abstruse-identity}.

  \section{The false theta function and its modular completion}\label[appendix]{app:false}

  In the main text we introduced a modification to the standard level-$k$ theta function~\eqref{eq:levelk-theta}. We now establish its relation to false-theta functions studied in the mathematics literature. To orient the reader, we rewrite the definition~\eqref{eq:false-theta-function} again here
  \begin{equation}
  \Psi_{k,\ell}(\dualz \mid \tau)
  = \sum_{m\in \mathbb{Z}} \, \mathop{\text{sgn}}(m + u_{k,\ell} )
  \, \dualy^{k\,\abs{m + u_{k,\ell}}}\,q^{k\,(m + u_{k,\ell})^2}\,.
  \end{equation}
  We introduced in the above a parameter $u_{k,\ell}$, which is defined as
  \begin{equation}
  u_{k,\ell} = \frac{\ell + \frac{1}{2}}{k} \,, \qquad 0 < u_{k,\ell} < 1\,.
  \end{equation}
  The constraint on $u_{k,\ell}$ follows from the allowed range of $\ell$.

  To understand the function $\Psi_{k,\ell}$ consider symmetrizing and antisymmetrizing it
  under $\dualz \to - \dualz$. For the antisymmetric combination, we see that one has the same information as the standard level-$k$ theta function~\eqref{eq:false-theta-asym}, viz.,
  \begin{equation}
  \Psi_{k,\ell}(\dualz \mid \tau) - \Psi_{k,\ell}(-\dualz \mid \tau)
  = \Theta_{k,\ell}(\dualz \mid \tau) - \Theta_{k,\ell}(-\dualz \mid \tau) \ .
  \end{equation}
  The symmetric combination is new, and can be written as
  \begin{equation}
  \begin{aligned}
  \Psi^{\mathrm{sym}}_{k,\ell}(\dualz \mid \tau)
   & =
  \Psi_{k,\ell}(\dualz \mid \tau) + \Psi_{k,\ell}(-\dualz \mid \tau)
  \\
   & = \sum_{m\in \mathbb{Z}} \,  \mathop{\text{sgn}}(m + u_{k,\ell} )
  \, \bqty{\dualy^{k\,\abs{m + u_{k,\ell}}} + \dualy^{-k\,\abs{m + u_{k,\ell}}}  }
  \,q^{k\,(m + u_{k,\ell})^2}\,.
  \end{aligned}
  \end{equation}
  Therefore, for understanding the nature of the function $\Psi_{k,\ell}$, it suffices to identify the symmetric part.

  \paragraph{A special case ($k=2$):}  First, focus on $k = 2$,
  whence only $\ell = 0$ is allowed. This implies
  \begin{equation}
  \begin{aligned}
  \Psi^{\mathrm{sym}}_{2,0}(\dualz \mid \tau)
   & =
  \sum_{m\in \mathbb{Z}}\, \mathop{\text{sgn}}(2\,m + \tfrac{1}{2}) \,
  \, \bqty{ \dualy^{\abs{2\,m + \tfrac{1}{2}}} + \dualy^{-\abs{2\,m +\tfrac{1}{2}}} }\,
  q^{\frac{1}{2}\, (2\,m+ \tfrac{1}{2})^2}
  \\
   & =
  \sum_{r \in \mathbb{Z}_{\geq 0}}\, (-1)^{r}\, \bqty{\dualy^{r + \tfrac{1}{2}} +
    \dualy^{-r - \frac{1}{2}}}\, q^{\frac{1}{2}\, \pqty{r + \tfrac{1}{2}}^2} \,.
  \end{aligned}
  \label{eq:Psi-sym-simplification}
  \end{equation}
  To obtain the second line, we simply redefined $r = 2\,m$ for $m \geq 0$ or
$r = -(2m +1)$ for $m < 0$.

  Consider now the false theta function in two variables defined in~\cite{Bringmann:2019vyd},\footnote{We have dropped a factor of $i$ from the definition of the function $\psi(\dualz \mid \tau)$, just as we did for the Jacobi theta function $\vartheta_1(\dualz \mid \tau)$. }
  \begin{equation}
  \psi(\dualz \mid \tau) = \sum_{n \in \mathbb{Z}}\, \mathop{\text{sgn}}(n + \tfrac{1}{2})\,
  (-1)^n \, \dualy^{n + \tfrac{1}{2}}\, q^{\frac{1}{2}\, (n + \tfrac{1}{2})^2}\,.
  \label{eq:BN-false-theta}
  \end{equation}
  This can be rewritten by separating $n \geq 0$ and $n < 0$ terms and relabeling the summation variable appropriately as the last line of~\eqref{eq:Psi-sym-simplification}. Therefore, the symmetric part of our modified theta function is the prototype false theta function considered in~\cite{Bringmann:2019vyd}. To wit,
  \begin{equation}
  \Psi^{\mathrm{sym}}_{2,0}(\dualz \mid \tau)
  = \psi(\dualz\mid \tau) \,.
  \end{equation}

  The function $\Psi_{2,0}^{\mathrm{sym}}(\dualz \mid \tau)$ is not modular. It is covariant under $\tau \to \tau +1$ by construction (transforms by a phase $e^{i \frac{\pi}{4}}$), but not under $\tau \to -\frac{1}{\tau}$. The modular completion is established in~\cite[Theorem 1.1]{Bringmann:2019vyd} following~\cite{Zwegers:2002kt}. Define for $\dualz \in \mathbb{R}$, a new object
  \begin{equation}
  \widehat{\psi}(\dualz \mid \tau; w)
  = \sum_{n \in \mathbb{Z}}\,
  (-1)^n\,  \dualy^{n + \tfrac{1}{2}}\, q^{\frac{1}{2}\, \pqty{n + \tfrac{1}{2}}^2}\,
  \mathop{\text{erf}}\bigg(-i\,\sqrt{\pi \, i\, (w -\tau)} \, ( n + \tfrac{1}{2})\bigg)\,.
  \end{equation}
  This function is shown to be modular under simultaneous $\mathrm{PSL}(2,\mathbb{Z})$ transformations of both $\tau$ and $w$. Furthermore, $\widehat{\psi}(\dualz\mid \tau; w = \tau + i\, t+ \epsilon) \to \psi(\dualz \mid \tau)$ as $t \to \infty$. In our application, rather than focus on the full modular completion, we are exploiting a particular smoothing section by setting $w = \tau + i\, \Im(\tau)$, analogous to~\cite{Zwegers:2002kt}.

  For our applications, using  $\mathop{\text{sgn}}(z) \, \mathop{\text{erf}}(\abs{z}) = \mathop{\text{erf}}(z)$, one can show,
  \begin{equation}
  \widehat{\psi}(\dualz \mid \tau) - \psi(\dualz \mid \tau)
  = \sum_{n \in \mathbb{Z}}\, \mathop{\text{sgn}}(n + \tfrac{1}{2})\, (-1)^n\,
  \dualy^{n+ \tfrac{1}{2}}\, q^{\frac{1}{2}\, \pqty{n + \tfrac{1}{2}}^2}\,
  \mathop{\text{erfc}}\big(\sqrt{\pi \,\tau_2}\, \abs{n + \tfrac{1}{2}}\big) \ .
  \end{equation}

  \paragraph{The level-$k$ false theta:} The general level-$k$ false-theta function is not defined as a function of two variables in~\cite{Bringmann:2019vyd}. Instead, they define
  \begin{equation}
  F_{j,N} (\tau) =
  \sum_{\substack{n \in \mathbb{Z}\\
    n \equiv j \; (\mathrm{mod}\; 2N)}}\,
  \mathop{\text{sgn}}(n)\, \, q^{\frac{1}{4\,N}\, n^2}\,.
  \end{equation}
  This can be seen to be related to $\Psi_{k,\ell}$ via
  \begin{equation}
  \Psi_{k,\ell}(0\mid \tau) = F_{2\,\ell +1, k} (\tau)\,,
  \end{equation}
  suggesting that the symmetric part of the level-$k$ false theta function is precisely the natural generalization $F_{j,N}(\dualz\mid \tau)$. One may proceed as above, using the observation that the modular completion replaces the sign function by an appropriate error function. One may then argue that the requisite modular completed function is
  \begin{equation}
  \widehat{\Psi}_{k,\ell}(\dualz \mid \tau)
  =
  \sum_{m\in \mathbb{Z}}\, \mathop{\text{erf}}\,
  \pqty{ (m + u_{k,\ell})\,
    \sqrt{2\pi\,k\,\Im(\tau)}}\,
  \dualy^{k\, \abs{m + u_{k,\ell}}} \,
  q^{k\, (m + u_{k,\ell})^2} \,.
  \end{equation}
  %

  \section{Toy models for the worldsheet moduli space integration}\label[appendix]{app:toymodels}

  Consider the integral
  \begin{equation}
  I(\dualtau)
  =\sqrt{k} \Im \dualtau\,
  \int_{-\frac{1}{2}}^{\frac{1}{2}}\dd\tau_1\int_0^\infty \frac{\dd\tau_2}{\tau_2^{3/2}}\,
  \mathrm{e}^{ -\frac{k\pi (\Im \dualtau)^2}{\tau_2}}\,\frac{\eta(\tau)^3}{\vartheta_1(\dualtau\mid \tau)} \, F(\tau)\,,
  \label{eq:toy-example}
  \end{equation}
  which is~\eqref{eq:EG-after-specialization} without the anti-holomorphic false theta function and with a general function $F(\tau)$ (that may depend on other parameters such as $\dualtau$ or $\dualz$), up to an overall minus sign, which we reinstate when we return to the elliptic genus. We only assume for now that $F(\tau)$ has a regular $q$ expansion converging in the upper half plane, $F(\tau)=\sum_{N=0}^\infty a_N q^N$.

  \paragraph{Derivation 1:} The reason for factoring out $\frac{\eta(\tau)^3}{\vartheta_1(\dualtau\mid \tau)}$ in the integrand is that we can make use of the expansions~\eqref{eq:theta1Fourier} with the constraint~\eqref{eq:wcells}.
  Using this expansion allows us to treat the poles of $\vartheta_1(\dualtau \mid \tau)$ on the integration region in a simple way. We subdivide the integration region of~\eqref{eq:toy-example} and use this expansion on each subregion, which leads to
  \begin{equation}
  I(\dualtau)
  =\sqrt{k} \Im \dualtau
  \sum_{w=0}^\infty \sum_{n \in \mathbb{Z}}\, \sum_{d\geq  w +1}\sum_{N=0}^\infty a_N (-1)^{d}\, \dualq^{n - \frac{1}{2}} \int_{-\frac{1}{2}}^{\frac{1}{2}}\dd\tau_1\int_{\frac{\Im \dualtau}{w+1}}^{\frac{\Im \dualtau}{w}} \frac{\dd\tau_2}{\tau_2^{3/2}}\,
  \mathrm{e}^{ -\frac{k\pi (\Im \dualtau)^2}{\tau_2}}\, q^{\frac{d}{2}(d+1-2n)+N} \,.
  \end{equation}
  The integral over $\tau_1$ imposes level-matching as usual, and hence requires vanishing of the $q$-exponent. We solve this condition for $n$, which is essentially the exponent of $\dualq$; this gives us the condition
  \[
  n\, d = \frac{d(d+1)}{2}+N \,.
\]
Solvability of this constraint requires the RHS to be divisible by $d$ (since $n \in \mathbb{Z}$). We can implement the constraint by introducing another sum
\[
  \frac{1}{d} \sum_{b=0}^{d-1} \mathrm{e}^{\frac{2\pi ib}{d}\big(\frac{d(d+1)}{2}+N\big)}\,.
\]
Putting the above together leads to the following expression for the integral
\begin{equation}
I(\dualtau)=\sqrt{k} \Im \dualtau\sum_{d=1}^\infty \frac{1}{d} \sum_{b=0}^{d-1} \sum_{N=0}^\infty a_N (-1)^{b+d}\mathrm{e}^{2\pi i\frac{ \dualtau+b}{d} \big(\frac{d^2}{2}+N\big)}\sum_{w=0}^{d-1} \int_{\frac{\Im \dualtau}{w+1}}^{\frac{\Im \dualtau}{w}} \frac{\dd\tau_2}{\tau_2^{3/2}}\,
\mathrm{e}^{ -\frac{k\pi (\Im \dualtau)^2}{\tau_2}}\,.
\end{equation}
In writing this formula we have exchanged the order of the sums; specifically, we have first evaluated the integral for fixed $d$, and then  summed over the different values of $d$.

We now realize that the sum over $N$ can be easily performed. It reassembles the summand into the function $F(\frac{\dualtau+b}{d})$. In addition, the sum over $w$ telescopes. We thus end up with the formula
\begin{equation}
\begin{aligned}
I(\dualtau)
 & =
\sqrt{k} \Im \dualtau\sum_{d=1}^\infty \,\frac{1}{d} \,
\sum_{b=0}^{d-1} \,  (-1)^{b+d} q^{\frac{d^2}{2}} \,
F(\tau)\bigg|_{\tau=\frac{\dualtau+b}{d}}
\, \int_{\frac{\Im \dualtau}{d}}^{\infty} \frac{\dd\tau_2}{\tau_2^{3/2}}\,
\mathrm{e}^{ -\frac{k\pi (\Im \dualtau)^2}{\tau_2}}
\\
 & =
\sum_{d=1}^\infty \,\frac{1}{d}\,
\sum_{b=0}^{d-1} \, (-1)^{b+d} \erf(\sqrt{k\pi d \Im \dualtau}) \,
q^{\frac{d^2}{2}}\, F(\tau)\bigg|_{\tau=\frac{\dualtau+b}{d}} \,,
\end{aligned}
\end{equation}
where we evaluated the integral explicitly in the last line in terms of an error function.

\paragraph{Derivation 2:} There is in fact a much quicker way to arrive at the same answer. This follows once one notices that
\begin{equation}
\frac{(-1)^{b+d}}{d} q^{\frac{d^2}{2}} F(\tau)\bigg|_{\tau=\frac{\dualtau+b}{d}}=-2\pi i \Res_{\tau=\frac{\dualtau+b}{d}} \frac{\eta(\tau)^3}{\vartheta_1(\dualtau \mid \tau)} \, F(\tau)\,,
\end{equation}
where the sign originates from $\vartheta_1'(0 \mid \tau)=2\pi i\, \eta(\tau)^3$ in our conventions, combined with $\partial_\tau(\dualtau+b-d\,\tau)=-d$. We may thus write the result also as
\begin{equation}
I(\dualtau)=-2\pi i \sum_{d=1}^\infty \sum_{b=0}^{d-1}  \erf(\sqrt{k\pi d \Im \dualtau}) \, \Res_{\tau=\frac{\dualtau+b}{d}} \frac{\eta(\tau)^3}{\vartheta_1(\dualtau \mid \tau)} \, F(\tau)\,. \label{eq:toy-model-final-result}
\end{equation}

The form~\eqref{eq:toy-model-final-result} makes it clear that we can also give a more conceptual derivation of the same result, which can be done as follows. For this, we first compute the $\tau_1$-integral in~\eqref{eq:toy-example}, which is
\begin{equation}
\int_{-\frac{1}{2}}^{\frac{1}{2}} \dd \tau_1 \, \frac{\eta(\tau)^3}{\vartheta_1(\dualtau \mid \tau)}\, F(\tau)=\int_{i\tau_2-\frac{1}{2}}^{i \tau_2+\frac{1}{2}} \dd \tau\,  \frac{\eta(\tau)^3}{\vartheta_1(\dualtau \mid \tau)}\, F(\tau)\,.
\end{equation}
Here, we consider $\tau_2$ to be fixed and no longer necessarily equal to $\Im \tau$ once we start to deform the contour.
We reinterpreted the $\tau_1$-integral as a holomorphic contour integral. We can compute this integral by  pushing the contour \emph{down} to small values of $\Im \tau$. In the process, we pick up all the residues of the integrand with $\Im \tau<\tau_2$, which are located at $\tau=\frac{\dualtau+b}{d}$ for $d > \frac{\Im \dualtau}{\tau_2}$. This potentially leaves a residual  boundary term
\begin{equation}
\lim_{\tau_2 \to 0} \int_{i \tau_2-\frac{1}{2}}^{i \tau_2+\frac{1}{2}}\dd \tau \frac{\eta(\tau)^3}{\vartheta_1(\dualtau \mid \tau)} \, F(\tau) \,.
\label{eq:boundary term}
\end{equation}
We will at first neglect this boundary term. The formula will hold under the condition that it vanishes.
Since pushing the contour downwards traverses the boundary of the enclosed region in the clockwise sense, the residues are picked up with a factor of $-2\pi i$; this is opposite to the upward deformation employed in~\eqref{eq:boundary-contribution-residue-theorem} below. This leads to
\begin{equation}
\begin{aligned}
I(\dualtau)
 &
=-2\pi i\, \sqrt{k}\, \Im \dualtau \int_0^\infty \,
\frac{\dd \tau_2}{\tau_2^{3/2}}\,
\mathrm{e}^{ -\frac{k\pi (\Im \dualtau)^2}{\tau_2}} \,
\sum_{d\ge \lceil\frac{\Im \dualtau}{\tau_2}\rceil} \,
\sum_{b=0}^{d-1} \, \Res_{\tau=\frac{\dualtau+b}{d}}
\frac{\eta(\tau)^3}{\vartheta_1(\dualtau \mid \tau)}\, F(\tau)
\\
 & =
-2\pi i\, \sqrt{k}\, \Im \dualtau \sum_{d=1}^\infty\,
\sum_{b=0}^{d-1}\,
\Res_{\tau=\frac{\dualtau+b}{d}} \,
\frac{\eta(\tau)^3}{\vartheta_1(\dualtau \mid \tau)}\, F(\tau)
\,\int_{\frac{\Im \dualtau}{d}}^\infty\frac{\dd \tau_2}{\tau_2^{3/2}}\,
\mathrm{e}^{ -\frac{k\pi (\Im \dualtau)^2}{\tau_2}} \,.
\end{aligned}
\end{equation}
Upon evaluating the remaining integral in terms of the error function, we
recover~\eqref{eq:toy-model-final-result}.

\paragraph{Analysis of boundary term:}
Let us now turn to analyze whether our assumption to drop the boundary term~\eqref{eq:boundary term} was justified. The most convenient way to compute the boundary term is actually by comparing to the result that we would have obtained when deforming the contour upwards. Doing so gives the boundary term
\begin{equation}
\frac{a_0}{\mathrm{e}^{\pi i \dualtau}-\mathrm{e}^{-\pi i \dualtau}}=\int_{i \infty-\frac{1}{2}}^{i \infty+\frac{1}{2}} \dd \tau \frac{\eta(\tau)^3}{\vartheta_1(\dualtau \mid \tau)}\, F(\tau)\ .
\end{equation}
Thus, we have by the residue theorem
\begin{equation}
\begin{aligned}
\lim_{\tau_2 \to 0} \int_{i \tau_2-\frac{1}{2}}^{i \tau_2+\frac{1}{2}}\dd \tau \frac{\eta(\tau)^3}{\vartheta_1(\dualtau \mid \tau)} \, F(\tau)
 & =
\frac{a_0}{\mathrm{e}^{\pi i \dualtau}-\mathrm{e}^{-\pi i \dualtau}}+2\pi i \lim_{d_\text{max} \to \infty} \sum_{d=1}^{d_\text{max}}\sum_{b=0}^{d-1} \LargeRes_{\tau=\frac{\dualtau+b}{d}} \frac{\eta(\tau)^3}{\vartheta_1(\dualtau \mid \tau)} \, F(\tau)
\\
 & =
\frac{a_0}{\mathrm{e}^{\pi i \dualtau}-\mathrm{e}^{-\pi i \dualtau}}+\lim_{d_\text{max} \to \infty}\sum_{d=1}^{d_\text{max}}\sum_{b=0}^{d-1} (-1)^{(b+1)(d+1)} \frac{\mathrm{e}^{\pi i d \dualtau}}{d} F\big(\tfrac{\dualtau+b}{d}\big)\ .
\label{eq:boundary-contribution-residue-theorem}
\end{aligned}
\end{equation}
In this form, we can analyze the boundary contribution explicitly, by using the series expansion for $F(\tau)$.
Assume first that the sums over $d$, $b$, as well as $N$, in the series expansion of $F$ converge absolutely. Then we may freely reorder the sum and in particular collect the sum participating in the coefficient $a_N$, which enters as
\begin{align}
a_N \sum_{d=1}^{\infty} \,\frac{1}{d}\,
\sum_{b=0}^{d-1} \, (-1)^{(b+1)(d+1)} \,
\mathrm{e}^{\pi i d \dualtau}\mathrm{e}^{\frac{2\pi i N}{d}(\dualtau+b)}
= a_N
\sum\limits_{\begin{subarray}{c} d=1 \\ N+\frac{d(d+1)}{2} \,\equiv\, 0 \bmod d \end{subarray}}^{\infty} (-1)^{d+1} \, \mathrm{e}^{\pi i (d+\frac{2N}{d}) \dualtau}\,.
\end{align}
The sum over $b$ projects to the condition $N+\frac{d(d+1)}{2} \equiv 0 \bmod d$ (and in particular $d \le 2N$). For $N>0$, it is easy to check that if $d$ satisfies this condition, then so does $d'=\frac{2N}{d}$. Furthermore, $d' \ne d$ and $(-1)^{d+1}=-(-1)^{d'+1}$. Thus, the two contributions for $N>0$ pairwise cancel. For $N=0$, the sum over $b$ projects to odd $d$ and
\begin{equation}
a_0 \lim_{d_\text{max} \to \infty} \sum_{d=1,\, \text{$d$ odd}}^{d_\text{max}} \mathrm{e}^{\pi i d \dualtau}=-\frac{a_0}{\mathrm{e}^{\pi i \dualtau}-\mathrm{e}^{-\pi i \dualtau}}\ ,
\end{equation}
which cancels the corresponding term in~\eqref{eq:boundary-contribution-residue-theorem}.

Thus, the boundary term vanishes if the sum is absolutely convergent, which amounts to the condition
\begin{equation}
\sum_{d=1}^\infty \sum_{N=1}^\infty |a_N|\mathrm{e}^{-\pi (d+\frac{2N}{d}) \Im \dualtau} <\infty\,.
\end{equation}
For fixed $N$, the sum over $d$ is maximized by choosing $d$ close to $\sqrt{2N}$. We can estimate the sum over $d$ from above by replacing it by an integral and applying a saddle-point approximation. So a sharp sufficient criterion is
\begin{equation}
\sum_{N=1}^\infty |a_N| N^{\frac{1}{4}} \mathrm{e}^{-2\sqrt{2N}\pi \Im \dualtau}<\infty\ .
\end{equation}

\paragraph{Convergence  of the sums:}  To better understand the convergence of the sums in an example closer to the elliptic genus evaluation, consider the choice
\begin{align}
F(\tau)=\frac{\vartheta_1(\dualtau \mid \tau)}{\eta(\tau)^3}\ .
\end{align}
We also put $k=1$ for simplicity.
The integral~\eqref{eq:toy-example}  is now trivial to evaluate and results in
$I(\dualtau)=1$. Let us instead proceed with our evaluation as above and see what we find. We are stymied at the first step, for
\begin{equation}
F(\tfrac{\dualtau+b}{d})=\frac{\vartheta_1(\dualtau \mid \frac{\dualtau+b}{d})}{\eta(\frac{\dualtau+b}{d})^3}=0\,,
\end{equation}
so commuting the limit with the sum over $d$ is illegal. Let us therefore be more careful and use a regulator $\varepsilon$. Set
\begin{equation}
F(\tau)=\frac{\vartheta_1(\dualtau +\varepsilon\mid \tau)}{\eta(\tau)^3}\ ,
\end{equation}
and expand in $\varepsilon$. We now have
\begin{align}
F(\tfrac{\dualtau+b}{d})=\frac{\vartheta_1(\dualtau +\varepsilon\mid \frac{\dualtau+b}{d})}{\eta(\frac{\dualtau+b}{d})^3}
= (-1)^{b+d+bd} \, \mathrm{e}^{-2\pi i d \varepsilon-\pi i d \dualtau} \,
\frac{\vartheta_1(\varepsilon \mid \frac{\dualtau+b}{d})}{\eta(\frac{\dualtau+b}{d})^3}\,.
\end{align}
This then leads to the answer
\begin{align}\label{eq:toy-exampleII-sumA}
I(\dualtau)
=\sum_{d=1}^\infty  \,\frac{1}{d} \,\sum_{b=0}^{d-1}\,
\mathrm{e}^{-2\pi i d \varepsilon} \frac{\vartheta_1(\varepsilon \mid \frac{\dualtau+b}{d})}{\eta(\frac{\dualtau+b}{d})^3} \, \erf(\sqrt{\pi d \Im \dualtau})\,.
\end{align}

From this expression it is clear that convergence in $d$ becomes very slow. The sum converges as long as $\Im \varepsilon<0$. We can expand around this point---to leading order $\frac{\vartheta_1(\varepsilon \mid \frac{\dualtau+b}{d})}{\eta(\frac{\dualtau+b}{d})^3} \sim 2\pi i \varepsilon$. The sum will then be dominated by large $d$, since the sum has to supply a $1/\varepsilon$ divergence to compensate for term by term vanishing. Since this cannot arise from the error function, we can replace $\erf(\sqrt{\pi d \Im \dualtau}) \to 1$. Doing so leads us to
\begin{equation}\label{eq:toy-exampleII-result}
I(\dualtau) =
2\pi i \,\varepsilon \, \sum_{d=1}^\infty\, \mathrm{e}^{-2\pi i d \varepsilon} \to 1\ ,
\end{equation}
as expected.

We can try to understand the convergence of the sum a bit better as follows. Once we sum over $w$, we have a sum over $d\in \mathbb{Z}_{\geq 1}$, which as we saw above required regulating. To make this explicit, let us introduce a formal parameter $p$ as suggested by the DMVV analogy. Consider instead of~\eqref{eq:toy-exampleII-sumA} the following sum
\begin{equation}
I(p,\dualtau)=\sum_{d=1}^\infty \, \frac{p^d}{d} \sum_{b=0}^{d-1}\,
\mathrm{e}^{-2\pi i d \varepsilon} \,
\frac{\vartheta_1(\varepsilon \mid \frac{\dualtau+b}{d})}{\eta(\frac{\dualtau+b}{d})^3}
\, \erf(\sqrt{\pi d \Im \dualtau})\,.
\end{equation}
Evaluating the sum over $b$ as before, after simplifying the small $\varepsilon$ limit of the summand, we have
\begin{equation}
I(p,\dualtau)= 2\pi i  \varepsilon\,
\sum_{d=1}^\infty \, (p\, e^{-2\pi i \varepsilon })^d
\erf(\sqrt{\pi d \Im \dualtau}) + \order{\varepsilon^3}\,.
\end{equation}
As long as $\abs{p} <1$ the sum is convergent and gives
\begin{equation}
I(p,\dualtau)=  2\pi i  \varepsilon\, \frac{p\, e^{-2\pi i \varepsilon }}{1- p\, e^{-2\pi i \varepsilon }} +
2\pi i  \varepsilon\, \sum_{d=1}^\infty \, (p\, e^{-2\pi i \varepsilon })^d
\bqty{
  \erf(\sqrt{\pi d \Im \dualtau}) -1} + \order{\varepsilon^3}\,.
\end{equation}
Since the error function deviates from unity at order $\frac{1}{\sqrt{d}}\, e^{-\pi d\, \Im \dualtau}$, the second term may be ignored. The first term clearly illustrates the issue: the limits $\varepsilon \to 0$ and $p\to 1^-$ do not commute. Indeed, for $\abs{p} < 1$ the limit $\varepsilon \to 0$ says that the sum evaluates to zero, but for $p = 1$, the limit attained is unity, as expected from~\eqref{eq:toy-exampleII-result}.

\paragraph{Generalization to non-holomorphic integrands:} The integrand of the elliptic genus computation is non-holomorphic, owing to the right-moving false theta function, cf.~\cref{eq:EG-after-specialization}. It  is straightforward to generalize the analysis above to such cases. For illustration, it suffices to include an additional $\bar{q}^{A}$ in the integrand. We assume $A>0$ for convergence. Here $A$ need not be integer, but the combination $F(\tau)\, \bar{q}^A$ has to be invariant under $\tau \to \tau+1$.

The contour deformation argument presented for the holomorphic integrands goes through for this case.  For fixed $\tau_2$, we can write $\bar{q}^A=\mathrm{e}^{-4\pi A\tau_2} q^{-A}$ and effectively make
$q^{-A}$ part of $F(\tau)$. For the left-over boundary term, the inclusion of
$q^{-A}$ has no effect. Indeed, on the contour $\Im \tau=\tau_2$ one has $\abs{q^{-A}}=\mathrm{e}^{2\pi A \tau_2}\to 1$ as $\tau_2 \to 0$, so that
\[
  \abs{\int_{i \tau_2-\frac{1}{2}}^{i \tau_2+\frac{1}{2}}\dd \tau \,
    \frac{\eta(\tau)^3}{\vartheta_1(\dualtau \mid \tau)}\, q^{-A}\, F(\tau)}
  \;\le\;
  \mathrm{e}^{2\pi A \tau_2}\int_{-\frac{1}{2}}^{\frac{1}{2}} \dd \tau_1 \,
  \abs{\frac{\eta(\tau)^3}{\vartheta_1(\dualtau \mid \tau)}\, F(\tau)}\,.
\]
The bound on the right is independent of $A$, so verifying it is the same computation as in the case $A=0$. The formula will therefore hold for any $A>0$ under the assumption that the boundary term~\eqref{eq:boundary term} vanishes with the absolute value taken inside the integral.\footnote{Here, as in~\eqref{eq:boundary term}, the limit $\tau_2 \to 0$ is understood along a sequence avoiding the heights $\tau_2=\frac{\Im \dualtau}{d}$, at which the contour meets the poles of $\vartheta_1(\dualtau\mid \tau)^{-1}$.} One gets instead
\begin{align}
I(\dualtau)
 & =
-2\pi i\,\sqrt{k} \Im \dualtau
\sum_{d=1}^\infty \,\sum_{b=0}^{d-1} \,
\Res_{\tau=\frac{\dualtau+b}{d}}
\bigg(\frac{\eta(\tau)^3}{\vartheta_1(\dualtau \mid \tau)}\, q^{-A} F(\tau) \bigg)\,
\int_{\frac{\Im \dualtau}{d}}^\infty\,
\frac{\dd \tau_2}{\tau_2^{3/2}}\,
\mathrm{e}^{ -\frac{k\pi (\Im \dualtau)^2}{\tau_2}-4\pi\, A \,\tau_2} \,.
\label{eq:generalized-toy-model}
\end{align}
To complete the final integral one defines `light-cone' coordinates
\[
  x_\pm=\sqrt{4\pi \,A \,\tau_2}\mp \frac{\sqrt{k \,\pi}\,\Im \dualtau}{\sqrt{\tau_2}}\,,
\]
to rewrite the integrand as
\begin{equation}
\sqrt{k} \Im \dualtau \, \frac{\dd \tau_2}{\tau_2^{3/2}}\,
\mathrm{e}^{ -\frac{k\pi (\Im \dualtau)^2}{\tau_2}-4\pi\, A\, \tau_2} =\frac{1}{\sqrt{\pi}} \sum_{\beta=\pm} \beta \, \dd x_\beta \,
\mathrm{e}^{-x_\beta^2 - 4\pi\, \beta\,\sqrt{k\, A} \,\Im \dualtau}\,.
\end{equation}
With this change the integrals are standard error function integrals (the integral depends only on the endpoints of $x_\beta$, which suffices, even though $x_-$ is not monotonic in $\tau_2$). One ends up with
\begin{equation}
\sqrt{k} \Im \dualtau \,
\int_{\frac{\Im \dualtau}{d}}^\infty \, \frac{\dd \tau_2}{\tau_2^{3/2}}\,
\mathrm{e}^{ -\frac{k\pi \,(\Im \dualtau)^2}{\tau_2}-4\pi\, A\, \tau_2}
=
\frac{1}{2}\sum_{\beta=\pm} \beta\,
\mathrm{e}^{-4\pi \,\beta \,\sqrt{A k}\, \Im \dualtau}\,\mathrm{erfc}
\bigg(\sqrt{\frac{\pi \Im \dualtau}{d}} (2\, \sqrt{A}-\beta \,d \,\sqrt{k})\bigg)\,,
\label{eq:error-function-integral}
\end{equation}
where $\mathrm{erfc}(x)=1-\mathrm{erf}(x)$ is the complementary error function.
This expression can be massaged to obtain the combination appearing in~\eqref{eq:EG-nonholo-Tdef}.

As explained in~\cref{sec:eg-pitfalls} we can use this expression for completing the integral~\eqref{eq:EG-after-specialization}, but the resulting representation is suboptimal. The evaluation in~\cref{sec:eg-good} on the other hand was carried out to ensure that we have a convergent sum, which as illustrated there, can be explicitly evaluated.


\providecommand{\href}[2]{#2}\begingroup\raggedright\endgroup

\end{document}